\documentclass[11pt]{article}

\usepackage{hyperref}
\usepackage{graphicx}
\usepackage{amssymb}
\usepackage{amsbsy}
\usepackage{float}
\usepackage{amsmath}
\usepackage{subcaption}
\usepackage{multirow}
\usepackage{setspace}
\usepackage{grffile}
\usepackage{verbatim}
\usepackage{enumerate}
\usepackage[normalem]{ulem}
\usepackage{color}
\usepackage{fancyhdr}
\usepackage[utf8]{inputenc}
\usepackage{graphicx}
\usepackage[table,xcdraw]{xcolor}
\usepackage{color, colortbl}
\usepackage{gensymb}
\usepackage[nottoc]{tocbibind}
\usepackage{mathtools}
\usepackage{amsmath}
\usepackage{booktabs}
\usepackage{multirow}
\usepackage{changepage}
\usepackage{pdflscape}
\usepackage[]{parskip}
\usepackage{listings}
\usepackage{mathtools}
\usepackage[justification=centering]{caption}
\usepackage{placeins}
\usepackage{url}
\usepackage{authblk}
\usepackage[margin=1.5cm]{geometry}
\usepackage[citestyle=numeric, sorting=none, maxbibnames=99]{biblatex}
\usepackage[export]{adjustbox}
\usepackage{makecell}

\newcommand{\ie}{{i.e.\ }}
\newcommand{\eg}{{e.g.\ }}
\newcommand{\no}{{no.\ }}

\newcommand{\figref}[1]{Figure \ref{#1}}
\newcommand{\equref}[1]{Equation \ref{#1}}
\newcommand{\tabref}[1]{Table \ref{#1}}
\newcommand{\appref}[1]{Appendix \ref{#1}}

\newcommand{\secref}[1]{Section \ref{#1}}

\usepackage[colorinlistoftodos]{todonotes}
\definecolor{lightblue}{rgb}{.90,.95,1}
\definecolor{darkgreen}{rgb}{0,.5,0.5}

\graphicspath{{images/}}

\title{Poisson CNN: Convolutional neural networks for the solution of the Poisson equation on a Cartesian mesh}

\author[1]{Ali Girayhan {\"O}zbay \footnote{Corresponding author. Email: aligirayhan.ozbay14@imperial.ac.uk}}%
\author[1]{Arash Hamzehloo}
\author[1]{Sylvain Laizet}
\author[2]{Panagiotis Tzirakis}
\author[2]{Georgios Rizos}
\author[2]{Bj{\"o}rn Schuller}

\affil[1]{Department of Aeronautics, Imperial College London, South Kensington Campus, London, SW7 2AZ, UK}

\affil[2]{Department of Computing, Imperial College London, South Kensington Campus, London, SW7 2AZ, UK}
\begin{document}
\maketitle

\noindent\hrulefill
\begin{abstract}
The Poisson equation is commonly encountered in engineering, for instance in computational fluid dynamics (CFD) where it is needed to compute corrections to the pressure field to ensure the incompressibility of the velocity field. In the present work, we propose a novel fully convolutional neural network (CNN) architecture to infer the solution of the Poisson equation on a 2D Cartesian grid with different resolutions given the right hand side term, arbitrary boundary conditions and grid parameters. It provides unprecedented versatility for a CNN approach dealing with partial differential equations. The boundary conditions are handled using a novel approach by decomposing the original Poisson problem into a homogeneous Poisson problem plus four inhomogeneous Laplace sub-problems. The model is trained using a novel loss function approximating the continuous $L^p$ norm between the prediction and the target. Even when predicting on grids denser than previously encountered, our model demonstrates encouraging capacity to reproduce the correct solution profile. The proposed model, which outperforms well-known neural network models, can be included in a CFD solver to help with solving the Poisson equation. Analytical test cases indicate that our CNN architecture is capable of predicting the correct solution of a Poisson problem with mean percentage errors below 10\%, an improvement by comparison to the first step of conventional iterative methods.  Predictions from our model, used as the initial guess to iterative algorithms like Multigrid, can reduce the RMS error after a single iteration by more than 90\% compared to a zero initial guess.
\end{abstract}

\noindent\hrulefill

\clearpage

\section{Introduction}
Partial differential equations (PDEs) describe complex systems in many fields of engineering and science, ranging from fluid flows to options pricing. Despite their ubiquity, they can be very costly to solve accurately. One of the most important PDEs in engineering is the Poisson equation, expressed mathematically as
\begin{equation}
    \label{eq:comp_poisson}
    \nabla^2  \phi = f,
\end{equation}
where $f$ is a forcing term and $\phi$ is the variable for which a solution is sought. Appearing in a diverse range of problems including heat conduction, gravitation, simulating fluid flows and electrodynamics, the Poisson problem plays a central role in the design of many modern technologies. However, solving the Poisson equation numerically for large problems involving millions of degrees of freedom is only tractable by employing iterative methods. For example, it is well known that the treatment of incompressibility is a real difficulty to obtain solutions of the incompressible Navier-Stokes equations, the mathematical model used to describe the motions of a turbulent flow. Any algorithm must ensure a divergence-free flow field at
any given time during the calculation. The unavoidable step of solving the Poisson equation, as introduced by a fractional step method per \cite{chorin1967numerical,chorin1968numerical} can be computationally very expensive (it can represent up to 90\% of the computational effort), since it is often based on the said complex iterative techniques.

One of the fastest algorithms for solving the Poisson equation iteratively, with $O(n)$ complexity per \cite{multigrid}, is the multigrid algorithm whereby the problem is solved using a number of successively coarser grids to eliminate low wavenumber error in a few iterations. However, even the multigrid algorithm can take prohibitively long amounts of time to iterate for very large problems. Convolutional neural networks (CNNs) are well positioned to accelerate such iterative methods; they are already in use in various areas of engineering and applied mathematics for complex regression and image-to-image translation tasks, have $O(n)$ runtime complexity\footnote{Convolutional neural networks have $O(n)$ (where $n$ is the product of the input size across each dimension) runtime complexity, as convolution is in fact equivalent to a banded matrix-vector product, where the banded matrix has the (flattened) convolution weights in each row  stored across the appropriate diagonals, which is known to be an $O(n)$ operation.} and can be run very efficiently on graphics processing units (GPUs) and other floating point acceleration hardware.

Leveraging these strengths, our principal motivation is to develop a convolutional neural network (CNN) based Poisson equation solver that does not require re-training to perform inference on inputs with different types of boundary conditions (BCs) within a given range of grid resolutions and sizes in the context of Cartesian grids. {In this work we present a novel CNN architecture capable of handling arbitrary right-hand side (RHS) functions $f$ and BCs of a given type on rectangular two-dimensional grids of different aspect ratios but uniform grid spacing $\Delta$. The proposed model is also included in a CFD solver to demonstrate its potential to provide an accurate initial guess (more accurate than the first step of conventional iterative methods), so that the rate of convergence can be dramatically increased.}

The outline of this paper is as follows: in \secref{sec:related_work}, we provide a brief summary of the recent advancements and ongoing challenges in applying neural networks (NNs) to solve the Poisson equation and more broadly to solve general PDEs. Then, we propose a novel BC handling strategy in \secref{sec:bc_strategy} and the related model architecture in \secref{sec:architecture}. The details of our dataset generation method, novel loss and training process are provided in \secref{sec:dataset_generation}, \secref{sec:loss} and \secref{sec:training}, respectively. The results and performance of the new model are showcased and discussed in \secref{sec:results}. Finally, conclusions are drawn and plans for future work are outlined.

\section{Related work}
\label{sec:related_work}
Interest in solving PDEs using NN-based methods has a relatively long history, beginning in the 90s with efforts by \cite{nn_pde_history_hyuk}, \cite{nn_pde_history_dissanayake} and \cite{nn_pde_history_lagaris}. A significant proportion of early works on using NNs to approximate the solutions of PDEs focus on approximating the solution $\phi$ given the variables it depends on, such as spatial coordinates, treating the NN as a continuous function. Training is performed to minimize the solution residuals inside the domain and on the boundaries. For example, \cite{nn_pde_history_lagaris} utilise a single layer perceptron (constrained by the computational power limitations of the time) to augment trial functions known to satisfy the equation in question to solve a number of benchmark problems in numerical analysis, such as the Poisson equation subject to both (fixed) Dirichlet and Neumann BCs using various RHS functions.

This rather intuitive application of NNs greatly benefits from their differentiability via the backpropagation algorithm since it provides a very accurate method to compute the residuals. Combined with other factors such as the accurate results achieved for specific problems using relatively few parameters, the computational constraints of the era and the relative infancy of more advanced NN architectures in use today, this method became the dominant approach in the early attempts to bring the fields of PDEs and NNs together. However, it suffered from the rather severe drawback of each set of trained weights being able to handle only a specific RHS function and set of BCs.

With the increase in computation power through the 2000s, more complicated models with more parameters and multiple layers became the norm. \cite{nn_pde_history_smaoui} used the greater amounts of computational power available to investigate deeper models, utilising multilayer perceptrons to predict the proper orthogonal decomposition modes of the one-dimensional  Kuramoto-Sivashinsky equation and the two-dimensional (2D) Navier-Stokes equation. Similarly, \cite{nn_pde_history_baymani}  used multilayer perceptrons to compute the solution of the Stokes equation by decomposing it into multiple Poisson problems and solving the Poisson problems with a procedure similar to the works discussed above. Furthermore, research into different neural computation methods became more popular, leading to works such as those using Radial Basis Function NNs by \cite{nn_pde_history_rbf_li} and by \cite{nn_pde_history_rbf_maiduy}. A more detailed survey of the methods of the era was published by \cite{nn_pde_history_kumar}.

Research into the applications of NNs to solve PDEs is still gaining significant momentum, driven by a number of factors providing a supporting ecosystem for deep learning research in general. Substantial improvements in the hardware used to run NN training and inference, discovery of better practices to train NNs, a culture of making publications available in open-access repositories and releasing source code using free and open-source licences in the research community help maintain a substantial growth rate. A brief overview of notable approaches being investigated in a modern setting is provided in the following subsections.

\subsection{Modern approaches to solving PDEs using NNs}

The above factors led to great strides in models using the older, `continuous' paradigm of training multilayer perceptron-style NNs to minimize solution residuals. Such models can now tackle a much greater variety of PDEs, solved in complex domains with a variety of BCs. A prominent  example of advances in this area in recent years is the advent of `Physics Informed Neural Networks (PINNs)' by \cite{pinn} who demonstrated the use of such a methodology to solve the Schr\"odinger and Burgers equations, the latter of which is of particular note due to the presence of discontinuities. Based on that paradigm, \cite{deepxde} developed the DeepXDE library, capable of solving a wide range of differential equations including partial- and integro-differential equations, providing a more user-friendly way of using NNs in this context. Furthermore, \cite{mpinn} proposed a method to help PINNs predict the correct values in a problem when given a combination of abundant yet less accurate `low-fidelity' plus sparse yet more accurate `high-fidelity' data.

Another subject that attracted substantial attention in the recent years is using NNs to `discover' the underlying PDE describing (\eg a physical system) from an existing dataset. In such a system, the NN tries to find the numerical coefficients of a previously assumed PDE form by minimizing residuals. Notable examples include works by \cite{discovering_pdes_pde_net}, who adopted a convolutional methodology to first discover the coefficients for and then predict the time evolution of a 2D linear diffusion equation, and also a number of works by \cite{pinn2,discovering_pdes_raissi_2018}, with examples of applications to, among others, the Burgers and Navier-Stokes equations. It is further noteworthy that interest in this subject extends beyond the use of NNs, as evidenced by works such as those from \cite{discovering_pdes_rudy} using sparse regression.

Meanwhile, separate from the initial approaches to applying NNs to PDEs described above, fully convolutional models that capitalize on significant strides made in computer vision using CNNs began to gain traction. Utilising common techniques in computer vision by treating the known terms and solutions of PDEs on rectangular grids as though they were images, such methods approach the task of solving a PDE as an image-to-image translation problem similar to the pix2pix method by \cite{CGAN_image_to_image}.

\subsection{CNNs and the Poisson equation}

A significant proportion of the efforts to use CNNs to solve PDEs has focused on the Poisson equation, considering its status as a well-understood benchmark problem with applications to many fields as mentioned earlier. 

In the fluid mechanics community, works of \cite{fluidnet} and \cite{poisson_cnn_xiao} pioneered the usage of CNNs to solve the Poisson equation. While both use their models within the framework of a complete CFD solver to simulate the motion of smoke plumes around objects, the architectures used and training methodologies are different. The former developed a model which takes a 3D array containing the velocity divergence and geometry information on cubic $128 \times 128 \times 128$ grids while the latter adopt a multigrid-like strategy with multiple discretizations to make predictions on larger $384 \times 384 \times 384$ grids. Moreover, the former approach trains the NN to minimize the divergence of the velocity field only while the latter adopts a more direct strategy by trying to instead minimize a linear combination of the L2 norms of the velocity divergence and the discrepancy between the predicted and ground truth pressure correction values by leveraging the additional data available from the specific methodology. Building on the strategy developed in these works, \cite{ajuria2020towards} used a CNN to handle the Poisson solver step of CFD simulations of plumes and flows around cylinders, demonstrating stable and accurate time evolution even for Richardson numbers greater than in the training data when applied in combination with several Jacobi iterations. Outside fluid mechanics, \cite{poisson_cnn_shan} investigated the application of a fully convolutional NN to predict the electric potential on cubic $64 \times 64 \times 64$ grids given the charge distributions and (constant) permittivities, claiming average relative errors below $3\%$ and speedups compared to traditional methods.

In general, all of these works attempt to predict the solution of the Poisson equation given an array containing the values of the RHS function on a grid. {In the present study, we propose a fully-convolutional NN architecture that can handle arbitrary BCs, on grids with different aspect ratios and uniform grid spacing. The mathematical formulation and the neural architecture of the proposed approach are discussed in the next two sections.}

\section{Mathematics of the proposed NN architecture}
\label{sec:bc_strategy}
Since the Poisson problem does not have a unique solution when BCs are absent, a way to include and process BC information alongside the RHS is required to obtain a model that is able to solve the Poisson problem with arbitrary BCs (as opposed to training different models tailored for one specific set of BCs). Matrix-based methods such as successive over-relaxation and direct solution handle the BCs by augmenting the RHS vector. Conversely, an NN must process boundary information more explicitly by integrating it into the model architecture. Following from \cite{pdes} and assuming arbitrary Dirichlet or Neumann BCs, the proposed methodology involves splitting the original (2D) Poisson problem into one Poisson problem with homogeneous (zero Dirichlet) BCs plus four Laplace problems where each Laplace problem has three homogeneous BCs plus one inhomogeneous Dirichlet or Neumann BC identical to one of the BCs in the original problem on the corresponding boundary. 

Formally, for Dirichlet BCs, if we consider the original problem on a rectangular domain $D$, denoting an edge of the domain with the index $k$ as $[\partial D]_k$ and the corresponding boundary condition on the edge as $g_k$,
\begin{equation}
    \label{eq:bc_strategy_decomposition_0}
    \nabla^2 \phi = f \, \land \, \phi([\partial D]_1) = g_1  \, \land \, \phi([\partial D]_2) = g_2 \, \land \, \phi([\partial D]_3) = g_3 \, \land \, \phi([\partial D]_4) = g_4,
\end{equation}
we can rewrite it in the following form, exploiting the linearity of the Laplace operator
\begin{equation}
    \label{eq:bc_strategy_decomposition_1}
    \nabla^2 (\phi_h + \phi_{BC0} + \phi_{BC1} + \phi_{BC2} + \phi_{BC3}) = f,
\end{equation}
such that
\begin{align}
    \label{eq:bc_strategy_decomposition_2}
    \nabla^2 \phi_h =& f \, \land \, \phi_h(\partial D) = 0,  \\
    \label{eq:bc_strategy_decomposition_3}
    \nabla^2 \phi_{BC k} =& 0 \, \land \,\phi_{BC k} ([\partial D]_j) =  \begin{cases} 
      g_k & j = k \\
      0 & j \neq k
   \end{cases}, \, j,k \in [0,1,2,3].
\end{align}
Thus, it is possible to solve the original problem by first solving the Poisson equation with the original RHS but zero BCs to find $\phi_h$, then solving Laplace problems for each BC to find $\phi_{BC k}$ and summing these results. Similarly, the proposed NN architecture will be composed of two parts -- one which solves the homogeneous Poisson problem and another which solves the Laplace problem given one BC. {It will be shown in \secref{sec:ablation} that this approach is superior to designing a single NN for an inhomogeneous Poisson problem. It also offers more flexibility for handling various types of BCs. Such a feature is not available in the geometry-restricted strategies seen in previous works. It means that the proposed model can tackle any Poisson problem with the appropriate BCs.}

%\clearpage
\begin{figure}[h!]
    \centering
    \includegraphics[width = \textwidth]{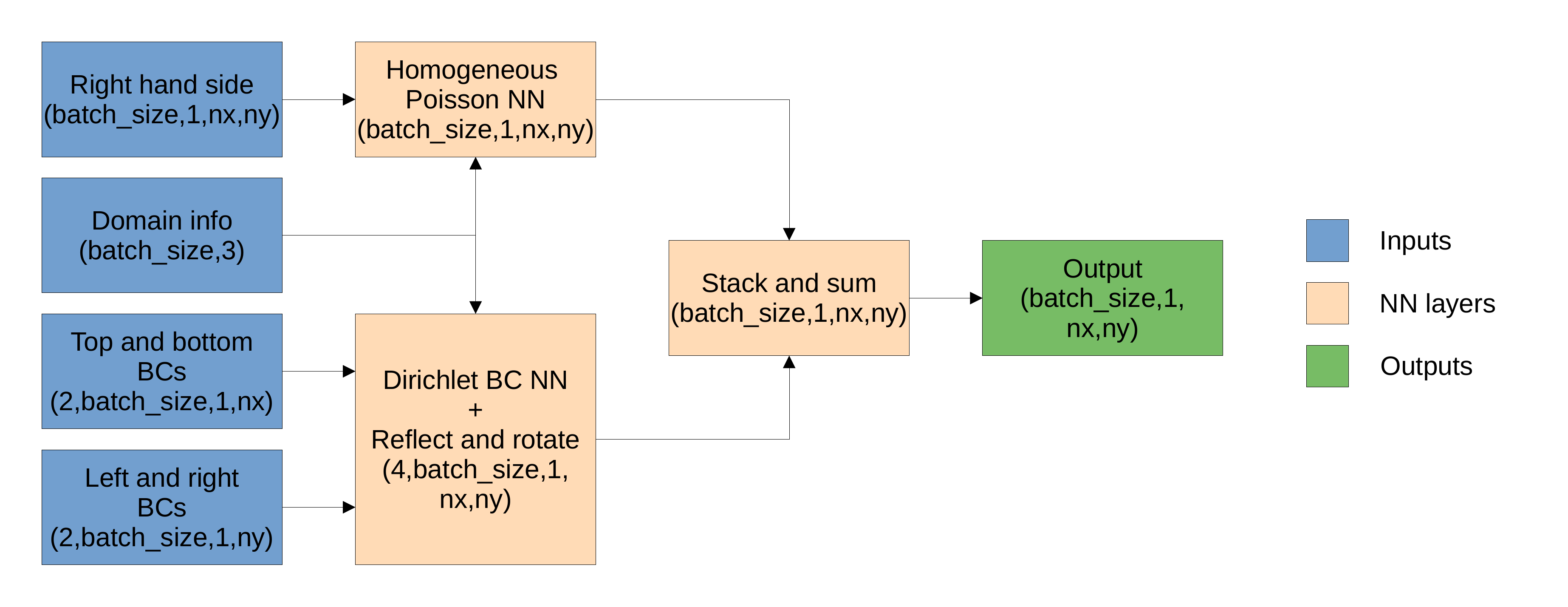}
    \caption{High-level diagram of Poisson CNN. Variable names in parentheses in each block indicate the shape of the output of the block}
    \label{fig:Poisson_CNN_diagram}
\end{figure}
%\clearpage

\section{Proposed NN architecture}
\label{sec:architecture}

\figref{fig:Poisson_CNN_diagram} provides an overview of the high-level structure of the NN architecture proposed in the current study. The NN components of the model are the blocks marked as Dirichlet BC NN (DBCNN), which approximates the solution of the Laplace equation with one inhomogeneous Dirichlet boundary, and the Homogeneous Poisson NN (HPNN) which approximates the solution of the Poisson equation with homogeneous BCs. The architecture mirrors the decomposition in Equations \ref{eq:bc_strategy_decomposition_0} to \ref{eq:bc_strategy_decomposition_3}. First, the DBCNN sub-model makes predictions for the four Laplace problems and applies reflection and rotation operations such that the inhomogeneous boundaries align with the corresponding boundary in the original problem. Then, the HPNN model makes a prediction for the Poisson problem with homogeneous BCs. Finally, these results are summed to obtain the final prediction for a Poisson problem with four inhomogeneous Dirichlet BCs. 

\subsection{Homogeneous Poisson NN (HPNN)}
\label{sec:Homogeneous_Poisson_NN}
\begin{figure}[h!]
    \centering
    \includegraphics[width = 0.99\textwidth]{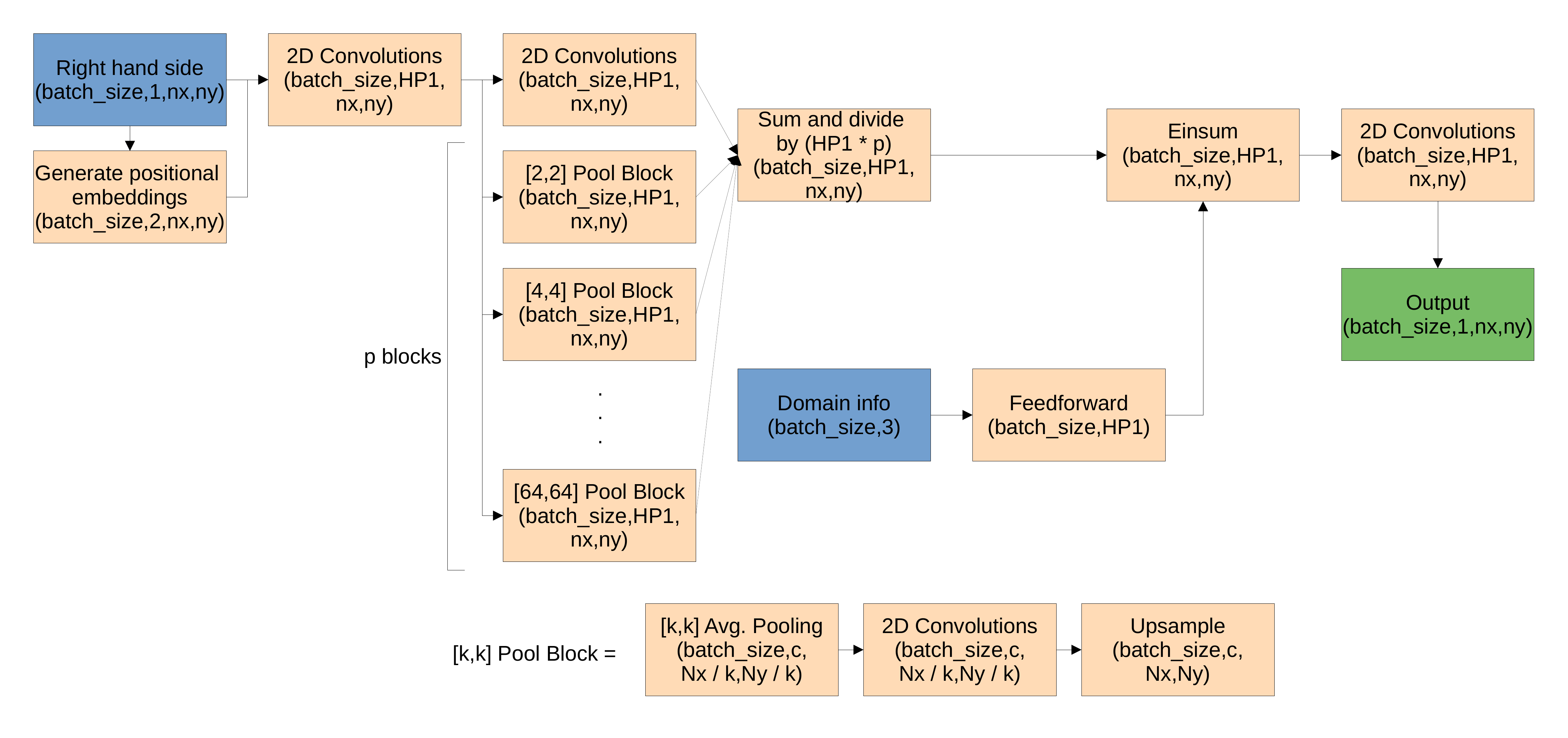}
    \caption{Homogeneous Poisson NN diagram. Variable names in parentheses in each block indicate the shape of the output of the block. Variables named `HP*' are hyperparameters}
    \label{fig:homogeneous_poisson_NN_diagram}
\end{figure}

The HPNN estimates the solution of the Poisson problem with homogeneous BCs and was inspired by the Fluidnet architecture by \cite{fluidnet}. The model has two inputs: the RHS and the domain information (grid spacing, domain sizes per spatial dimension). \figref{fig:homogeneous_poisson_NN_diagram} provides an overview of the model architecture. Our model expands upon the Fluidnet model by adding ResNet blocks (first proposed by \cite{resnet}) after most convolutions, incorporating a substantially larger number of independent pooling operations and using dense layers as well as positional embeddings\footnote{Positional embeddings are created by concatenating a cosine wave $\cos(\pi x_i / L_{x_i})$, tiled to match the input grid shape, in each direction to the right hand side in the channel dimension} to handle different grid resolutions, grid sizes and aspect ratios.

In the present study, it was found that the additional model architecture features lead to substantial gains in accuracy for relatively little increases in runtime. {A more detailed investigation of the improvements afforded by the novel model features can be found in \secref{sec:ablation} with an ablation study.} Further investigations indicated that without incorporating a mechanism to handle domain and grid spacing information as with the positional embeddings and the feedforward layers, the model learns to merely reproduce the solution profile with an average peak magnitude value, substantially reducing its usefulness. The necessity to supply the grid spacing is evident from the fact that this is necessary for classical methods as well; on a grid, the finite difference approximation for the Laplacian operator depends on the $\Delta ^2$ factors in the denominator and hence this information is needed to reverse the operation. On the other hand, while there is no such similar absolute necessity to supply the domain size information (and indeed classical methods do not require this information at all), in practice it was found to greatly boost the performance of the model since it enables the model to adjust its output to different aspect ratios more easily.

When processing the RHS, first the data is passed through several convolutions. Then, the computation is split into multiple independent pooling operations, each of which applies average pooling of progressively larger pool sizes to capture lower-wavenumber modes and applies further convolutions to the pooled results. Then, the pooled results are upsampled using either a polynomial reconstruction method (in the case of the pooling threads with the largest two pool sizes) or transposed convolutions (for branches with smaller pool sizes). Using polynomial interpolation based upsampling methods for pooling branches with large pool sizes was observed to reduce artefacting in the output; upsampling \eg a $2 \times 2$ input generated using $128 \times 128$ pooling from a $200 \times 200$ source image requires using a stride\footnote{Please see \cite{conv_arithmetic} for further details} of 128 for the transposed convolution to upsample to the original size, which is excessively large. Finally, the results from all branches are summed and are divided by the number of branches times the number of channels in each branch.

The domain information (\ie $\Delta$, $L_x$ and $L_y$) is processed by several dense layers. The RHS and domain information branches are subsequently merged by multiplying every channel from the RHS branch by one of the outputs of the domain information branch, expressed using tensor notation (without the implicit summation) as
\begin{equation}
    \label{eq:rhs_domaininfo_merge}
    A_{ijkl} B_{ij} = C_{ijkl},
\end{equation}
where index $i$ is the batch dimension, index $j$ is the channel dimension and the remaining indices are for the spatial dimensions. Finally, several more convolutions are applied to the result from \equref{eq:rhs_domaininfo_merge}, reducing the final number of channels to 1 to produce the solution.

\subsection{Dirichlet Boundary Condition NN (DBCNN)}
\label{sec:Dirichlet_BC_NN}

\begin{figure}[h!]
    \centering
    \includegraphics[width = 0.95\textwidth]{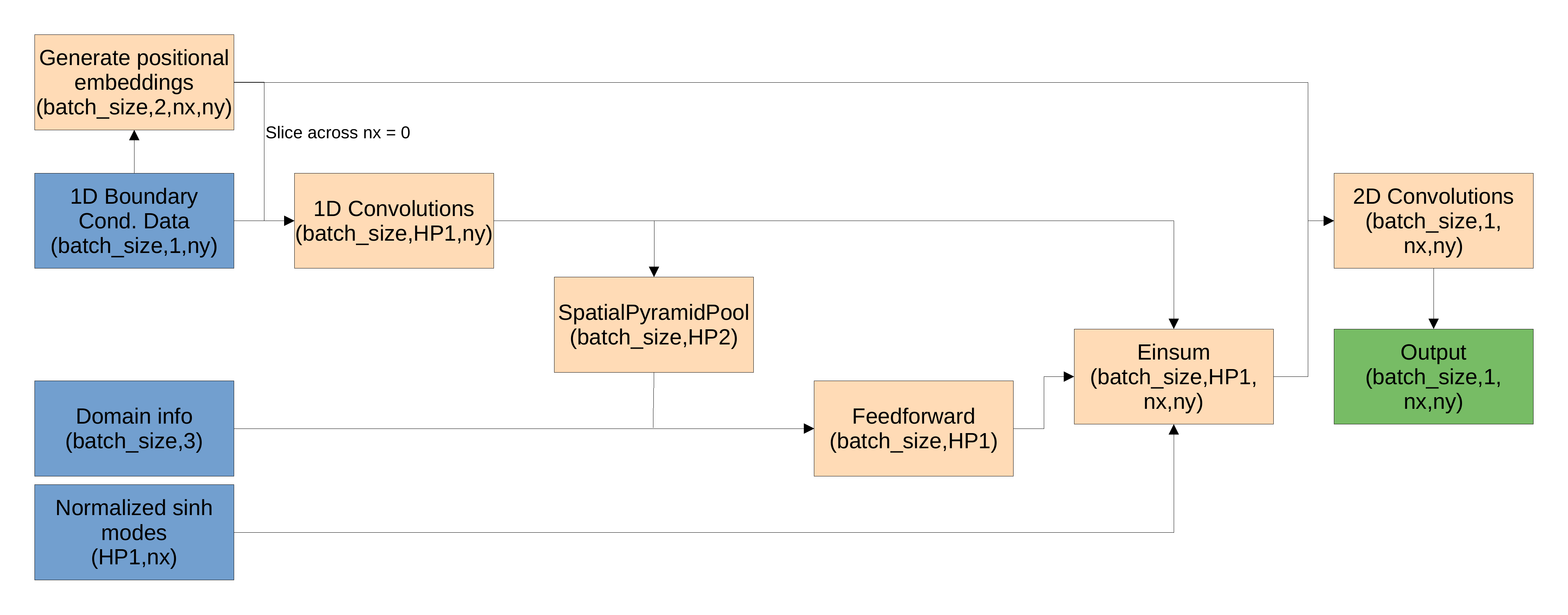}
    \caption{Dirichlet Boundary Condition NN diagram. Variable names in parentheses in each block indicate the shape of the output of the block. Variables named `HP*' are hyperparameters}
    \label{fig:Dirichlet_BC_NN_diagram}
\end{figure}

The Dirichlet Boundary Condition NN (DBCNN) estimates the solution of the Laplace equation with one inhomogeneous Dirichlet BC. As inputs, it takes one 1D array containing the BC information, the same domain information used for the HPNN and the number of grid points in the direction orthogonal to the provided BC. \figref{fig:Dirichlet_BC_NN_diagram} further details the operation of the model.

The difficulty of solving this problem is constructing the solution, which is a 2D scalar field, from the BC information which is only a 1D scalar field. To overcome this issue, a known mathematical property of the solutions of the Laplace equation in this setting is exploited; given homogeneous Dirichlet BCs on three boundaries plus one inhomogeneous Dirichlet BC on another on a rectangular domain, the solution of the Laplace equation on the domain $[0,L_x] \times [0,L_y]$ can be written as a series of the form
\begin{equation}
    \label{eq:laplace_equation_solution_series}
    \phi = \sum_{m=1}^{\infty} A_m \sinh\left(\frac{m\pi(x-L_x)}{L_x} \right) \sin\left(\frac{m \pi y}{L_y}\right),
\end{equation}
when the non-homogeneous boundary is placed at $x=0$. Thus, we can say that the solution along any constant $y=\Tilde{y}$ can be written as 

\begin{equation}
    \phi(x,\Tilde{y}) = \sum_{m=1}^{\infty} B_m \sinh\left(\frac{m\pi(x-L_x)}{L_x} \right).
\end{equation}

Thus, to reconstruct the solution, the $B_m$ values for each gridpoint in the $y$ direction are required. Naturally, as is common practice for numerical algorithms, a cutoff point $n_c$ is chosen for the series. The DBCNN model works by first performing a series of 1D convolutions on the BC data to increase the number of channels to $n_c$. Following that, spatial pyramid pooling as introduced by \cite{spatial_pyramid_pooling} is performed on this result and the fixed-size vector from this operation is concatenated with the domain/grid information. This vector is processed through a feedforward neural network, the final layer of which has $n_c$ units. As the third step, the $\sinh$ modes $\sinh(m \pi (x-L_x) / L_x)$ are constructed and are normalized such that the maximum absolute value in each mode is $1.0$. Finally, a tensor product between the results in the first three steps is performed to create a batch of $(n_c, n_x, n_y)$ sized tensors and a number of 2D convolutions are performed to reduce the number of `channels' $n_c$ to 1. 

It should be noted that this model returns results such that the inhomogeneous BC is always on the left boundary of the domain. However, a series of rotations and reflections are sufficient to modify the result such that any other boundary is the inhomogeneous boundary instead. 

\section{Dataset}
\label{sec:dataset_generation}
The dataset plays a fundamental role when training NNs, since NNs learn to reproduce the conditional probability distribution of the data as outlined by \cite{prml}. Hence, training on a dataset that does not reflect the probability distribution of the problem in question will give inaccurate results when performing inference. This is especially important when the model will be trained on synthetic data as with the our approach, since the onus is on the user to ensure the training set reflects the conditional PDF of the `real' data.

Furthermore, the Poisson problem in the most general case will have source functions and solutions that are finite yet unbounded, which present significant obstacles for NNs. Hence, there is a need to normalize the dataset in a way that does not lead to a loss of generality in the predictive performance of the model while constraining the range of the inputs and outputs to a well-defined and consistent interval. Fortunately, the linearity of the Laplace operator allows us to do this by merely scaling the RHS and the solution by some coefficient $\alpha$:
\begin{align}
    \nabla^2 \phi = f \implies \alpha \nabla^2 \phi = \nabla^2 (\alpha \phi) = \alpha f.
\end{align}{}

A similar argument may be made to show that scaling the Dirichlet boundary conditions for such a Poisson problem by a constant results in the solution of the corresponding Laplace problem (as shown in \equref{eq:bc_strategy_decomposition_3}) being scaled by the same constant. In this work, it was chosen to constrain the source functions and BCs such that their maximum absolute value is always $1.0$. Different training data generation methods were adopted for the two sub-models, detailed below.

\subsection{Training data for Homogeneous Poisson NN (HPNN)}
\label{sec:dataset_generation_hpnn}
The HPNN sub-model was trained on an analytically generated dataset, composed of synthetically generated solution-RHS pairs based on truncated Fourier and polynomial series. First, the solutions were generated as the sum of a Fourier series and a polynomial, both  with random coefficients 
\begin{equation}
    \phi = \phi_F + \phi_P.
\end{equation}

Given homogeneous Dirichlet BCs, the Poisson equation on a rectangular domain can be shown to have solutions of the form
\begin{equation}
    \label{eq:poisson_homogeneous_analytical_soln}
    \phi = \sum_{m,n} A_{mn} \sin \left(\frac{m \pi x}{L_x} \right) \sin\left(\frac{n \pi y}{L_y} \right),
\end{equation}
where \eg $L_x$ indicates the domain length in the $x$ direction. It should be noted that that the Laplacian of such a function will correspond to an RHS which is also the sum of products of sine waves. Using such a truncated Fourier series however results in RHS=0 for all grid points on the boundary -- a loss of generality. The RHS need not be zero on the boundaries when solving a Poisson problem analytically as an infinite Fourier series can be fitted for a discontinuous function, Gibbs' phenomenon notwithstanding. However, since a truly infinite series cannot be computed on a computer, the solution was instead augmented by a polynomial component 
\begin{equation}
    \phi_P = \prod_i p_{i}(x_i) = p_x(x) p_y(y).
\end{equation}

The multi-dimensional polynomial $\phi_P$ is generated by multiplying one-dimensional polynomials together, benefiting from the variable separable nature of the Poisson equation. Each polynomial component was created by randomly choosing roots $r$ within the extent of the domain in the respective dimension plus a random coefficient $B$, reserving two of the roots for the extrema of the domain
\begin{equation}
    p_{i}(x_i) = B_i (x_i-0)(x_i-L_i)(x_i-r_{i,0})(x_i-r_{i,1})...
\end{equation}

A similar methodology was followed for the Fourier series counterpart with random coefficients $C$, again utilising the variable separability
\begin{align}
    \phi_F = \prod_i s_i (x_i), \\
    s_i(x_i) = \sum_k C_{k,i} \sin \left(\frac{k \pi x_i}{L_i} \right).
\end{align}

Subsequently, the peak magnitudes of the two components $\max(|\phi_F|)$ and $\max(|\phi_P|)$ were equalized to ensure neither component dominates the generated dataset. The corresponding right hand sides were calculated by appropriately adjusting the coefficients in the case of $\phi_F$ and using autodifferentiation to compute $p_{i}''(x_i)$ in the case of $\phi_P$. 

As a final step, normalization was carried out on the dataset to improve model performance. Both the right hand side and the solution are divided by $\max(|\mathrm{RHS}|)$ to set the RHS' maximum absolute value to 1. Furthermore, the solutions were divided by an additional $\max([L_x, L_y])^2$ factor, \ie square of the longest domain extent, to ensure that both the RHS and the solution are properly nondimensionalized\footnote{The division by $\max(|\mathrm{RHS}|)$ nondimensionalizes the RHS but since the Laplace operator modifies the dimensionality of its argument by a factor of $[L]^{-2}$, this additional normalization is necessary to ensure that the solution is also nondimensionalized. This normalization was observed to reduce loss by over $50\%$.}.

{During the dataset synthesis process, the number of terms for each sample in the Fourier and Taylor series components were chosen randomly from a uniform distribution within a pre-determined range. Overall, this methodology to generate synthetic homogeneous Poisson problems provides four parameters per spatial dimension to control the roughness of the RHS -- two parameters for both series components prescribing the range to draw the number of terms from -- for a sum of eight parameters for the two dimensional problems investigated in \secref{sec:results}. The grid parameters were similarly chosen randomly from uniform distributions within pre-determined ranges. This necessitates two parameters to choose the range of grid spacings, plus a further two parameters per spatial dimension for the maximum and minimum number of grid points across each dimension, for a total of six parameters for a 2D problem. Note that grid spacings were randomly chosen per-batch instead of per-sample.}

\subsection{Training data generation for the Dirichlet BC NN}
A different approach to generating training data for the Dirichlet BC NN sub-model was chosen, eschewing the series based approach for the HPNN sub-model in favor of a purely numerical approach. This was motivated by the fact that the peak magnitude of the $\sinh$ component in the series for the solution of the Laplace equation shown in \equref{eq:laplace_equation_solution_series} grows exponentially as the series index $m$ grows. To avoid having to artificially skew the distribution of the coefficients, instead it was chosen to randomly generate the boundary conditions which necessitates the usage of a numerical method to obtain the solutions to the Laplace problems. 

However, since the boundary conditions of interest are continuous, generating \eg 200 random values for a 200 gridpoint boundary results in extremely noisy BCs that do not reflect the typical use case for Poisson solvers and are very difficult for NNs to learn from. To rectify this issue, it was instead chosen to generate lower resolution random BCs and then upscale these to the desired resolution using cubic interpolation. Owing to the $C^2$ continuity of cubic upscaling, this ensures that the resulting BCs are smooth when the ratio of the final resolution to initial resolution is high enough. The drawback of this method is the addition of this lower resolution to the list of user-defined parameters. To prevent the NN from overfitting for a specific value of this parameter, this value was {chosen randomly from a uniform distribution for each batch. This methodology requires $2(N_D-1)$ user-defined parameters to choose the minimum and maximum possible quantities of control points, where $N_D$ is the number of spatial dimensions. The number of parameters needed to control the grid size is identical to \secref{sec:dataset_generation_hpnn}.}

Solutions to the resulting linear systems were computed using the Python algebraic multigrid package \texttt{PyAMG} by \cite{OlSc2018}. Specifically, the `classical' Ruge-Stuben algebraic multigrid method with a tolerance of $10^{-10}$ was chosen. Calculations were done on a 32-core 64-thread AMD Threadripper 2990WX CPU, using 64 threads.  \tabref{tab:runtime} in \secref{sec:runtime} outlines the wall-clock runtime needed to solve single problems at various grid sizes, excluding the time needed to generate the RHS and BCs, along with a comparison to the multigrid method running on the GPU based on the \texttt{pyamgx} Python bindings by \cite{pyamgx} to the \texttt{AMGX} library by \cite{amgx}. In this work, the coarse grid sizes (across each dimension) were set between 2 and 10. To avoid overfitting on specific examples, the datasets are generated on-the-go before each batch is fed to the training routine. 

\section{Loss function and approximation of loss value}
\label{sec:loss}
In typical regression applications, it is common to use the mean squared error (MSE) as the loss function. However, as shown in multiple studies on image-to-image translation using convolutional models such as those by \cite{CGAN_image_to_image}, MSE loss does not typically lead to good results. As an alternative, since the targets are strictly smooth functions, the continuous version of the $L^p$ norm between the output and the target presents a more meaningful loss for a pair of continuous functions. The norm is defined as
\begin{equation}
    \label{eq:lp_norm}
    \left[ \frac{1}{A} \int_{A} (y - t)^p \, \mathrm{d}A \right]^{1/p},
\end{equation}
\noindent where $y$ is the prediction, $t$ is the target and $A$ is a finite region of $\mathbb{R}^n$. However, we know the functions' values only on a rectangular grid. In a naive approach, we could evaluate the expression in Equation \ref{eq:lp_norm} by using a polynomial reconstruction method and integrating the piecewise polynomial. In fact, MSE loss can be interpreted as doing this with the midpoint rule. A more accurate  alternative is Gauss-Legendre quadrature. In practice, this methodology can be written down for a function $h:\mathbb{R}^n \xrightarrow{} \mathbb{R}$ as
\begin{equation}
    \int_V h(x_1,x_2,...,x_n) \, \mathrm{d}x_1 ... \mathrm{d}x_n \approx \sum_{i_1=1}^{k_1} ... \sum_{i_n=1}^{k_n} \left(\prod_{j=1}^{n} w_{i_j} \right) h({\bar{x}_{1_{i_1}}},...,{\bar{x}_{n_{i_n}}}),
\end{equation}
where $k_l$ is the order of the quadrature in the $l^{th}$ direction, $\bar{x}_{l_{i_l}}$ is the $i_l^{th}$ quadrature point (\ie Legendre polynomial root) in the $l^{th}$ direction and $w$ is a vector containing the quadrature weights (hence $w_{i_j}$ denotes the $i_j^{th}$ weight in the $j^{th}$ direction). Applying this formula to Equation \ref{eq:lp_norm} for $y,t:\mathbb{R}^2 \xrightarrow{} \mathbb{R}$ with $k$ weights in both directions, we can write
\begin{equation}
    \label{eq:lp_norm_quad_1}
    \left[ \frac{1}{A} \int_{A} (y(x_1,x_2) - t(x_1,x_2))^p \, \mathrm{d}A \right]^{1/p} \approx \left[\frac{1}{A} \sum_{i}^k \sum_{j}^k w_i w_j [y(\bar{x}_{1_i}, \bar{x}_{2_j}) - t(\bar{x}_{1_i}, \bar{x}_{2_j})]^p \right]^{1/p}.
\end{equation}

This approach, however, relies on knowing the values of $y$ and $t$ at the quadrature points $\bar{x}_1,\bar{x}_2$ which are not equispaced and therefore cannot be made to coincide with the grid points using a simple affine transformation as is the common practice for utilising Gauss-Legendre quadrature for domains other than the canonical $[-1,1]$ domain. To overcome this issue, $y(\bar{x}_1,\bar{x}_2)$ and $t(\bar{x}_1,\bar{x}_2)$ were approximated using bilinear interpolation within the `cell' enclosing each quadrature point. In practice, for each quadrature point, the value is a linear combination of the 4 known values on the corners
\begin{equation}
    y(\bar{x}_1,\bar{x}_2) \approx \begin{bmatrix} b_{00} & b_{01}  & b_{10}  & b_{11} \end{bmatrix} \begin{bmatrix} y_{i,j} \\ y_{i+1,j} \\ y_{i,j+1} \\ y_{i+1,j+1}  \end{bmatrix} = \vec{b}^{\, \intercal} \vec{y},
\end{equation}
where the weight vector $\vec{b}$ depends on the coordinates of the quadrature point and the grid points. Denoting $\vec{\cdot}_{ij}$ as the vector containing the values of the variable in question for the $i$th quadrature point in the $x_1$ direction and the $j$th quadrature point in the $x_2$ direction, we can rewrite Equation \ref{eq:lp_norm_quad_1} as

\begin{equation}
     \left[ \frac{1}{A} \int_{A} (y(x_1,x_2) - t(x_1,x_2))^p \, \mathrm{d}A \right]^{1/p} \approx \left[\frac{1}{A} \sum_{i}^k \sum_{j}^k w_i w_j [\vec{b}^{\, \intercal}_{ij} (\vec{y}_{ij} - \vec{t}_{ij} )]^p \right]^{1/p}.
\end{equation}

During training, it was found that augmenting the integration with a mean absolute error (MAE) loss component sped up the loss minimization. Hence, the final expression for the loss is

\begin{equation}
    \label{eq:loss_func}
    L = \lambda_1 \left[\frac{1}{A} \sum_{i}^k \sum_{j}^k w_i w_j [\vec{b}^{\, \intercal}_{ij} (\vec{y}_{ij} - \vec{t}_{ij} )]^p \right]^{1/p} + \frac{\lambda_2}{N} \sum_i^N |y_i - t_i|.
\end{equation}

Experimentation during training indicated that values of $\lambda_1 = 0.4$, $\lambda_2 = 1.0$ and $p = 2$ provide good performance. {A benchmark of training the DBCNN model with and without this novel loss function can be found in \secref{sec:ablation}.}

Training the model by using the RMS solution residual as the loss as in works by \cite{pinn, pinn2} was also tried, both on its own and as an extra term in the loss expression in \equref{eq:loss_func}. It was found that when used by itself the residual loss leads to unstable training. This issue can be overcome by starting training with an MAE and/or integral loss and introducing the residual component after several epochs of training. Unfortunately this leads to higher MSE than the loss function in \equref{eq:loss_func}, albeit resulting in smoother solutions that may in some cases look qualitatively better. However, it was chosen to focus on numerical measures of error in this work.

\section{NN Training}
\label{sec:training}
The model was implemented using Tensorflow 2.3. The sub-models were trained independently; no further end-to-end training was carried out for the full model configuration shown in \figref{fig:Poisson_CNN_diagram}. Adam was chosen as the optimizer method, with an initial learning rate of $10^{-4}$ for the DBCNN and $10^{-5}$ for the HPNN. Learning rate was dynamically tapered to $10^{-7}$ as loss plateaued. To determine when to stop the training, early stopping based on the mean squared error with 20-epoch patience was used. \tabref{tab:numepochs} summarizes the information regarding the number of samples used to train each model.  

\begin{table}[h!]
\centering
\caption{Details of the number of samples used to train each sub-model}
\begin{tabular}{@{}lcccc@{}}
\toprule
Model       & Batch size & Batches per epoch   & Epochs & Samples \\ \midrule
HPNN        & 50         & \multirow{2}{*}{200} & 62     & 620,000   \\
DBCNN       & 50         &                     & 49     & 370,000   \\ \bottomrule
\end{tabular}
\label{tab:numepochs}
\end{table}

Furthermore, a number of best practice guidelines for both training and tweaking the model architecture were established for ensuring the best model performance:
\begin{itemize}%[\itemsep=0em]
    \item Final layers should have small kernels to prevent artefacting near the domain edges since smaller kernels require less padding to retain the input shape, while initial layers should have large kernels to propagate information from each gridpoint to every other gridpoint\footnote{Propagating information from each gridpoint to every other gridpoint is necessary for the Poisson problem due to the elliptic nature of the PDE.} using fewer convolutional layers.
    \item Randomizing input domain shapes by independently picking the size of each dimension from a uniform distribution leads to a nonuniform distribution of aspect ratios (ARs), as the ratio of two uniform distributions is known to result in a highly non-uniform distribution per \cite{marsaglia1965ratios}. This leads to poor performance with ARs which are less commonly encountered in the dataset. To rectify this, instead the ARs themselves can be picked from a uniform distribution and the grid sizes generated based on the ARs.
    \item Using large batch sizes is critical for ensuring good model performance. In the case of the HPNN sub-model, increasing the batch size from 13 to 50 led to a $45\%$ reduction in loss with an identical number of samples and a $78\%$ reduction at the end of training.
    \item Adding as many pooling levels to the HPNN as possible is crucial to allow the model to capture information at multiple scales. 
    \item Deconvolutional upsampling in the HPNN provides superior results when reversing pooling operations with small pool sizes, while bilinear/nearest neighbour are better for very large pool sizes (\eg $64\times 64$).
    \item Increasing the number of spatial pyramid pooling (SPP) levels in the DBCNN was found to reduce prediction accuracy beyond a certain level; increasing the number of output values from the SPP layer to 123 from 58 reduced loss by $41\%$, but further increasing it to 228 led to a $32\%$ increase relative to 123 outputs.
    \item Using Rectified Linear Unit (ReLU) activations in the intermediate layers for this task was observed to cause the dying ReLU problem. Other activations such as tanh or leaky ReLUs were observed to result in higher accuracy.
    \item Residual connections boost performance of both the DBCNN and the HPNN.
\end{itemize}{}

\section{Results and performance}
\label{sec:results}

In this section we show the performance of a Poisson CNN model, the submodels of which were trained independently on problems with grids containing 192 to 384 grid points in each direction and $\Delta$ values between 0.005 and 0.05. Sections \ref{sec:validationset} and \ref{sec:tgv} outline the test cases chosen and give greater detail regarding the performance of the model in each test case. Moving outside the `comfort zone' of the model, \secref{sec:previously_unseen_grids_perf} investigates the performance of the model with problem sizes outside the training data. \secref{sec:post-smoothing} details ways to work around artefacting that can show up in the model's predictions, and demonstrates the usage of Poisson CNN predictions as initial guesses for the multigrid algorithm. Building on that demonstration, \secref{sec:cfd} showcases the usage of the Poisson CNN within a CFD simulation and compares the accuracy obtained with a fully conventional simulation. Finally, \secref{sec:ablation} explores the impact of key innovations in the model architecture on the predictive performance of the model and \ref{sec:runtime} compares the wall-clock runtime of the model against a conventional iterative solver algorithm. Furthermore, to demonstrate the flexibility of our approach, we present in \appref{sec:appendix_mesh} a gallery of results with a greater range of aspect ratios.

\tabref{tab:parameter_number} outlines the number of parameters for the models with the specific choice of hyperparameters (\eg kernel sizes, \no of channels) made. It should be noted that our choice does not represent a necessarily optimal choice for these hyperparameters, as this work is meant as a feasibility study. 

\begin{table}[h!]
\centering
\caption{Summary of the number of parameters.}
\begin{tabular}{@{}lccc@{}}
\toprule
                 & Homogeneous Poisson NN & Dirichlet BC NN & Poisson CNN \\ \midrule
\# of parameters & 5,559,108              & 483,878         & 6,042,986     \\ \bottomrule
\end{tabular}
\label{tab:parameter_number}
\end{table}

\tabref{tab:results_summary_table} gives an overview of the performance of the models in terms of the MAE, mean absolute percentage error (MAPE) and the percentage of grid points in each case for which the model made a prediction within 10\% of the target. Supplementing the selected test cases, averaged results for larger sets of similar examples are also available where applicable. 

In addition to our models, we provide an overview of the performance of two baseline models trained on the same datasets using identical loss functions and optimizers to illustrate the strength of our individual sub-models on their respective tasks. A 4-level U-Net, proposed by  \cite{ronneberger2015unet}, with leaky ReLU activations and 7.5 million parameters serves as the baseline case for the HPNN sub-model. Meanwhile, the baseline case for the DBCNN sub-model is a stack of six bidirectional Long Short-Term Memory (LSTM) layers (see \cite{lstm}) with 100 units per layer and $\tanh$ activations containing approximately 440,000 parameters, plus bilinear upscaling applied to the final LSTM output to match the target shape.

Detailed results from the baseline experiments are omitted for brevity, however our models greatly outperformed the naive baseline implementations; both the HPNN and the DBCNN model achieved MAEs an order of magnitude smaller than their respective baselines. Particularly, the U-Net produced results with the kind of severe checkerboard artefacting patterns associated with transpose convolutions, in line with the observations of \cite{deconvolutioncheckerboard}, displaying the advantage of using both transpose convolutions and traditional upsampling techniques in the HPNN architecture. Meanwhile, compared to the LSTM stack, the DBCNN model's specifically tailored architecture leveraging the mathematical properties of the investigated problem produces results which exhibit substantially less high-frequency noise in the direction orthogonal to the boundary. Another striking result is the percentages of grid points for which the prediction is within 10\% of the target value, which are much higher for our models when compared to the baselines.

%\clearpage
%\begin{landscape}
\begin{table}[h!]
\centering
\caption{Summary of the results presented in this section, plus averaged figures for larger numbers of examples for each case where applicable. Note that grid points with absolute percentage errors above 200\% were excluded from the MAPE calculation due to MAPE values approaching infinity near the zero solution contours. Baseline cases are denoted in \textit{italics}. Average figures were computed over 600 randomly generated samples. Peak error figures were normalized by the maximal absolute value of the ground truth.}
\begin{tabular}{@{}llcccc@{}}
\toprule
\multicolumn{2}{c}{Case}                                                                                                                                                                     & MAE                   & MAPE & \begin{tabular}[c]{@{}c@{}}\% of gridpts\\ within 10\% \\ of target\end{tabular} & \begin{tabular}[c]{@{}c@{}}Normalized \\ peak \\ error\end{tabular} \\ \midrule
\multirow{10}{*}{\begin{tabular}[c]{@{}l@{}}Examples similar \\ to the training \\ set (\secref{sec:validationset})\end{tabular}} & HPNN (Fig.\ \ref{fig:hpnn_tset_example_1})                        & $4.73 \times 10^{-2}$ & 13.25     & 56.24 & $1.43 \times 10^{-1}$                                                                          \\
                                                                & HPNN (Fig.\ \ref{fig:hpnn_tset_example_2})                        & $3.24 \times 10^{-2}$ & 8.96     & 74.88    & $1.26 \times 10^{-1}$                                                                     \\
                                                                                                                              & HPNN -- Avg                                      & $3.50 \times 10^{-2}$ & 13.04     & 57.47    & $1.58 \times 10^{-1}$                                                                     \\ 
                                                                                                                              &
                                                                \textit{U-Net -- Avg}                                     & $6.98 \ \times 10^{-1}$ & 50.37      & 20.04 & $1.22 \times 10^{0}$        \\ \cmidrule(l){2-6}
                                                                                                                              & DBCNN (Fig.\ \ref{fig:dbcnn_tset_example1})                      & $7.78 \times 10^{-3}$ & 18.29     & 41.16  & $8.01 \times 10^{-1}$                                                                       \\
                                                                                                                              & DBCNN (Fig.\ \ref{fig:dbcnn_tset_example2}) & $8.03 \times 10^{-3}$ & 20.66      & 30.73 & $7.56 \times 10^{-1}$                                                                        \\
                                                                                                                              & DBCNN -- Avg                                     & $1.96 \times 10^{-2}$ & 19.26     & 40.54   & $8.10 \times 10^{-1}$                                                                      \\
                                                                  & \textit{LSTM -- Avg}                                   & $1.06 \ \times 10^{-1}$ & 53.03     & 10.04  &  $1.04 \times 10^{0}$      \\ \cmidrule(l){2-6}
                                                                                                                              & \textbf{Full model (Fig.\ \ref{fig:pcnn_tset_example_3})}          & $9.62 \times 10^{-2}$ & 9.81     & 66.44           &    $8.32 \times 10^{-1}$                                                           \\
                                                                                                                              & \textbf{Full model -- Avg}                                & $6.39 \times 10^{-2}$ & 8.48     & 71.70 & $8.76 \times 10^{-1}$                                                                        \\[0.8ex]
\midrule
\multirow{3}{*}{ \begin{tabular}[c]{@{}l@{}} Taylor-Green Vortex \\ (\secref{sec:tgv}) \end{tabular} }                                                                 & HPNN (Fig.\ \ref{fig:tgv_hpnn_pred})                            & $1.01 \times 10^{-2}$ & 12.91     & 63.48 & $6.91 \times 10^{-2}$                                                                         \\
                                                                                                                              & DBCNN (Fig.\ \ref{fig:tgv_dbcnn_pred})                          & $4.65 \times 10^{-3}$ & 15.80     & 43.28       & $3.86 \times 10^{-2}$                                                                  \\
                                                                                                                              & \textbf{Full model (Fig.\ \ref{fig:tgv_pcnn_pred})}                      & $1.51 \times 10^{-2}$ & 12.08     & 60.98    & $1.31 \times 10^0$                                                                     \\ \bottomrule
\end{tabular}
\label{tab:results_summary_table}
\end{table}
%\end{landscape}
%\clearpage

\subsection{Examples similar to the training set}
\label{sec:validationset}
It is crucial for machine learning (ML) algorithms to be able to make correct predictions on previously unseen data that comes from a dataset with the same conditional probability distribution between the inputs and outputs as the training data. In typical ML applications, this is tested for by splitting the dataset (of \eg photos) into a training and a validation set and observing if a model that performs well on the training set does equally well on the validation set. In our approach, since each batch of training data is generated from random noise synthetically just before being fed to the training loop, analogously we can test the model's performance on new random samples generated in an identical manner. This section presents the models' performance on examples generated in the same manner as the training data first for each submodel individually in Sections \ref{sec:tset_hpnn} and \ref{sec:dbcnn_tset_result}, finally presenting an example for a Poisson problem with four inhomogeneous Dirichlet BCs in \ref{sec:tset_pcnn}.

\subsubsection{Homogeneous Poisson NN}
\label{sec:tset_hpnn}
\begin{figure}[h!]
    \centering
    \begin{subfigure}[b]{0.99\textwidth}
        \centering
		\includegraphics[width=0.36\textwidth]{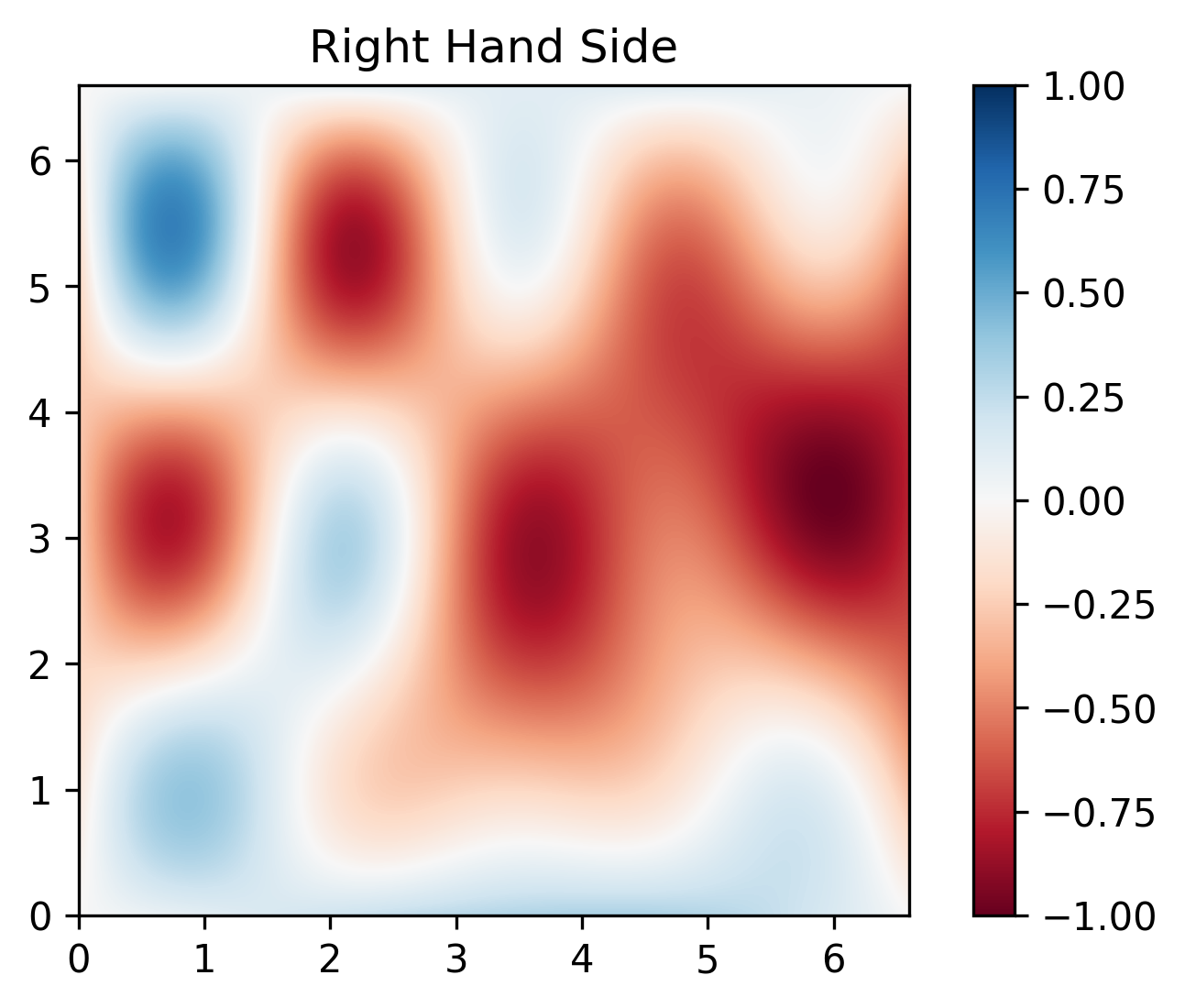}
		\includegraphics[width=0.35\textwidth]{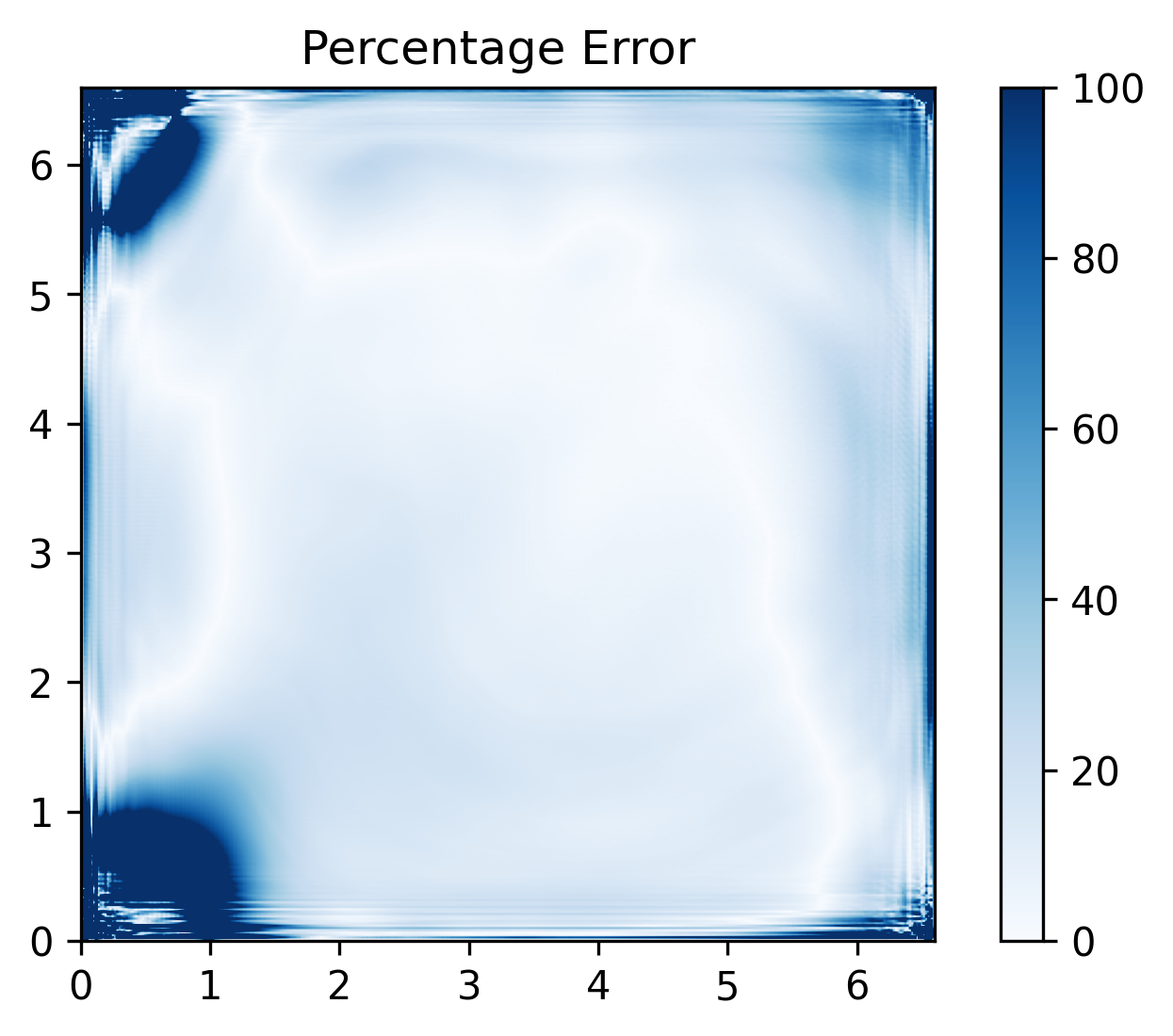}
    \end{subfigure}
%  \end{figure}
%  \begin{figure}\ContinuedFloat
    \begin{subfigure}[b]{0.99\textwidth}
        \centering
		\includegraphics[width = 0.295\textwidth]{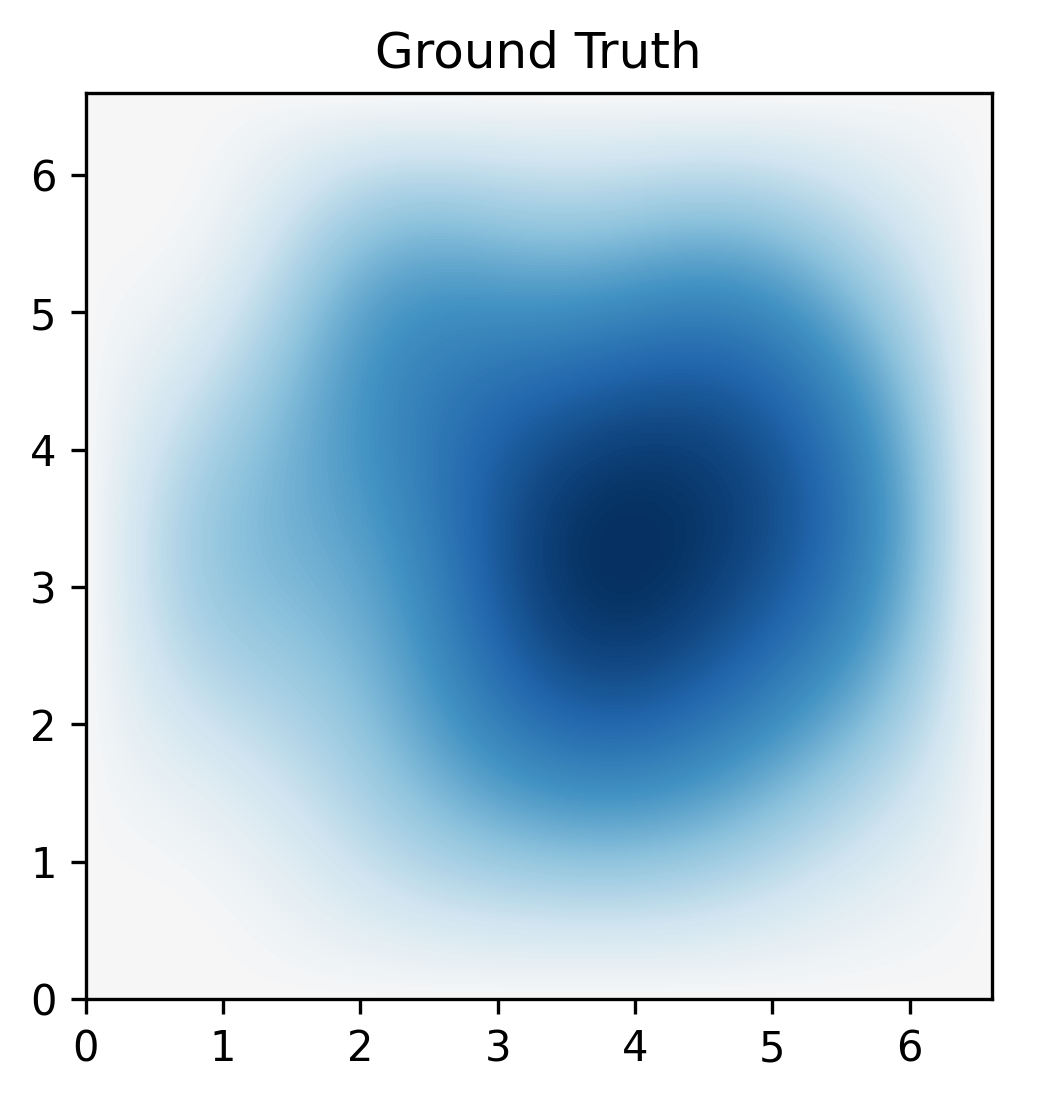}
        \centering
		\includegraphics[width = 0.36\textwidth]{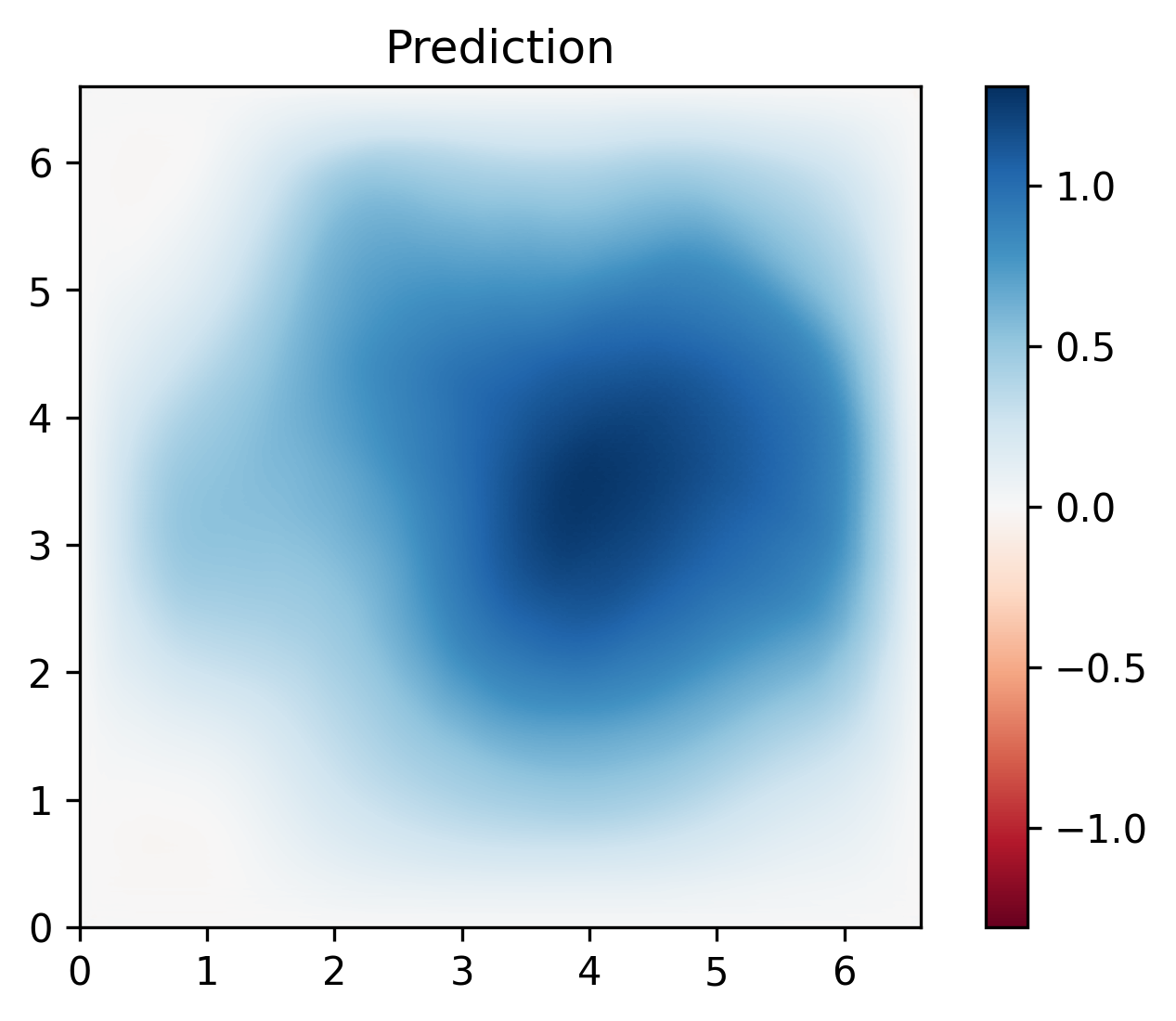}
    \end{subfigure}
    \caption[Performance of the Homogeneous Poisson NN model on an example with grid size $365 \times 365$ and $\Delta = 1.81 \times 10^{-2}$]{Performance of the Homogeneous Poisson NN model on an example with grid size $365 \times 365$ and $\Delta = 1.81 \times 10^{-2}$. The HPNN sub-model performs well, producing a prediction within 10\% of the target for over half of the grid points}
    \label{fig:hpnn_tset_example_1}
\end{figure}

\begin{figure}[h!]
    \centering
%    \begin{subfigure}[b]{0.49\textwidth}
%        \centering
		\includegraphics[width=0.49\textwidth]{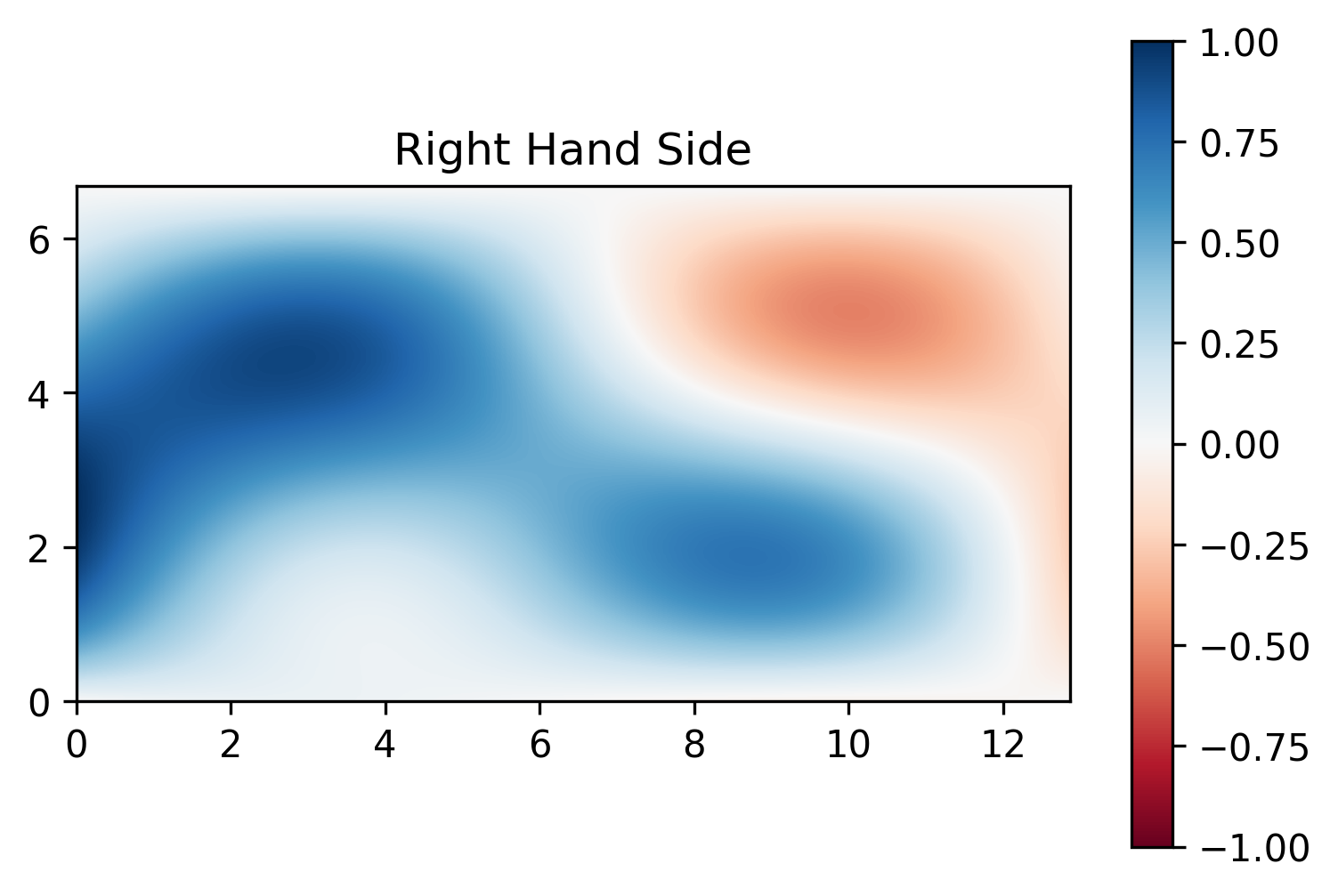}
%    \end{subfigure}
%    \begin{subfigure}[b]{0.465\textwidth}
%        \centering
		\includegraphics[width=0.465\textwidth]{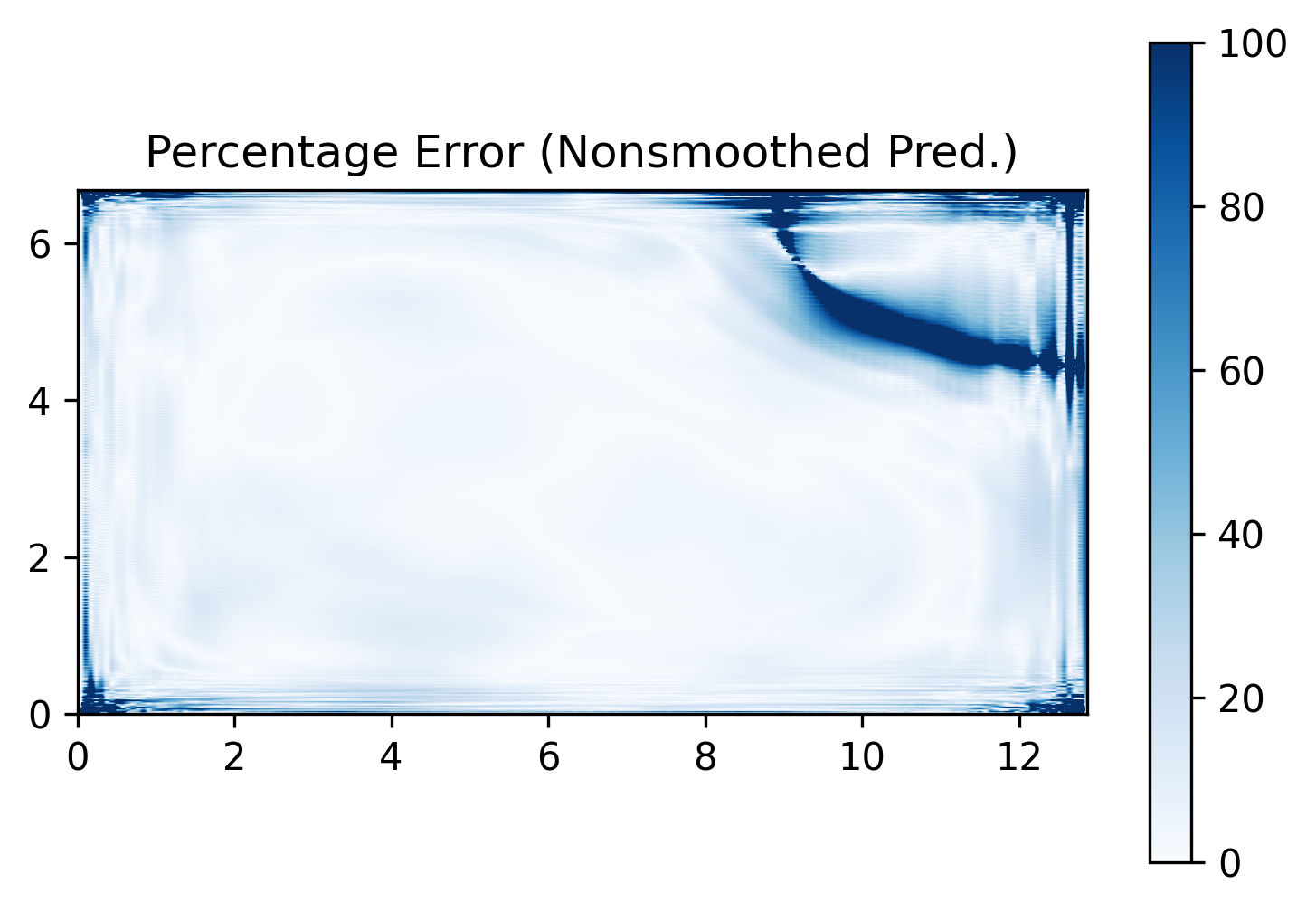}
%    \end{subfigure}
% \end{figure}
% \begin{figure}[h!]\ContinuedFloat
    \centering
%    \begin{subfigure}[b]{0.415\textwidth}
%        \centering
		\includegraphics[width = 0.415\textwidth]{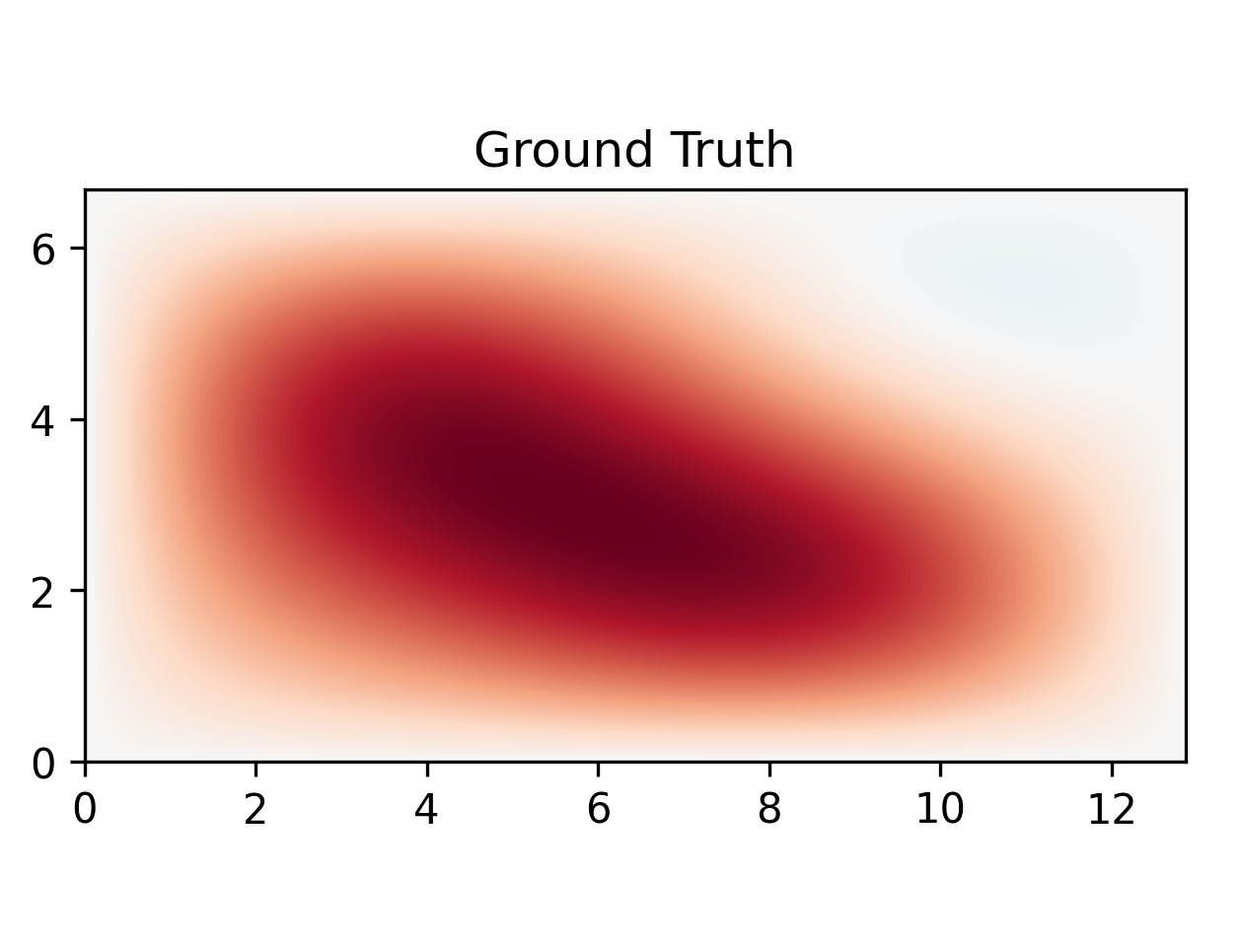}
%    \end{subfigure}
%    \begin{subfigure}[b]{0.49\textwidth}
%        \centering
		\includegraphics[width = 0.49\textwidth]{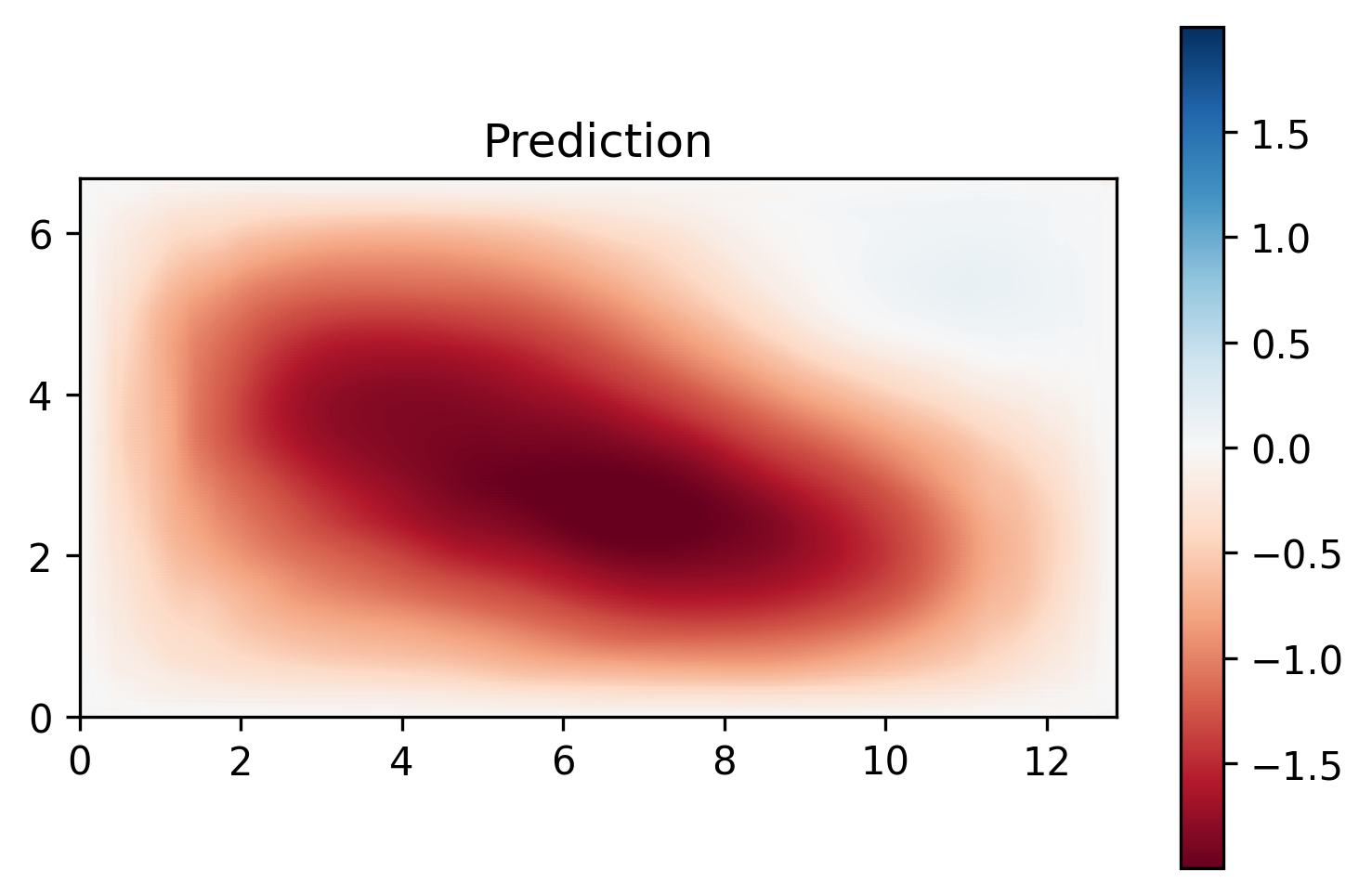}
%    \end{subfigure}
    \caption[Performance of the Homogeneous Poisson NN model on an example with grid size $384 \times 192$ and $\Delta = 3.47 \times 10^{-2}$]{Performance of the Homogeneous Poisson NN model on an example with grid size $384 \times 192$ and $\Delta = 3.47 \times 10^{-2}$. The HPNN sub-model can handle a variety of different aspect ratios well, including in this case where it predicts values within 10\% of the target for almost three-quarters of the grid points}
    \label{fig:hpnn_tset_example_2}
\end{figure}

Figures \ref{fig:hpnn_tset_example_1} and \ref{fig:hpnn_tset_example_2} depict the performance of the Homogeneous Poisson NN on two random examples generated in the same manner as the training data, with different ARs. The model displays good predictive accuracy with the majority of the predictions lying within 10\% of the target for the example in \figref{fig:hpnn_tset_example_1}, rising to over three quarters for the example in \figref{fig:hpnn_tset_example_2}. A substantial proportion of the predictions that lie outside this 10\% band are clustered around the contours where the solution $\phi = 0$, where high percentage errors occur despite good predictions in terms of the numerical value. The most notable inaccuracies (in terms of the absolute values) lie near points where the ground truth has a local maximum or minimum due to slight mispredictions of both the location and the value of these local extrema.

%\clearpage

The example in \figref{fig:hpnn_tset_example_2}, when juxtaposed against \figref{fig:hpnn_tset_example_1}, showcases the capability of the HPNN sub-model to deal with problems involving vastly differing ARs and grid sizes without retraining. This capability is crucial for Poisson CNN's intended aim as an acceleration step for iterative algorithms, as reusability for different grid parameters greatly enhances the attractiveness for this purpose.

\FloatBarrier
\subsubsection{Dirichlet BC NN}
\label{sec:dbcnn_tset_result}
\begin{figure}[h!]
 \centering
%    \begin{subfigure}[t]{0.395\textwidth}
%        \centering
		\includegraphics[width=0.395\textwidth]{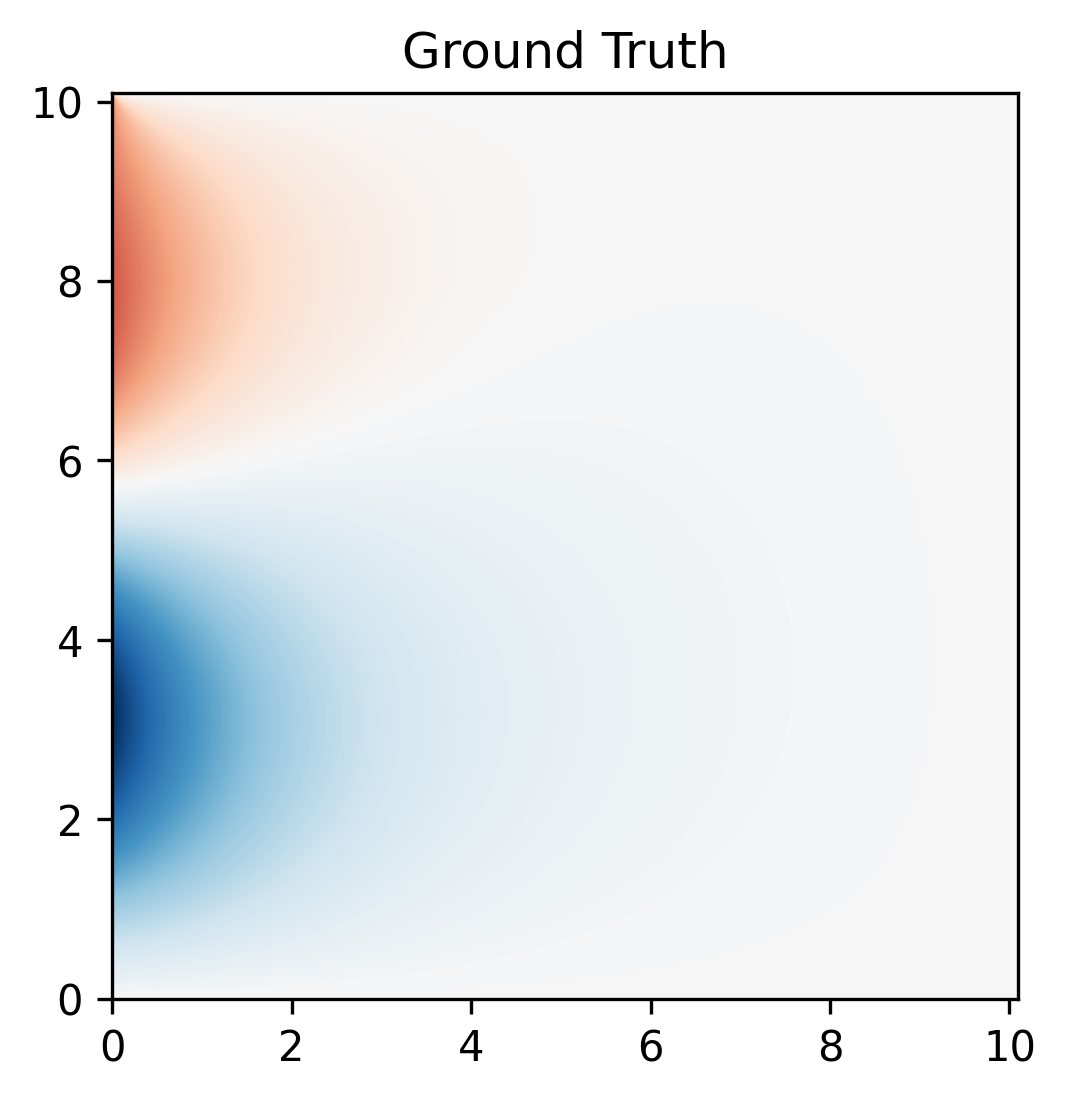}
%    \end{subfigure}
 %   \centering
 %   \begin{subfigure}[t]{0.49\textwidth}
 %       \centering
		\includegraphics[width=0.49\textwidth]{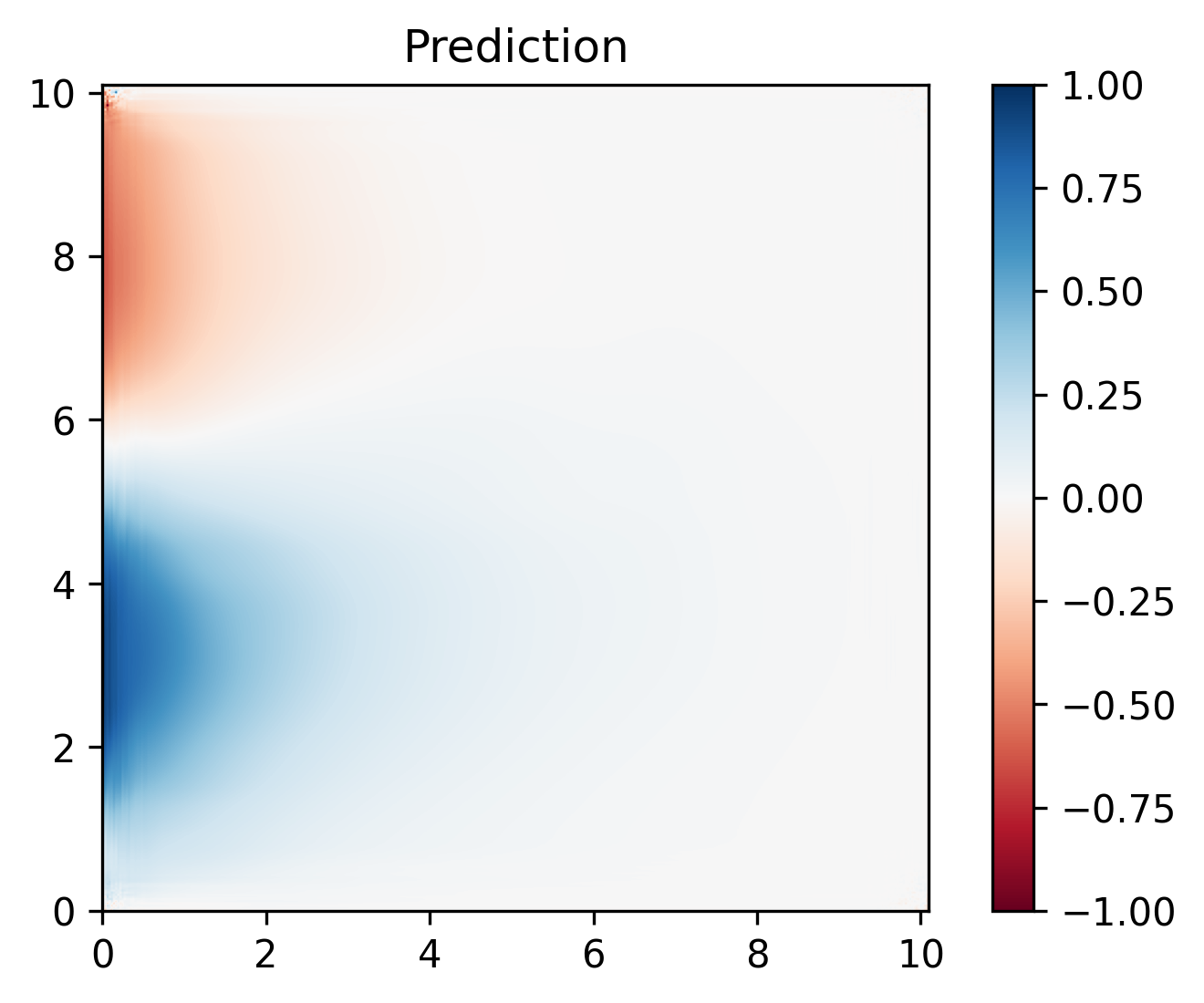}
 %   \end{subfigure}
% \end{figure}
% \begin{figure}\ContinuedFloat
%    \centering
%    \begin{subfigure}[t]{0.50\textwidth}
        \raisebox{6.5mm}{
%        \centering
		\includegraphics[width=0.49\textwidth]{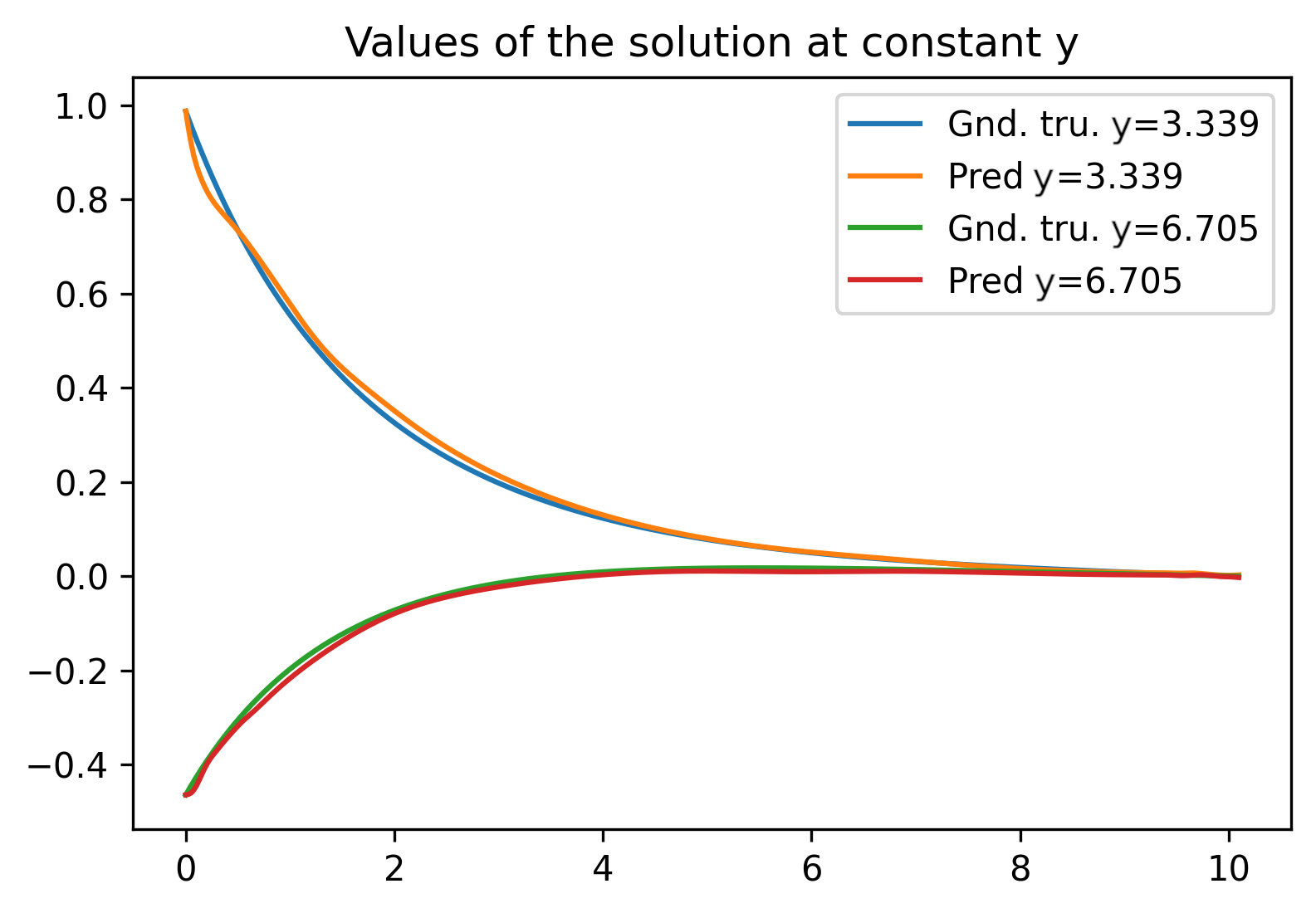}
		}
%    \end{subfigure}
%    \begin{subfigure}[b]{0.49\textwidth}
%        \centering
		\includegraphics[width=0.49\textwidth]{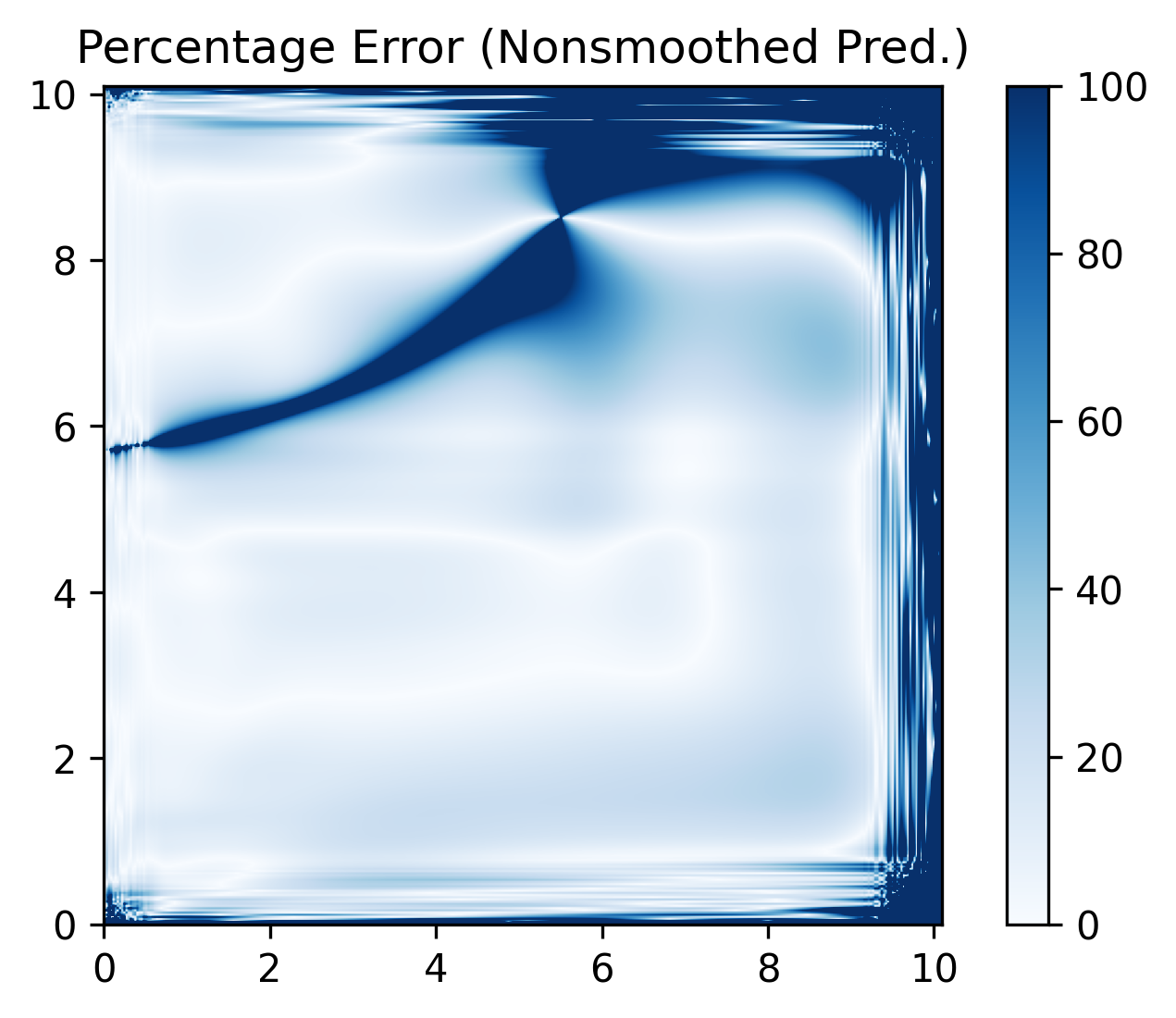}
 %   \end{subfigure}
	\caption{Performance of the DBCNN model on an example generated in the same manner as its training data, with a grid size of $384 \times 384$ and $\Delta = 2.63 \times 10^{-2}$. The DBCNN sub-model replicates the hyperbolic sine solution profile successfully}
    \label{fig:dbcnn_tset_example1}
\end{figure}

\begin{figure}[h!]
 %   \begin{subfigure}[t]{0.40\textwidth}
        \centering
		\includegraphics[width=0.4\textwidth]{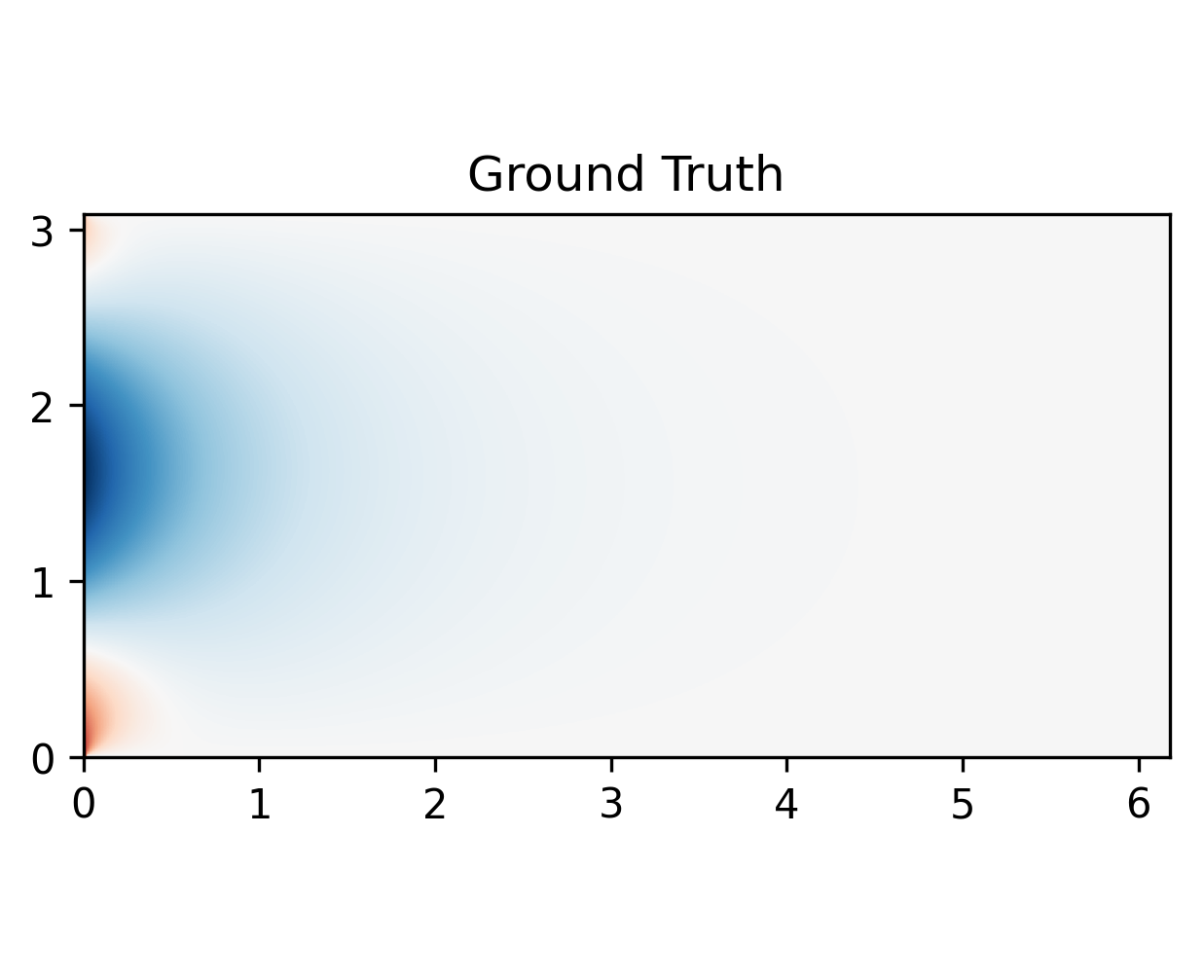}
%    \end{subfigure}
%    \centering
%    \begin{subfigure}[t]{0.49\textwidth}
%        \centering
		\includegraphics[width=0.49\textwidth]{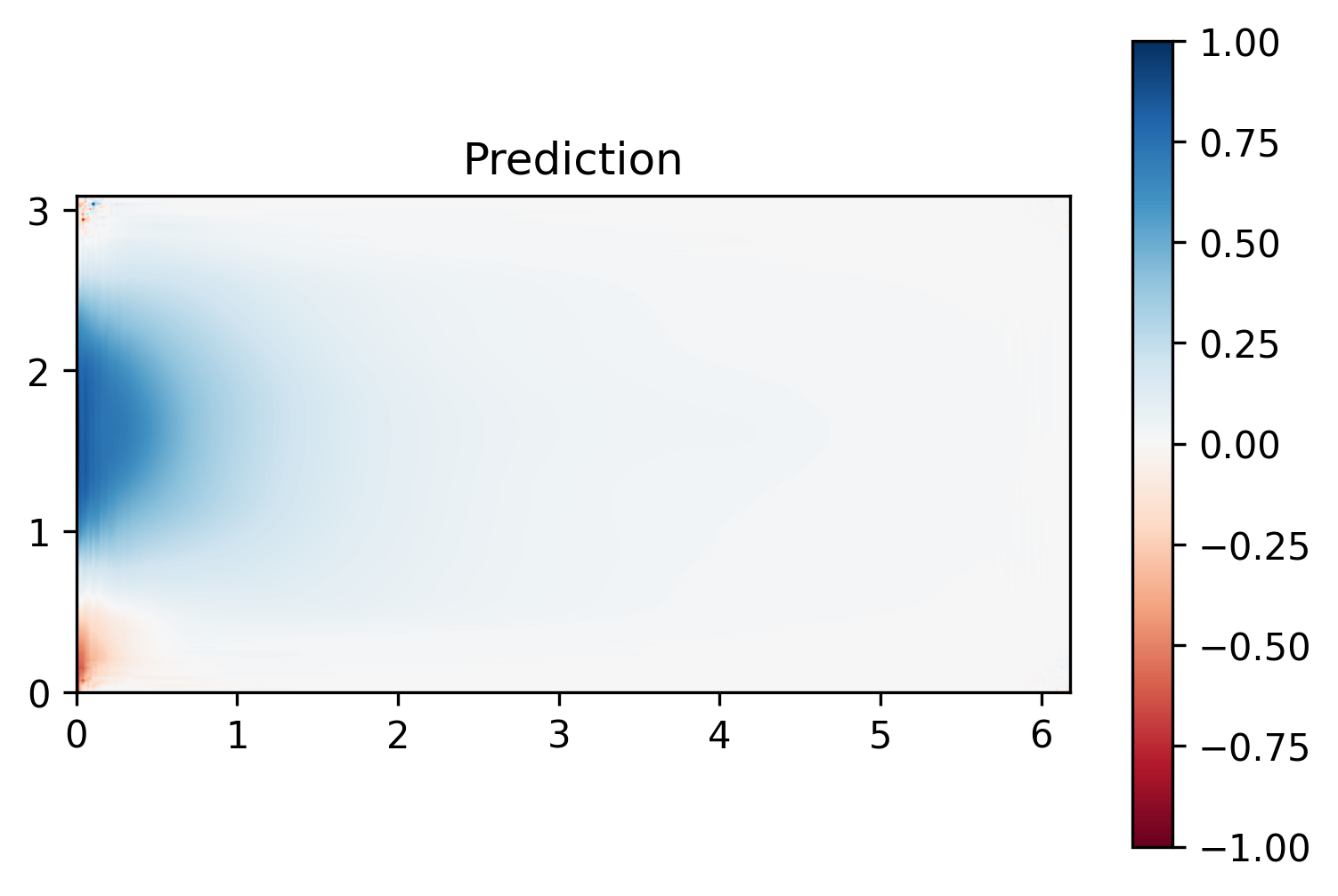}
 %   \end{subfigure}
 %   \centering
 %   \begin{subfigure}[t]{0.50\textwidth}
        %\raisebox{1mm}{
 %       \centering
		\includegraphics[width=0.49\textwidth]{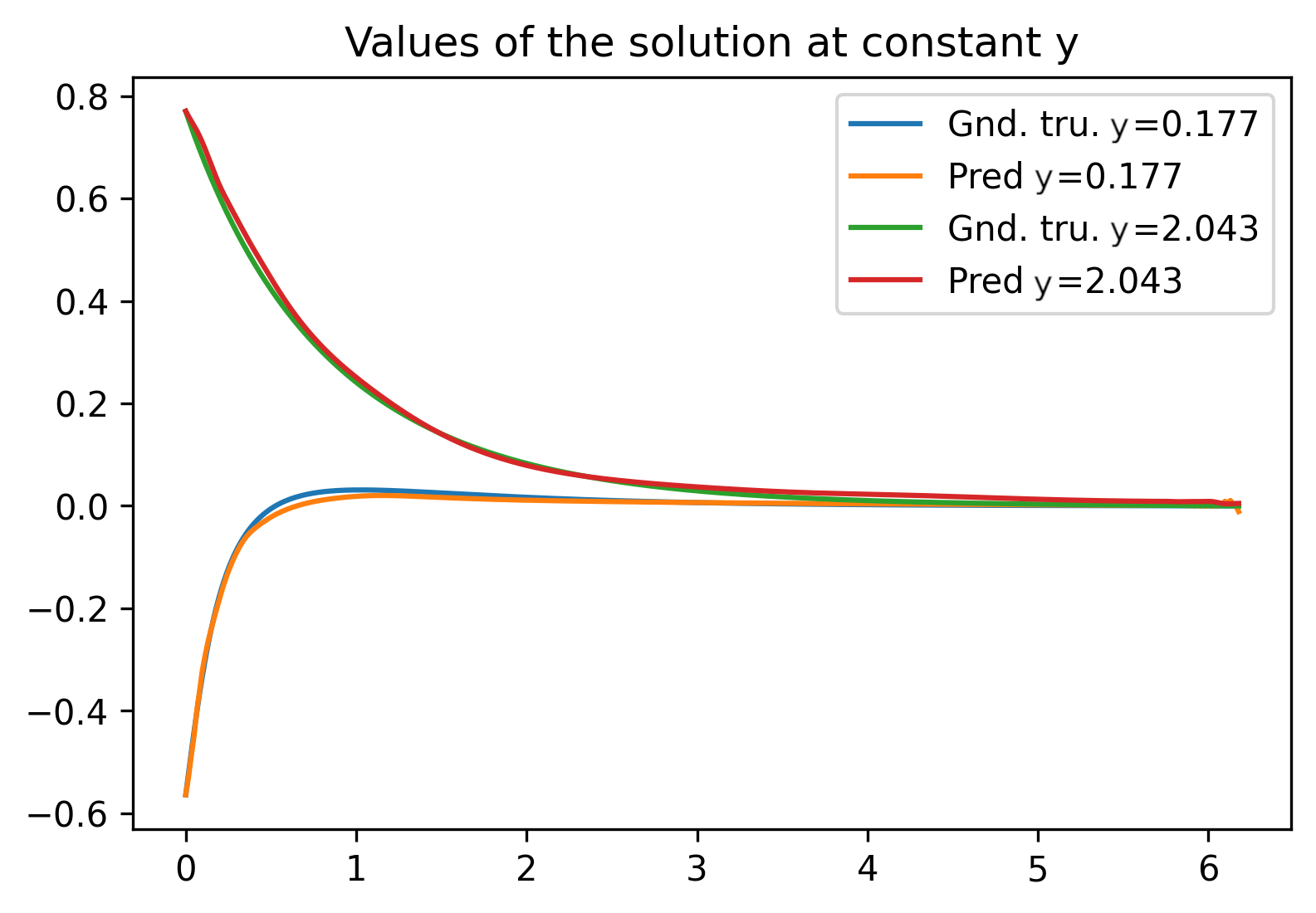}
		%}
 %   \end{subfigure}
 %   \begin{subfigure}[t]{0.49\textwidth}
 %       \centering
		\includegraphics[width=0.49\textwidth]{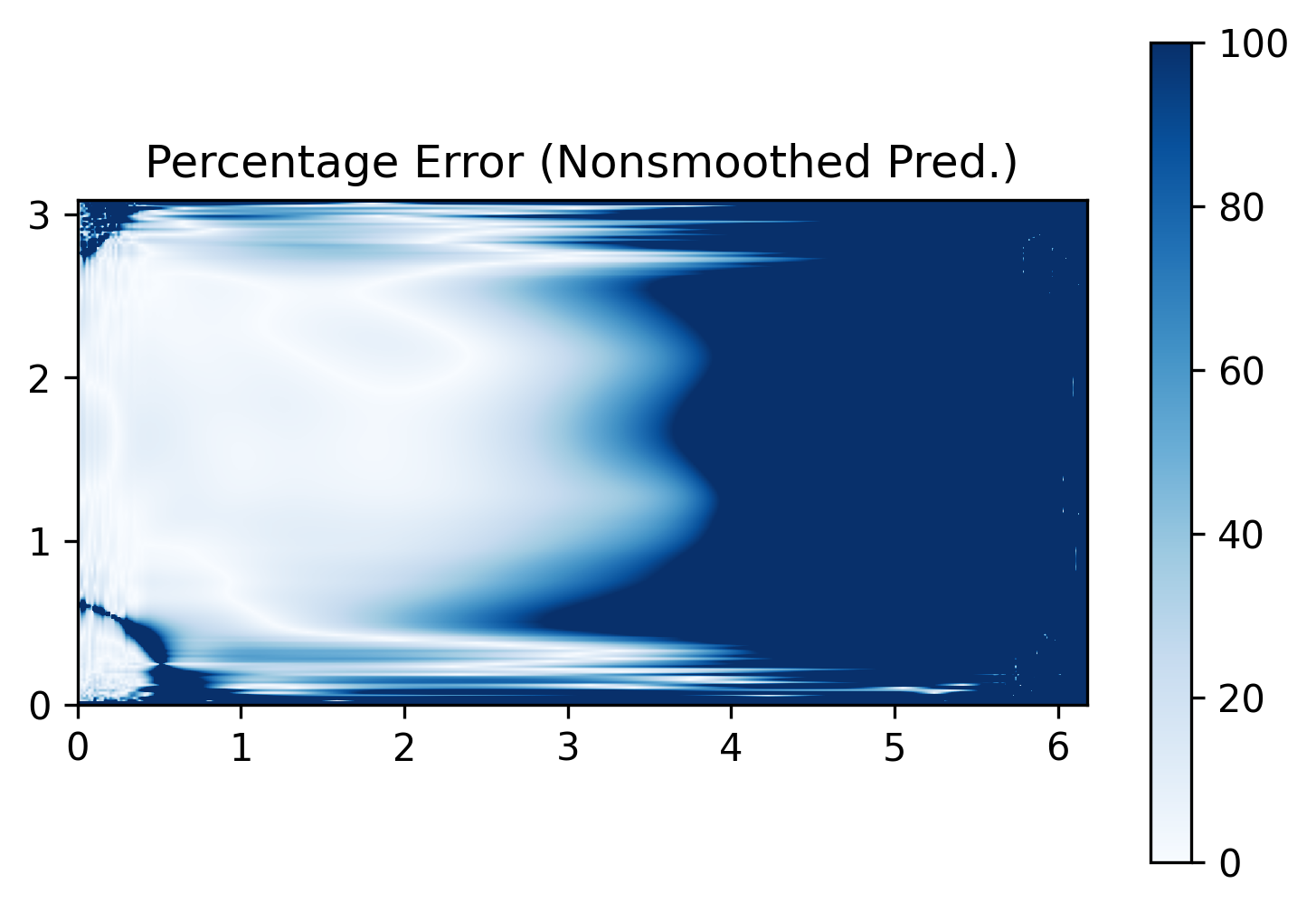}
 %   \end{subfigure}
	\caption{Performance of the DBCNN model on an example generated in the same manner as its training data, with a grid size of $382 \times 197$ and $\Delta = 1.59 \times 10^{-2}$. Similar to the HPNN, the DBCNN can handle different aspect ratios effectively, replicating the solution profile properly}
    \label{fig:dbcnn_tset_example2}
\end{figure}

As shown in Figures \ref{fig:dbcnn_tset_example1} and \ref{fig:dbcnn_tset_example2}, the Dirichlet BC NN reproduces the multigrid solution with a good degree of accuracy, replicating the hyperbolic sine solution profile well. The largest absolute errors occur due to some oscillatory behaviour near the left boundary, caused by the padding-related issues explained in \secref{sec:training} being magnified by the large solution magnitudes in this region. \secref{sec:post-smoothing} discusses the application of Jacobi post-smoothing to resolve this issue. Jacobi smoothing substantially reduces the peak normalized error down to $2.49 \times 10^{-1}$ and $1.96 \times 10^{-1}$ for Figures \ref{fig:dbcnn_tset_example1} and \ref{fig:dbcnn_tset_example2} respectively -- an almost three quarter reduction compared to the values in \tabref{tab:results_summary_table}. The performance in terms of the percentage of grid points with predictions within 10\% of the target are $41\%$ and $31\%$ for the two cases, slightly worse than the HPNN sub-model. Conversely, the MAE for the case in \figref{fig:dbcnn_tset_example1} is $84\%$ lower than the MAE for the HPNN sub-model in \figref{fig:hpnn_tset_example_1} as shown in \tabref{tab:results_summary_table}.

The seemingly contradictory MAPE and MAE values are explained by the observation that $\%$ errors increase progressively as we move in the positive horizontal direction. For the DBCNN, $\%$ errors rise towards the right boundary since the sinh-shaped solution profile in this direction rapidly reduces the solution magnitudes. Thus, absolute errors remain very small despite large relative errors. ARs further complicates this comparison as the solution decays quicker in terms of normalized horizontal coordinates $x/L_x$ for high AR domains as seen in \figref{fig:dbcnn_tset_example2}. Hence, MAPEs are relatively small and MAEs are relatively large for low AR cases and the opposite is true for higher ARs.

\clearpage
\subsubsection{Poisson CNN (Full model)}
\label{sec:tset_pcnn}
\begin{figure}[h!]
    \centering
    \begin{subfigure}[b]{0.48\textwidth}
        \raisebox{1mm}{
        \centering
		\includegraphics[width=0.99\textwidth]{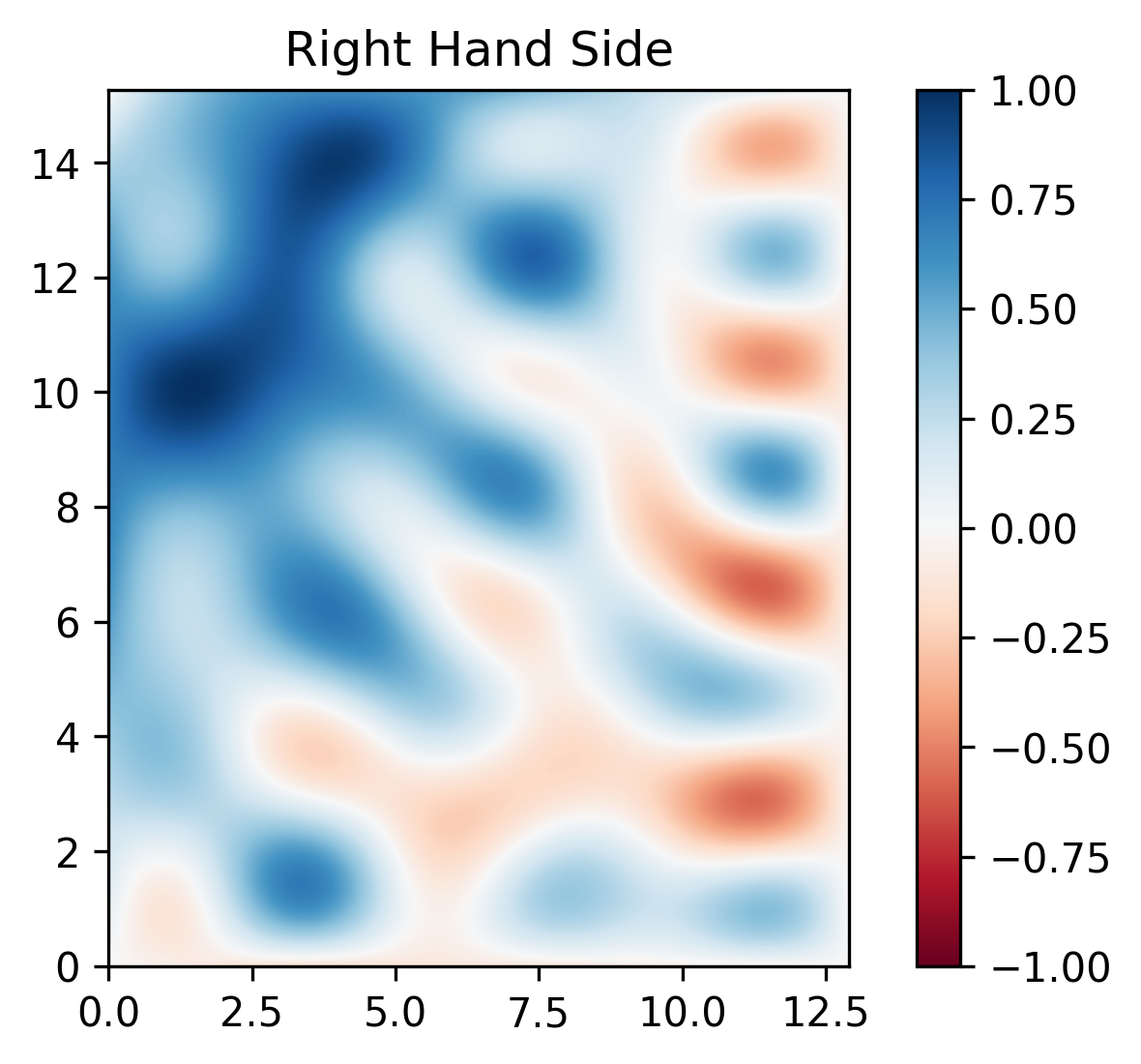}
		}
    \end{subfigure}
    \begin{subfigure}[b]{0.48\textwidth}
        \centering
		\includegraphics[width=0.99\textwidth]{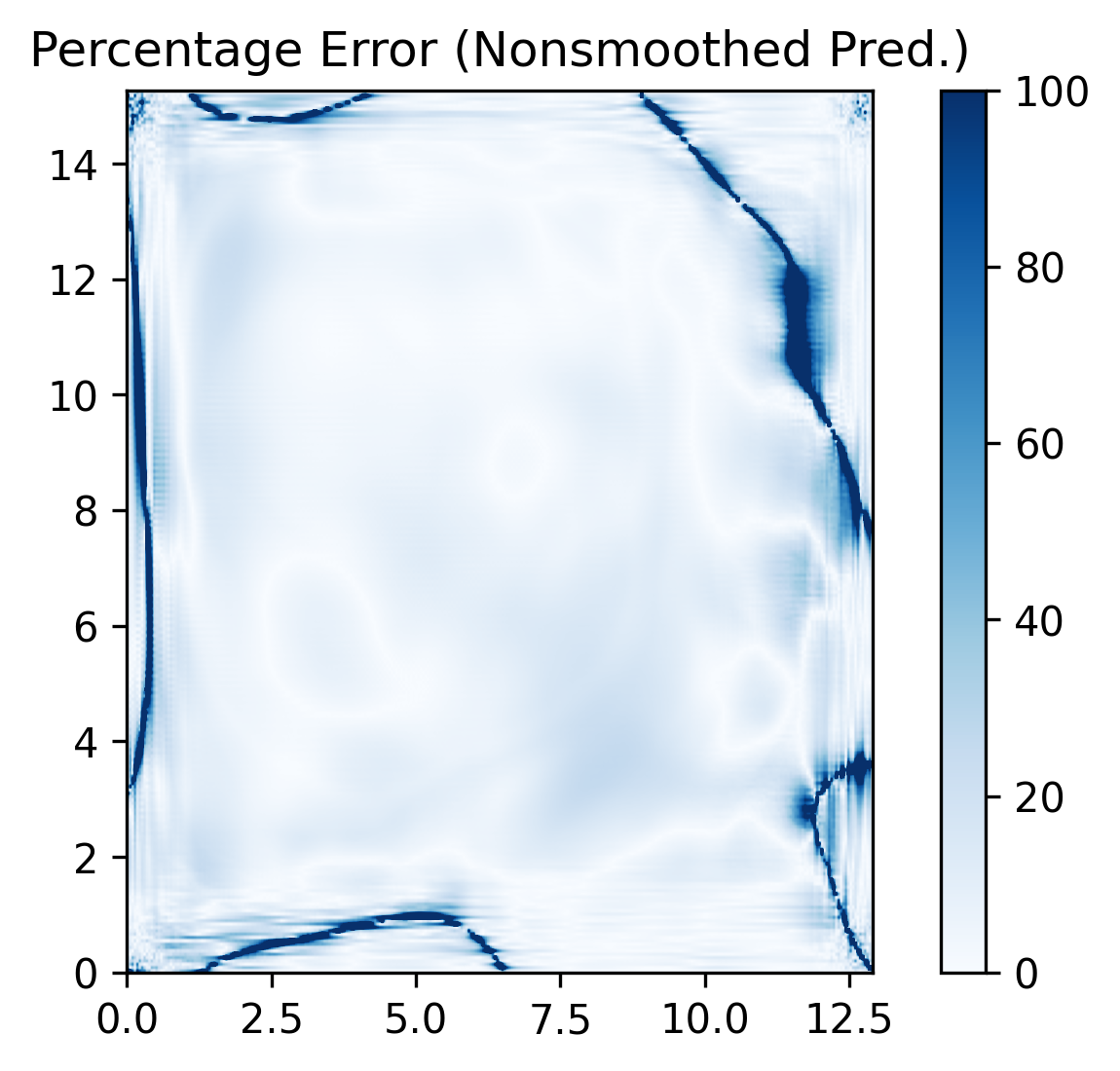}
    \end{subfigure}
% \end{figure}
% \begin{figure}\ContinuedFloat
    \centering
    \begin{subfigure}[b]{0.385\textwidth}
        \centering
		\includegraphics[width = 0.99\textwidth]{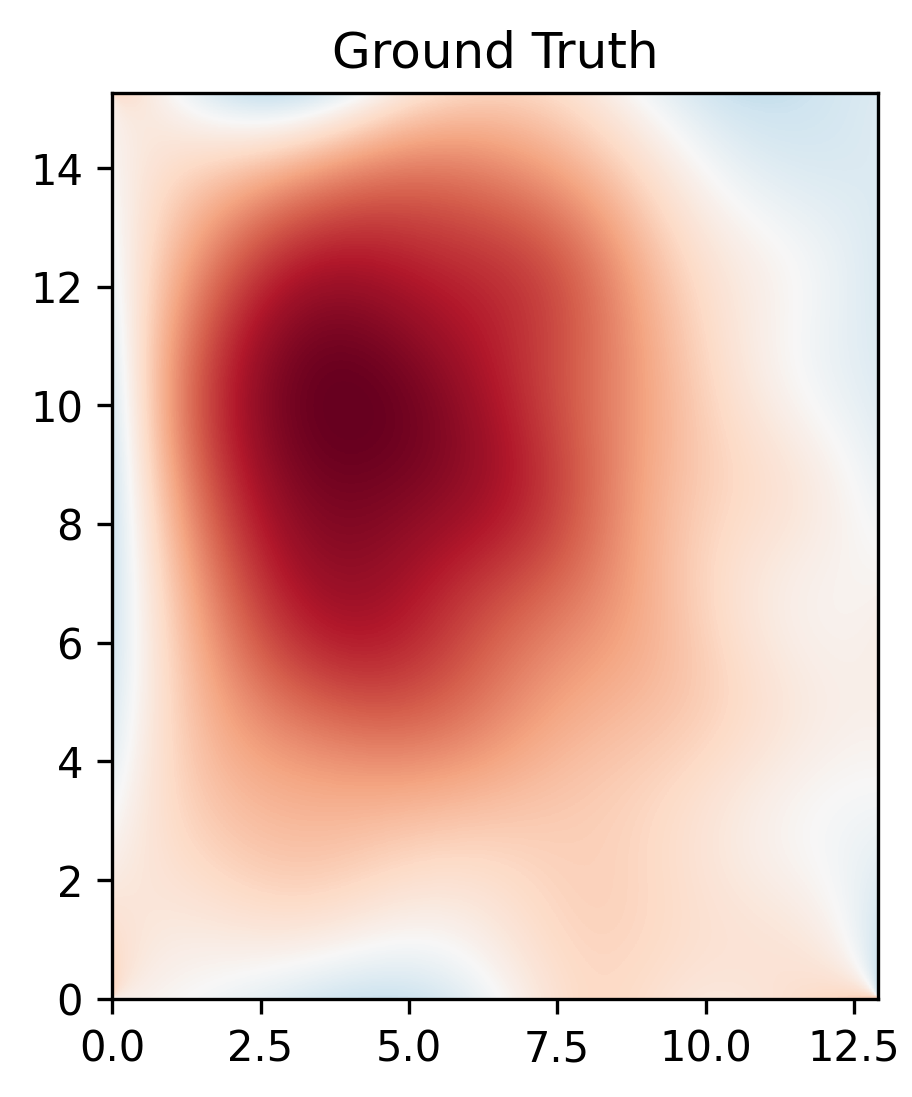}
    \end{subfigure}
    \begin{subfigure}[b]{0.47\textwidth}
        \centering
		\includegraphics[width = 0.99\textwidth]{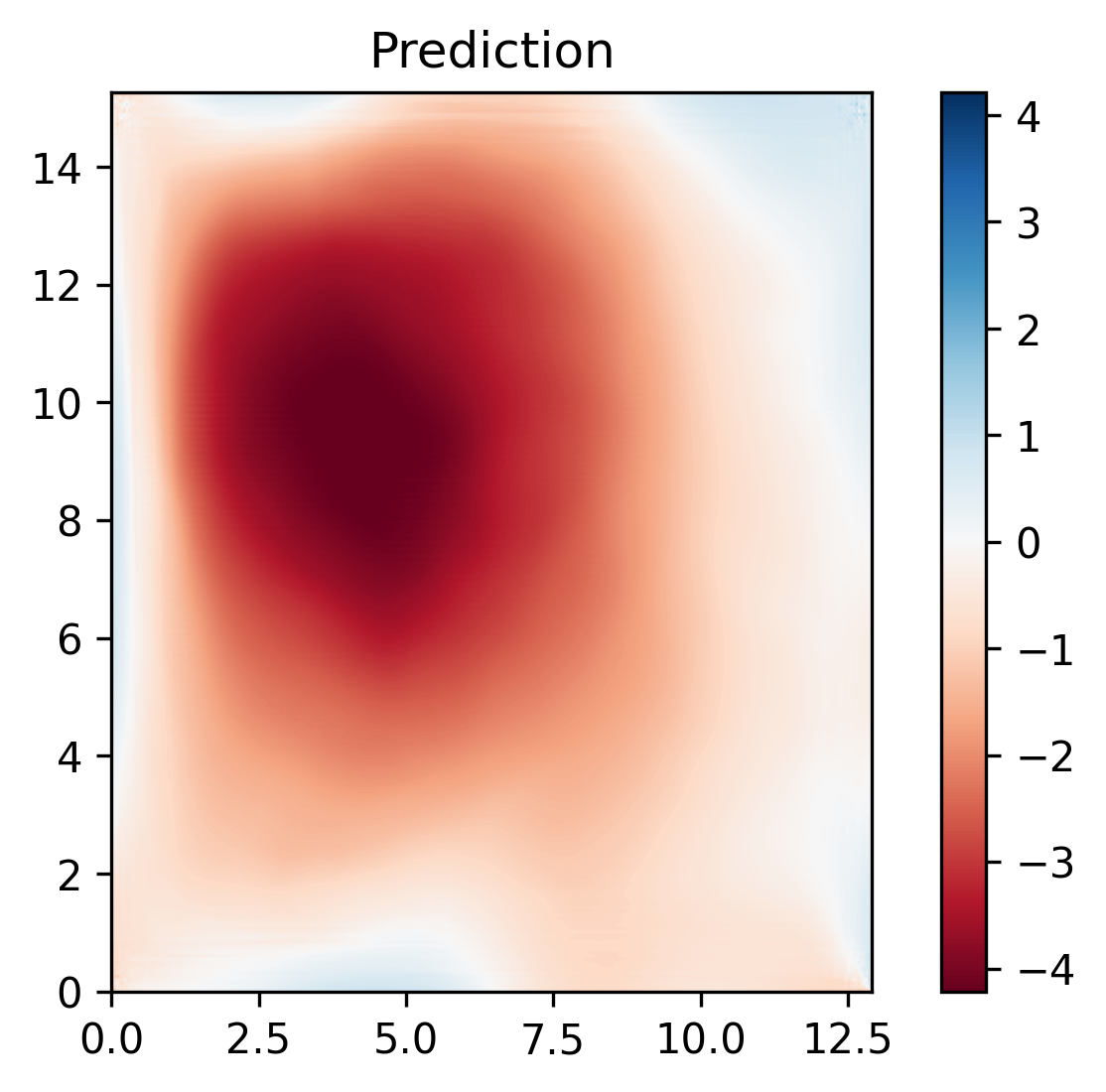}
    \end{subfigure}
    \caption[Performance of the Poisson CNN model on an example with grid size $263 \times 311$ and $\Delta = 4.91 \times 10^{-2}$]{Performance of the Poisson CNN model on an example with grid size $263 \times 311$ and $\Delta = 4.91 \times 10^{-2}$. The components predicted by the sub-models reconstruct the solution accurately, demonstrating the flexibility of the decomposition}
    \label{fig:pcnn_tset_example_3}
\end{figure}

\figref{fig:pcnn_tset_example_3} depicts the performance of the full model on a randomly generated Poisson problem. The performance of the model in terms of MAPE is $9.81\%$ and over two thirds of grid points have predictions within $10\%$ of the target. The decomposition of the Poisson problem as shown in \equref{eq:bc_strategy_decomposition_3} performs well and the Poisson CNN architecture proposed is capable of providing good estimates for solutions to Poisson problems with 4 inhomogeneous BCs, substantially exceeding the performance of the individual sub-models on their respective sub-problems.

\clearpage
\subsection{Taylor-Green Vortex problem}
\label{sec:tgv}
The Taylor-Green Vortex (TGV) in two-dimensions is a well-known analytical solution to the Navier-Stokes equations. The pressure component of the TGV, with the time-dependent term and density set as unity for simplicity, presents an easy-to-construct analytical test case. It is a benchmark frequently used in CFD community and is representative of Poisson equations encountered in practical applications. We can construct a Poisson problem in the domain $[0,\pi] \times [0,\pi]$ by setting the solution $\phi$ as the TGV pressure field
\begin{align}
    \phi = -\frac{1}{4}(\cos(2x) + \cos(2y)) \\
    \therefore \mathrm{RHS} = \nabla^2 \phi = \cos(2x)+\cos(2y),
    \label{equ:tgv_rhs}
\end{align}{}
with the BC along each boundary defined as
\begin{equation}
    b(t) = -(\cos(2t) + 1)/4,
    \label{equ:tgv_bc}
\end{equation}{}
where $t$ is the coordinate along the boundary. As a benchmark case, we investigate the performance of the model on this problem with a grid size towards the middle of the range the model has seen during training -- $255 \times 255$.

\begin{figure}[h!]
    \centering
    \begin{subfigure}[b]{0.42\textwidth}
        \centering
		\includegraphics[width=0.99\textwidth]{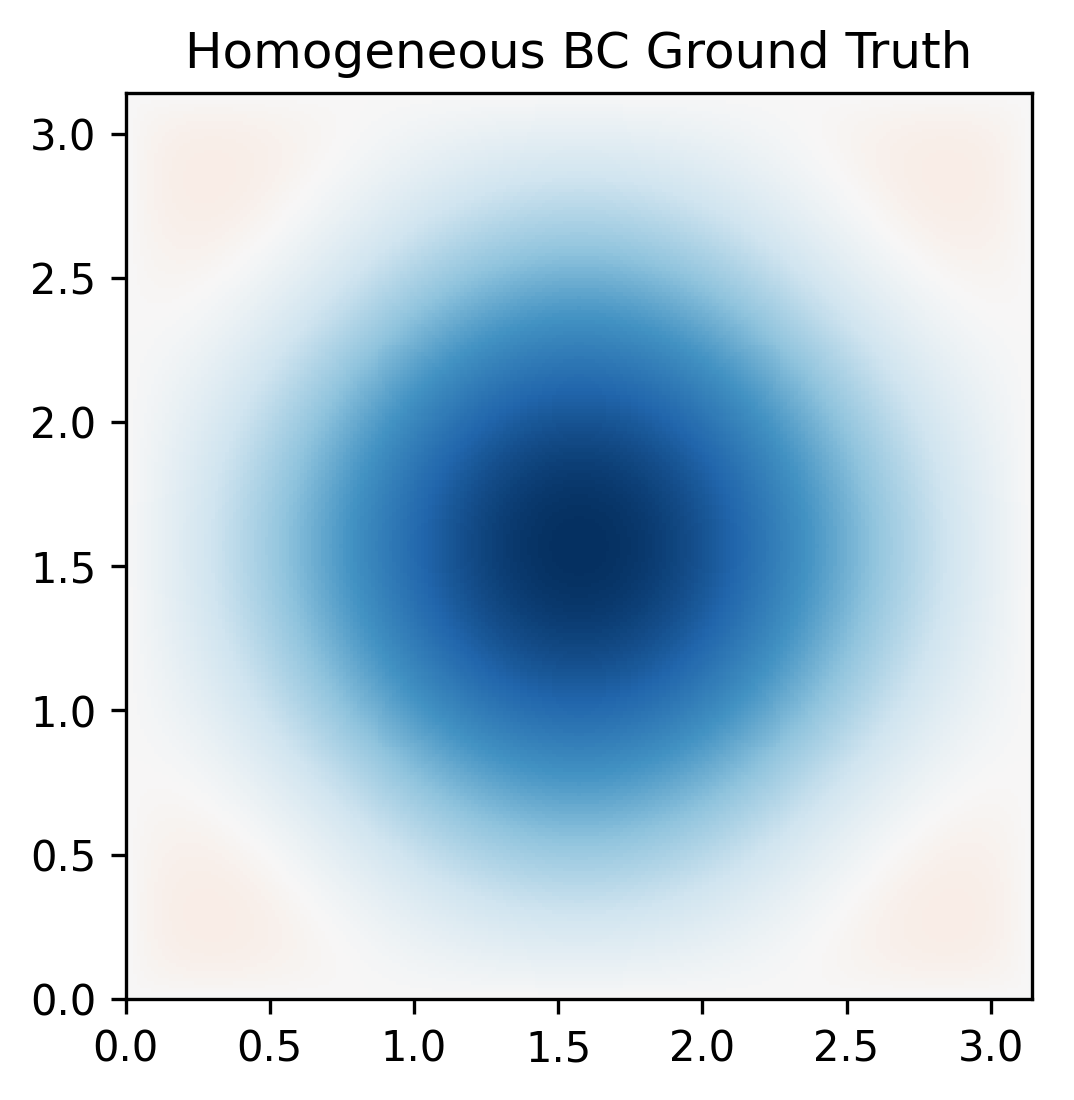}
    \end{subfigure}
    \begin{subfigure}[b]{0.51\textwidth}
        \centering
		\includegraphics[width=0.99\textwidth]{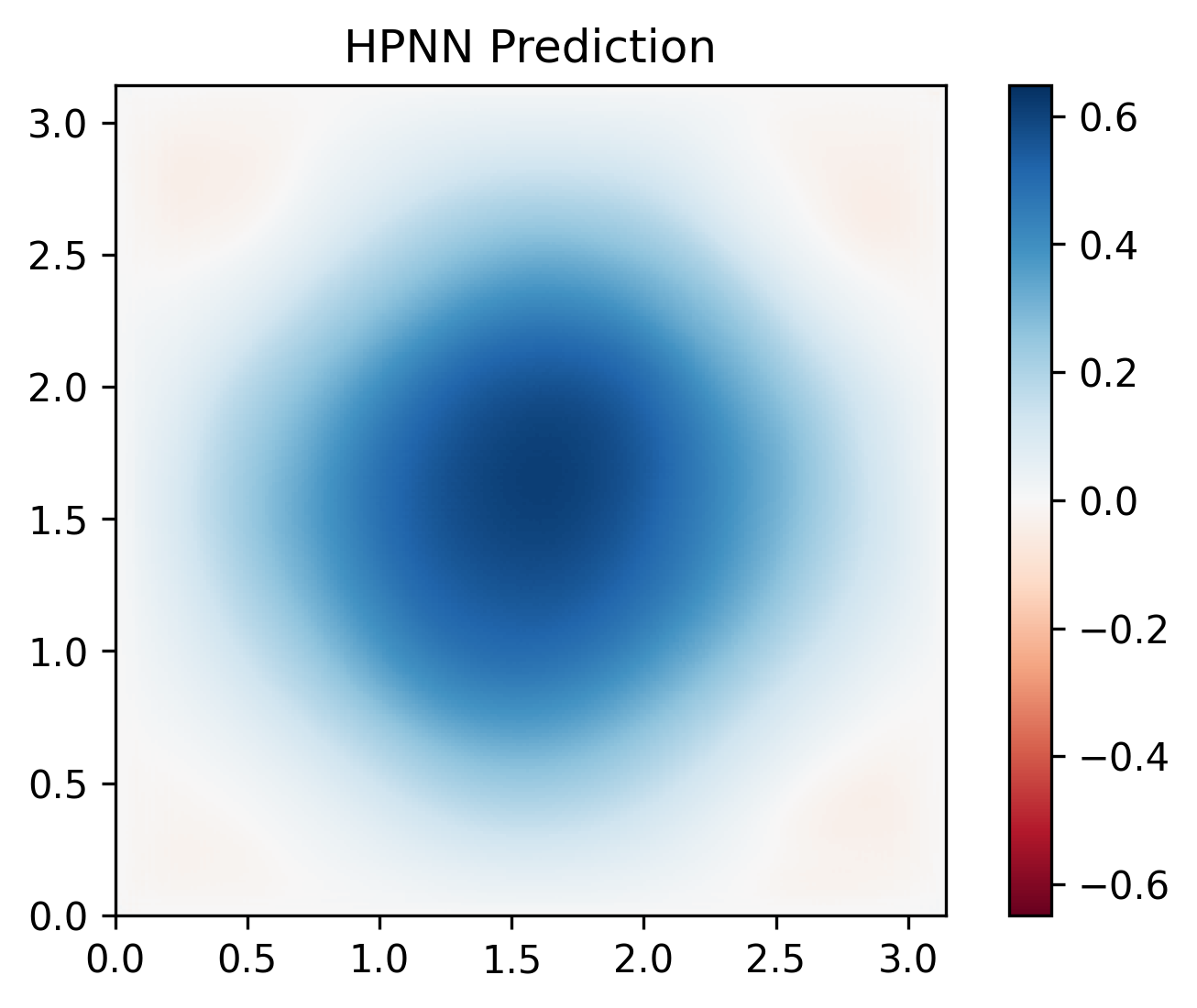}
    \end{subfigure}
% \end{figure}
% \begin{figure}\ContinuedFloat
    \centering
    \begin{subfigure}[b]{0.47\textwidth}
        \centering
		\includegraphics[width=0.99\textwidth]{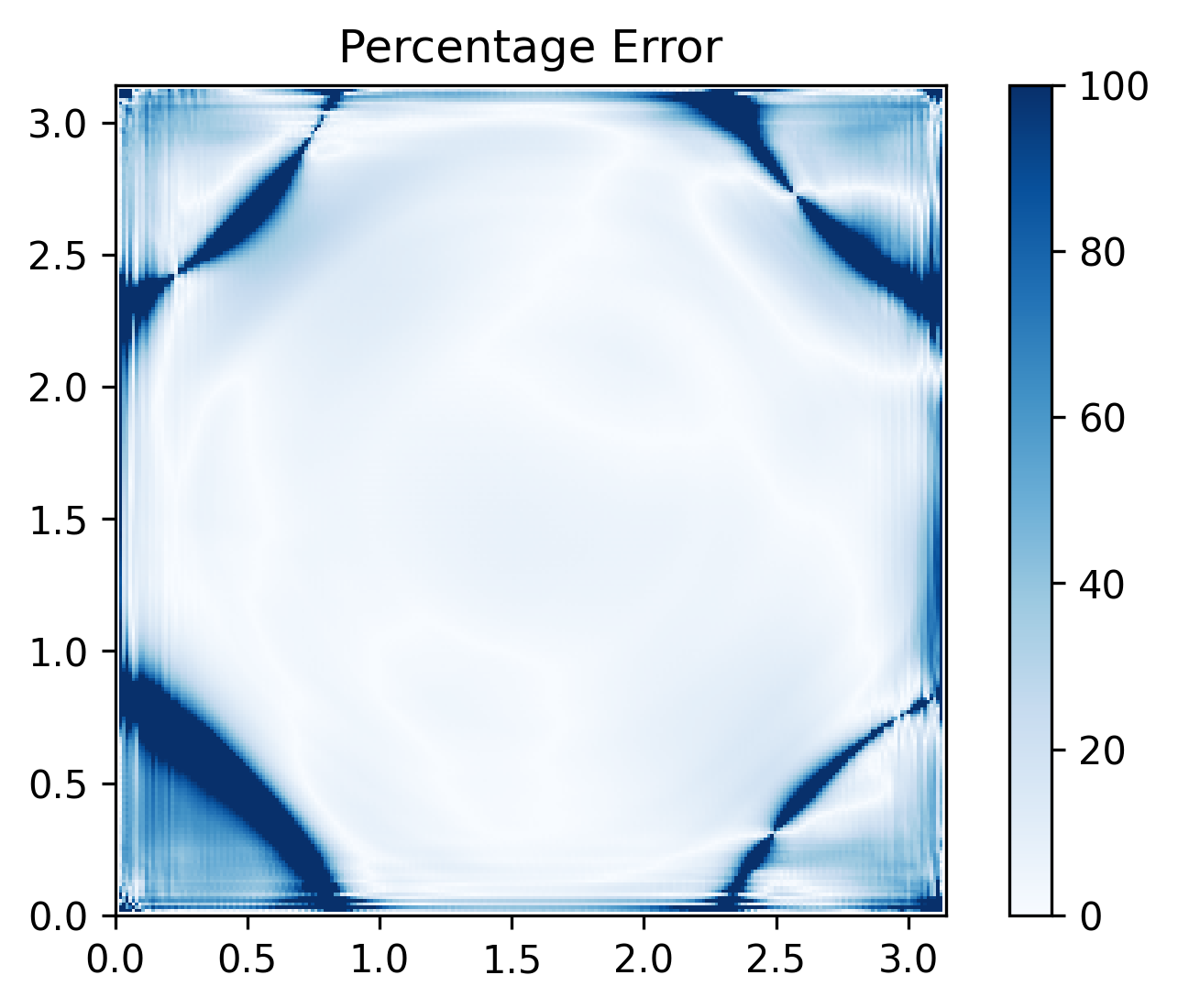}
    \end{subfigure}
    \begin{subfigure}[b]{0.47\textwidth}
        \centering
		\includegraphics[width=0.99\textwidth]{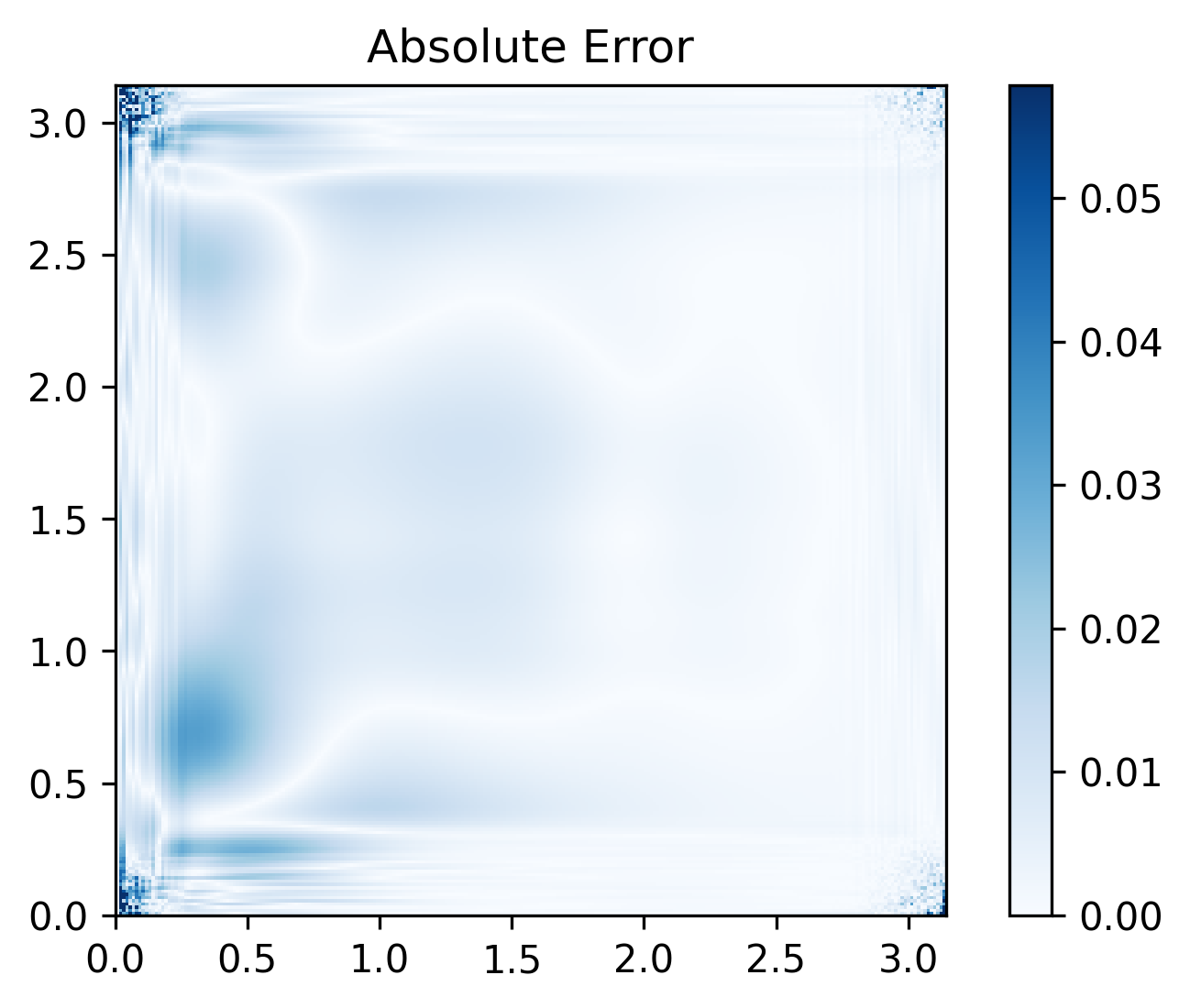}
    \end{subfigure}
    \caption{Prediction of the HPNN sub-model on the TGV case, compared to multigrid. Though the problem is materially dissimilar to the training set, the HPNN sub-model performs in line with the 600 example average in \tabref{tab:results_summary_table}}
    \label{fig:tgv_hpnn_pred}
\end{figure}

\begin{figure}[h!]
    \centering
    \centering
    \begin{subfigure}[b]{0.42\textwidth}
        \centering
		\includegraphics[width=0.99\textwidth]{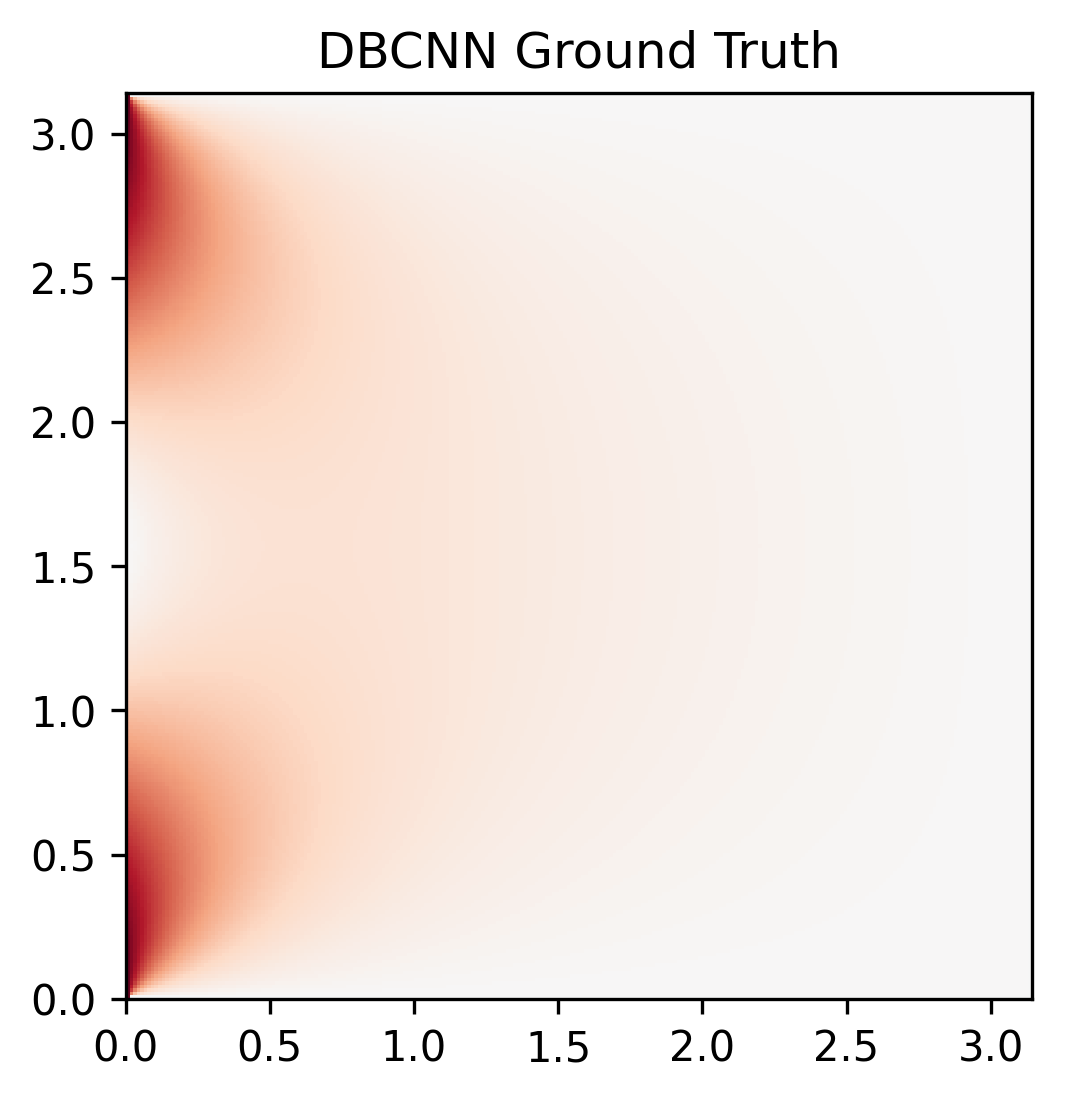}
    \end{subfigure}
    \begin{subfigure}[b]{0.51\textwidth}
        \centering
		\includegraphics[width=0.99\textwidth]{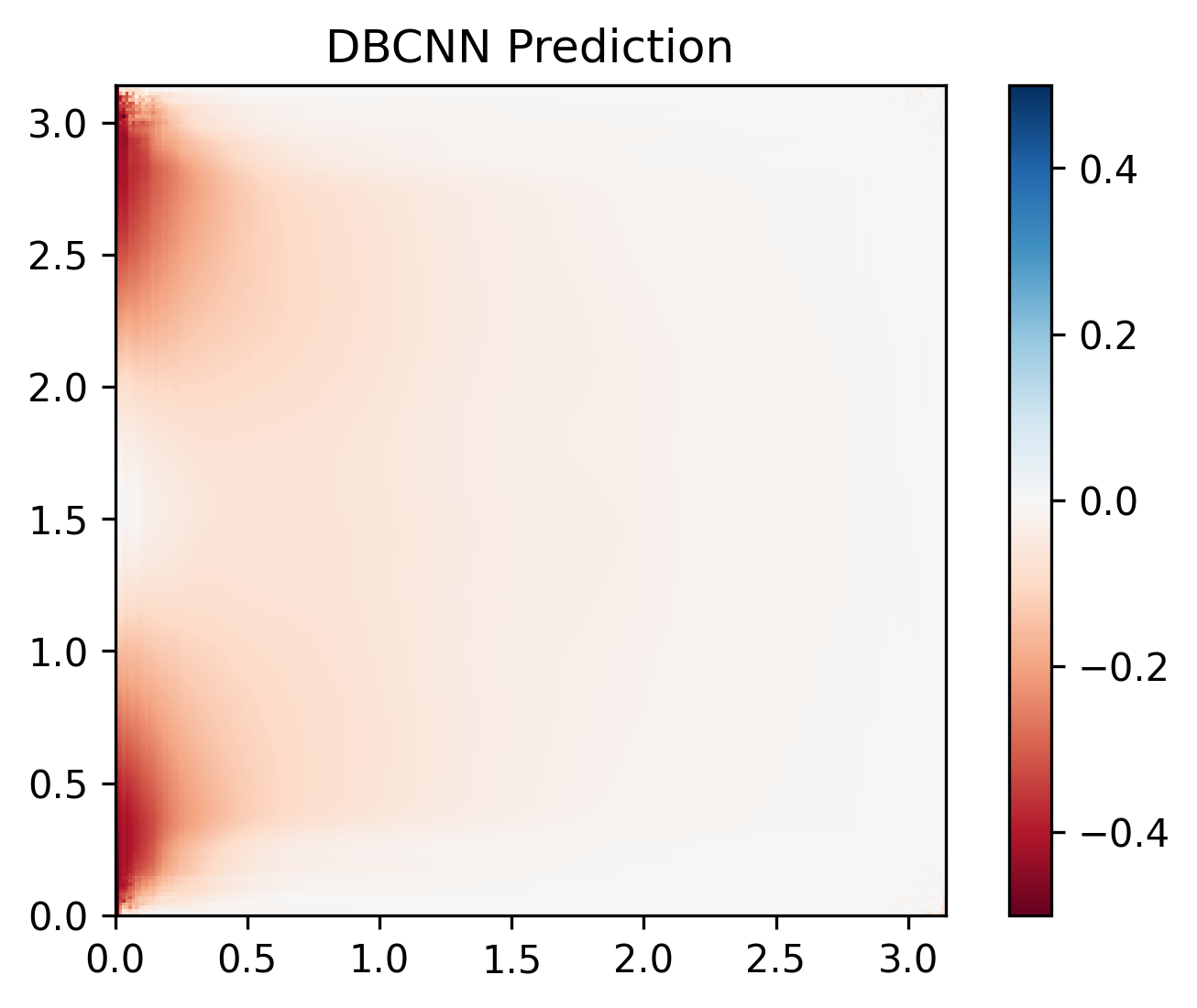}
    \end{subfigure}
% \end{figure}
% \begin{figure}\ContinuedFloat
    \begin{subfigure}[b]{0.47\textwidth}
        \centering
		\includegraphics[width=0.99\textwidth]{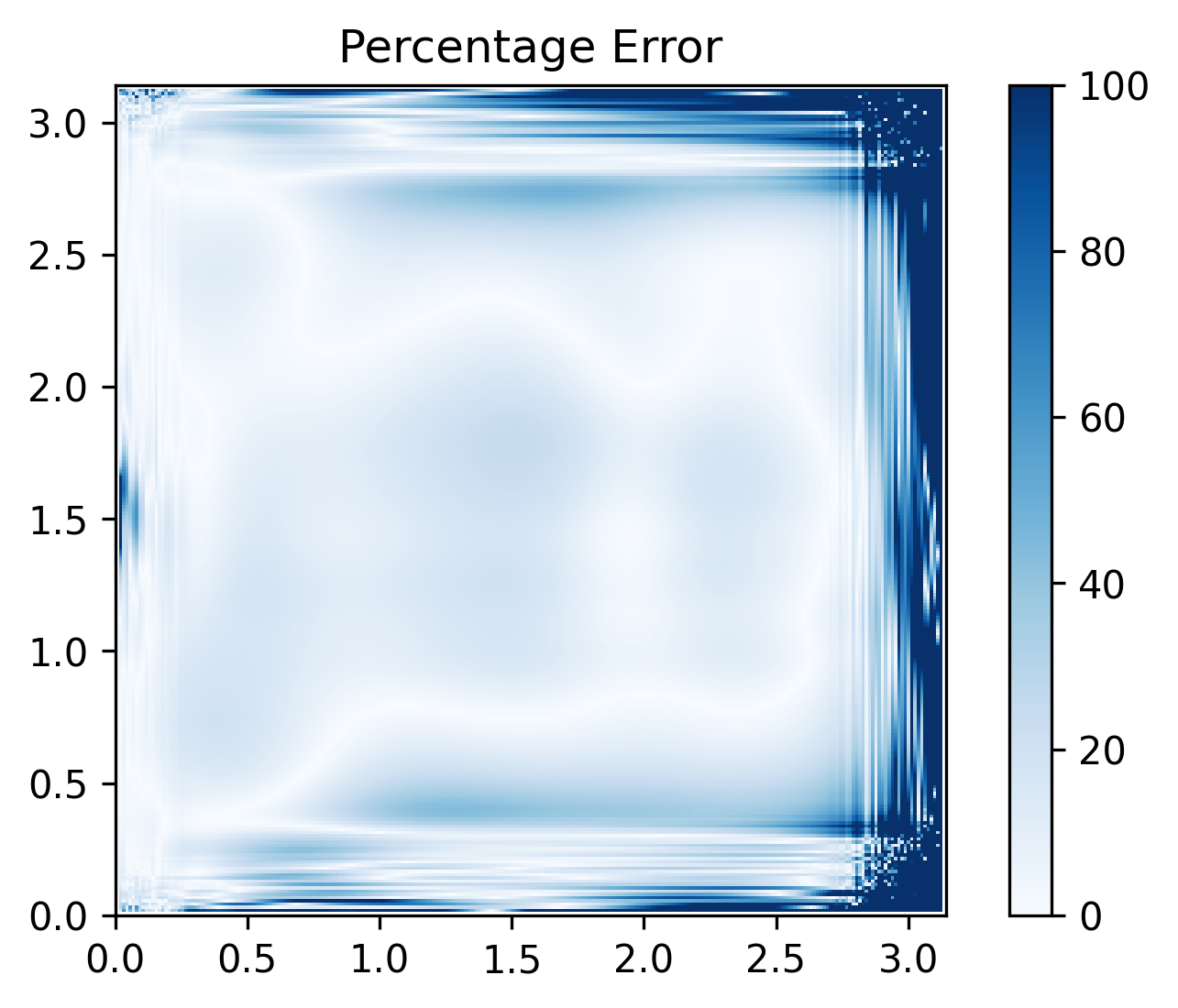}
    \end{subfigure}
    \begin{subfigure}[b]{0.47\textwidth}
        \centering
		\includegraphics[width=0.99\textwidth]{tgv_255x255_dx1.23e-02_dbcnn_errormap.png}
    \end{subfigure}
    \caption{Prediction of the DBCNN sub-model on the TGV case, compared to multigrid. The DBCNN sub-model performs well in this test case, marginally better than the 600 example average in \tabref{tab:results_summary_table}, although mild artefacting is visible near the corners}
    \label{fig:tgv_dbcnn_pred}
\end{figure}

\begin{figure}[h!]
    \centering
    \begin{subfigure}[b]{0.42\textwidth}
        \centering
		\includegraphics[width=0.99\textwidth]{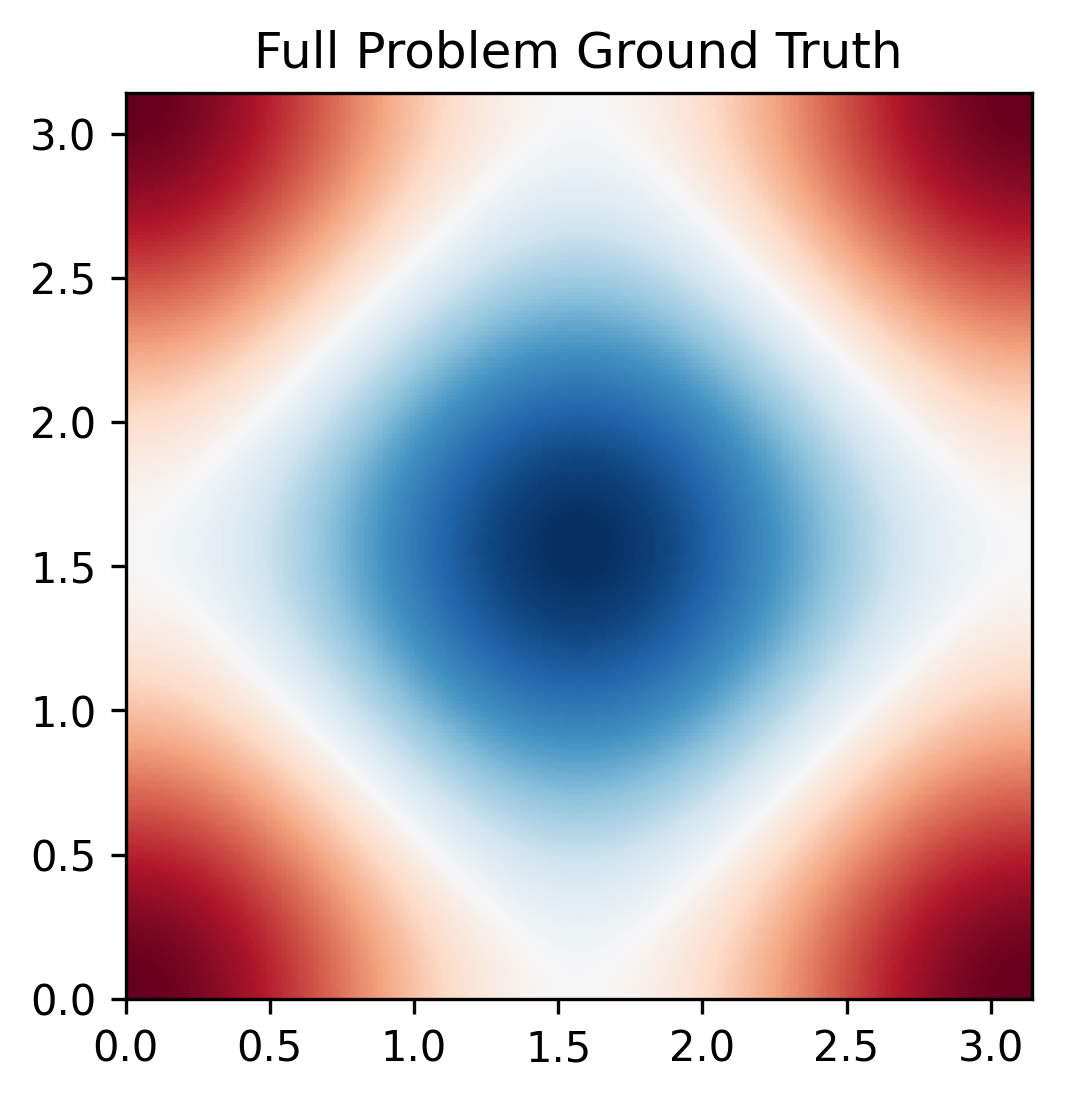}
    \end{subfigure}
    \begin{subfigure}[b]{0.51\textwidth}
        \centering
		\includegraphics[width=0.99\textwidth]{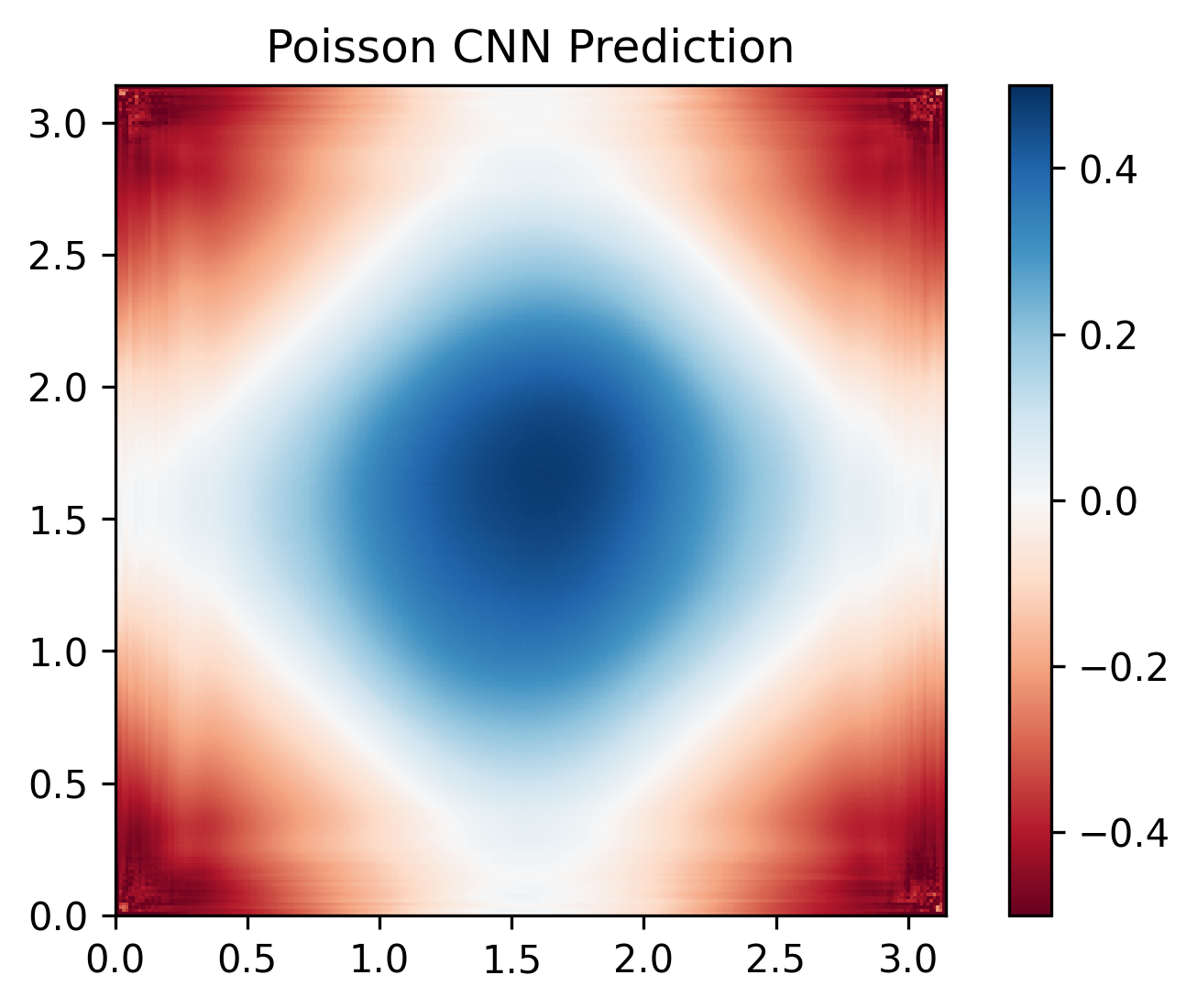}
    \end{subfigure}
% \end{figure}
% \begin{figure}\ContinuedFloat
    \begin{subfigure}[b]{0.47\textwidth}
        \centering
		\includegraphics[width=0.99\textwidth]{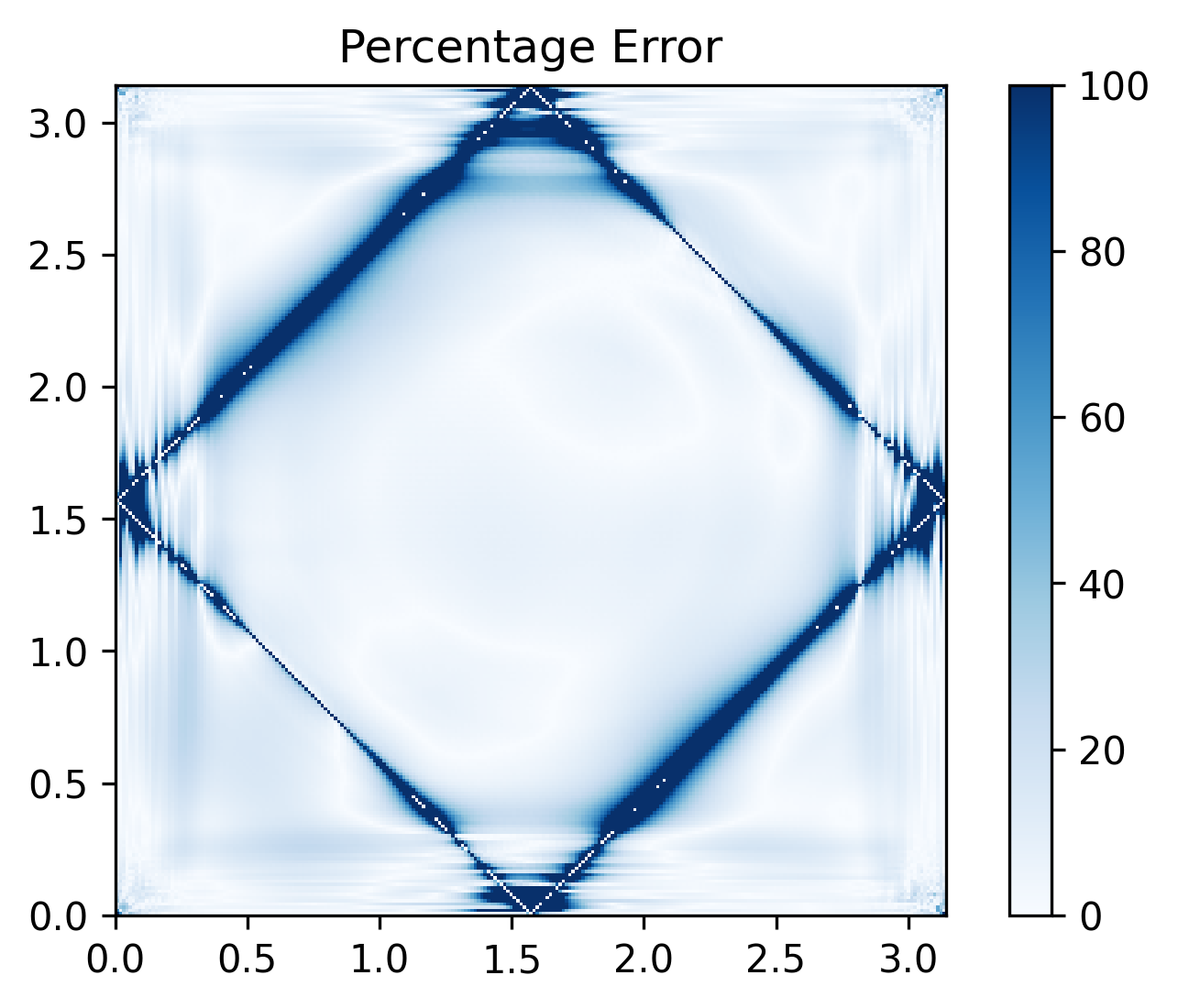}
    \end{subfigure}
    \begin{subfigure}[b]{0.47\textwidth}
        \centering
		\includegraphics[width=0.99\textwidth]{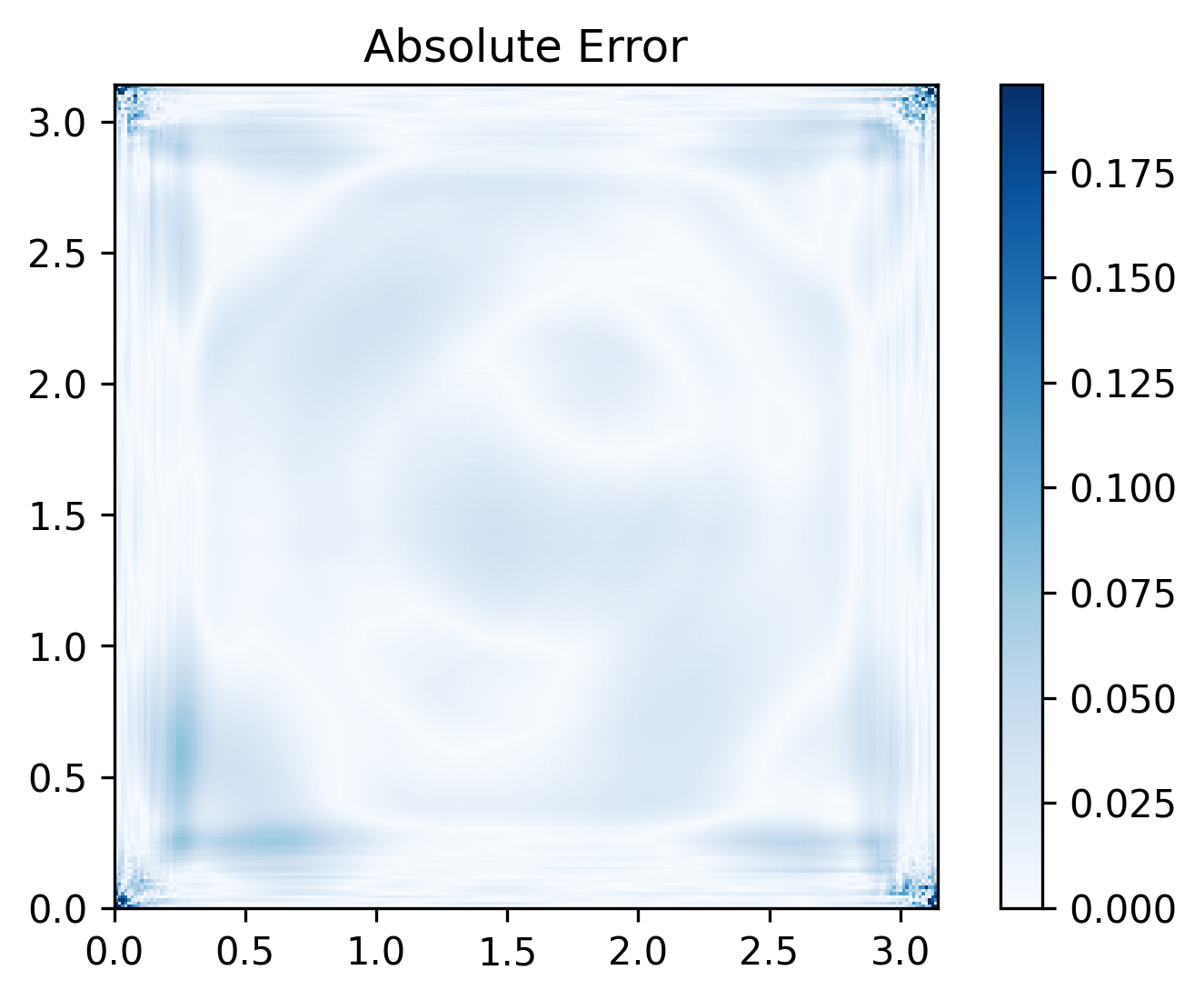}
    \end{subfigure}
    \caption{Prediction of the full model on the TGV case. Overall, the Poisson CNN exhibits good performance only slightly behind the performance shown on the problems similar to the ones encountered in the dataset, highlighting the generalization performance of the model}
    \label{fig:tgv_pcnn_pred}
\end{figure}

Figures \ref{fig:tgv_hpnn_pred}, \ref{fig:tgv_dbcnn_pred} and \ref{fig:tgv_pcnn_pred} show the performance of the HPNN, DBCNN and Poisson CNN models respectively. The HPNN sub-model performed at a level in line with the results displayed in Section \ref{sec:tset_hpnn} when predicting on the RHS function. The model achieved predictions within 10\% of the target at over half of the grid points and reproduced key solution features such as symmetricity about the $x=y$ line. Mirroring the previous cases, largest absolute errors are concentrated near local extrema, particularly near the local minimum in the middle of the domain, and largest percentage errors are near the $\phi=0$ contours.

%\clearpage
%\clearpage

The DBCNN sub-model performed slightly better in this case compared to the 600-sample average shown in \tabref{tab:results_summary_table}. Though providing somewhat grainy results and underpredicting the peak magnitude of the solution along the $y = (2k+1) \pi/2$ contours where the BC value approaches 0, the predictions along the mid-section of the domain are very good. Highest absolute errors are seen near the left boundary by the corners while the highest percentage errors are by the right boundary, mirroring the previous sample shown in \secref{sec:dbcnn_tset_result}.

The performance of the full Poisson CNN model exceeds that of the individual sub-models. The MAPE of the full model lies just above that of the two sub-models and is ajust above the previous 600-sample average shown in \tabref{tab:results_summary_table}. Important solution features such as the sharp contours along the $x+y=(2k+1)\pi/2$ lines are present. However, from a purely qualitative perspective, some artefacting is visible near the corners, stemming from the artefacting seen in the DBCNN prediction.

Overall, both sub-models as well as the overall Poisson CNN model demonstrate solid performance in this analytical test case, giving further evidence that the model did not overfit on the training data. 

%\clearpage
\clearpage
\subsection{Performance with previously unseen grids}
\label{sec:previously_unseen_grids_perf}
\begin{figure}[h!]
    \centering
    \includegraphics[width = 0.66\textwidth]{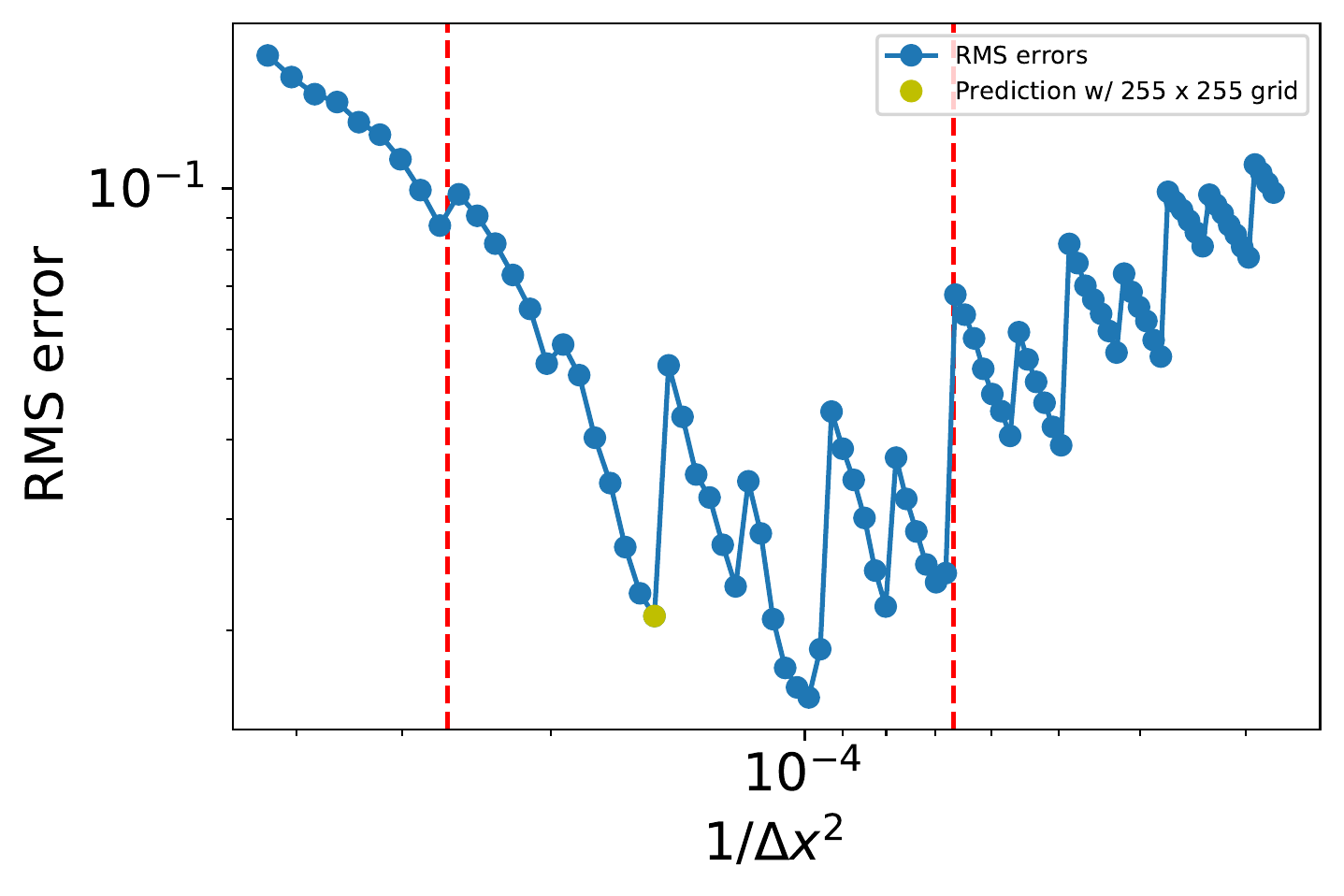}
    \caption{RMS error model output w.r.t. the analytical solution versus grid density for the TGV case. Interval of grid sizes encountered in training is marked by red lines. Yellow dot indicates the results presented in \secref{sec:tgv}}
    \label{fig:tgv_gridsize_vs_rms}
\end{figure}

An important aspect of the generalization performance of the model is whether it is able to handle grids with parameters (such as grid spacing and sizes) outside the range encountered during training. \figref{fig:tgv_gridsize_vs_rms} shows the RMS error of the model prediction relative to the analytical solution on the same TGV case shown in \secref{sec:tgv} with progressively denser grids.

{}

The results clearly indicate that while the model is struggling to handle grids that are coarser than the training data, the quality of the predictions degrade more gradually for larger grids. In addition, even within the range seen in the training data, the NN model behaves unlike a traditional numerical algorithm where a linear decrease of the RMS error with a slope equal to the order of the method is expected.

\begin{figure}[h!]
%\end{figure}
%\begin{figure}\ContinuedFloat
    \centering
    \begin{subfigure}[b]{0.30\textwidth}
        \centering
		\includegraphics[width=0.99\textwidth]{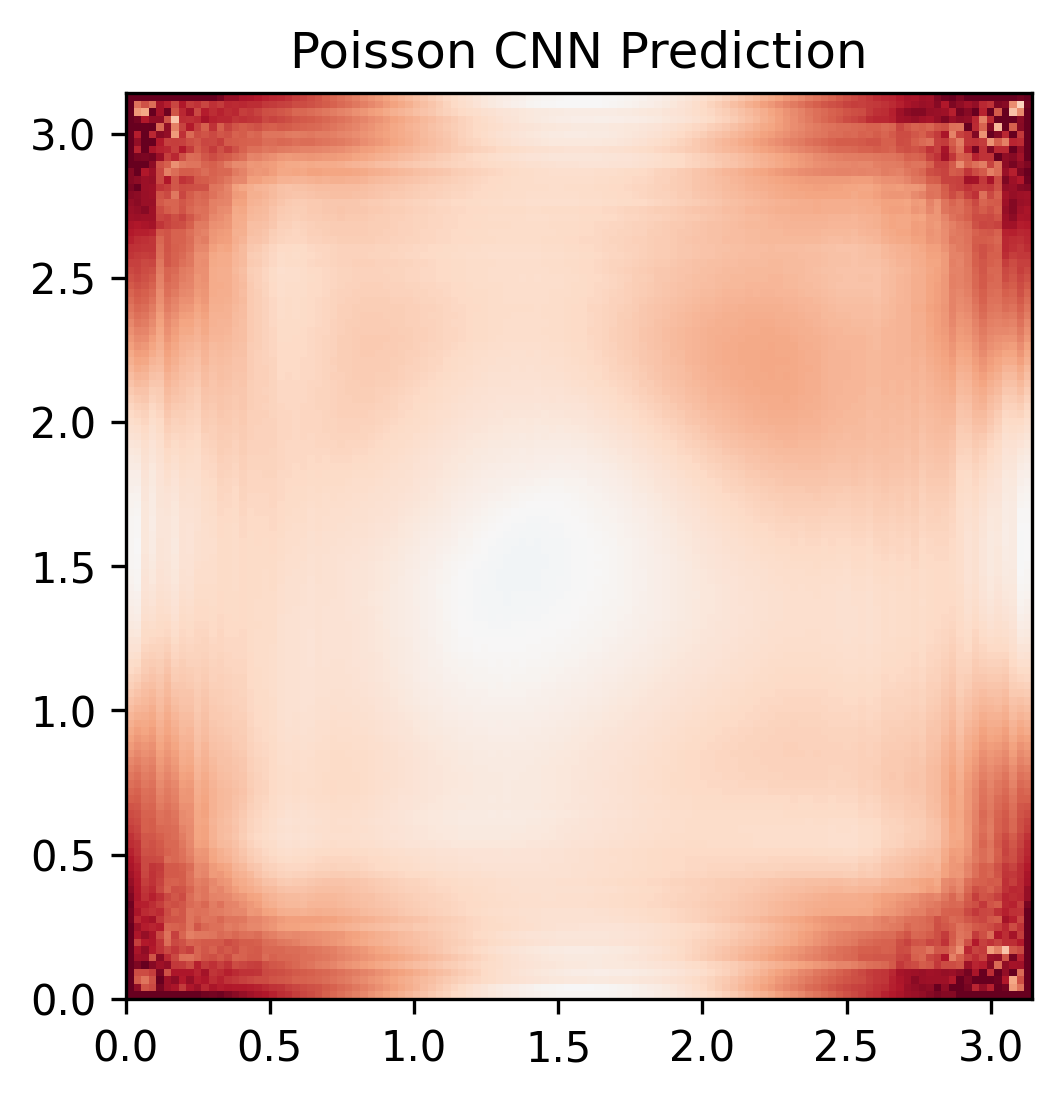}
    \end{subfigure}
    \begin{subfigure}[b]{0.30\textwidth}
        \centering
		\includegraphics[width=0.99\textwidth]{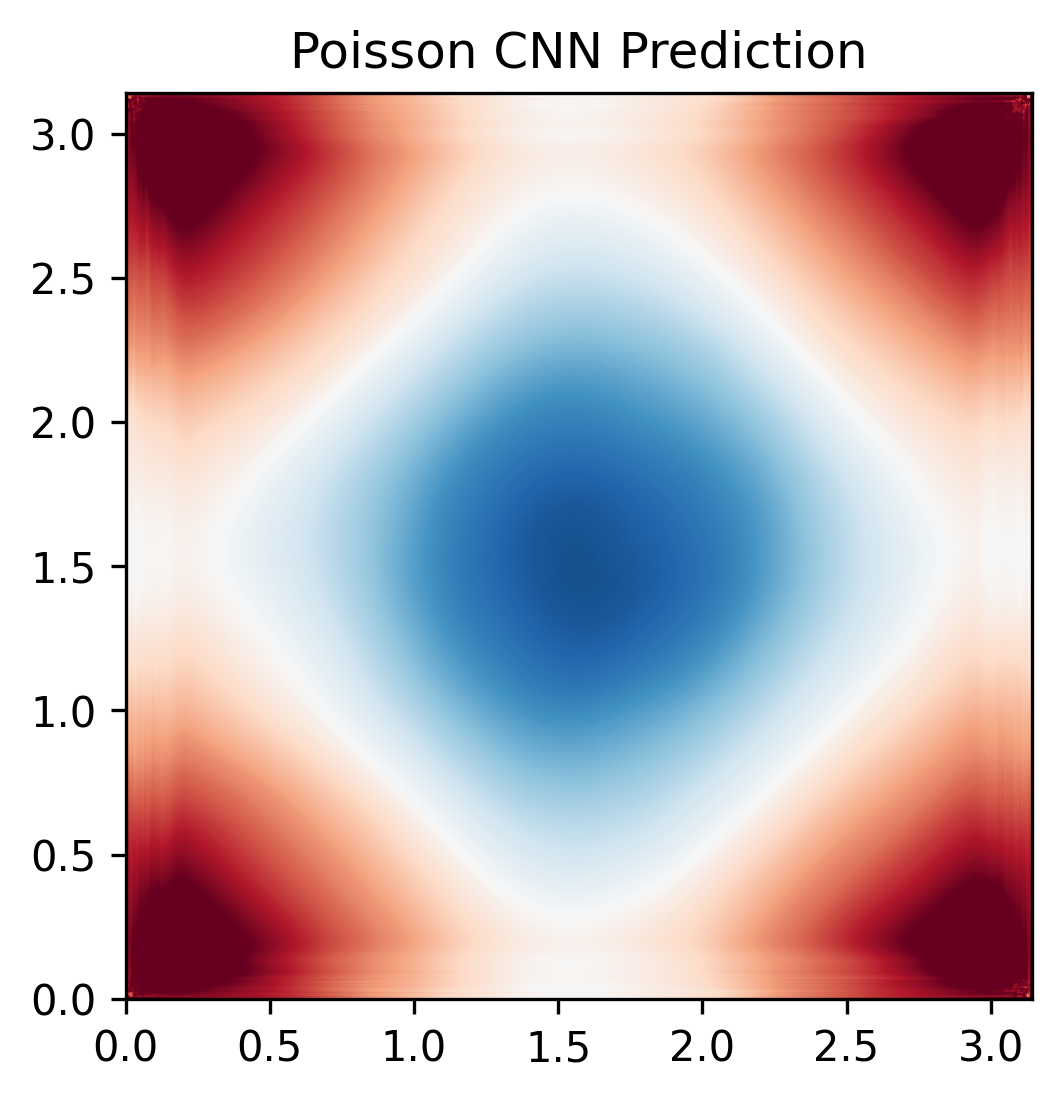}
    \end{subfigure}
    \begin{subfigure}[b]{0.37\textwidth}
        \centering
		\includegraphics[width=0.99\textwidth]{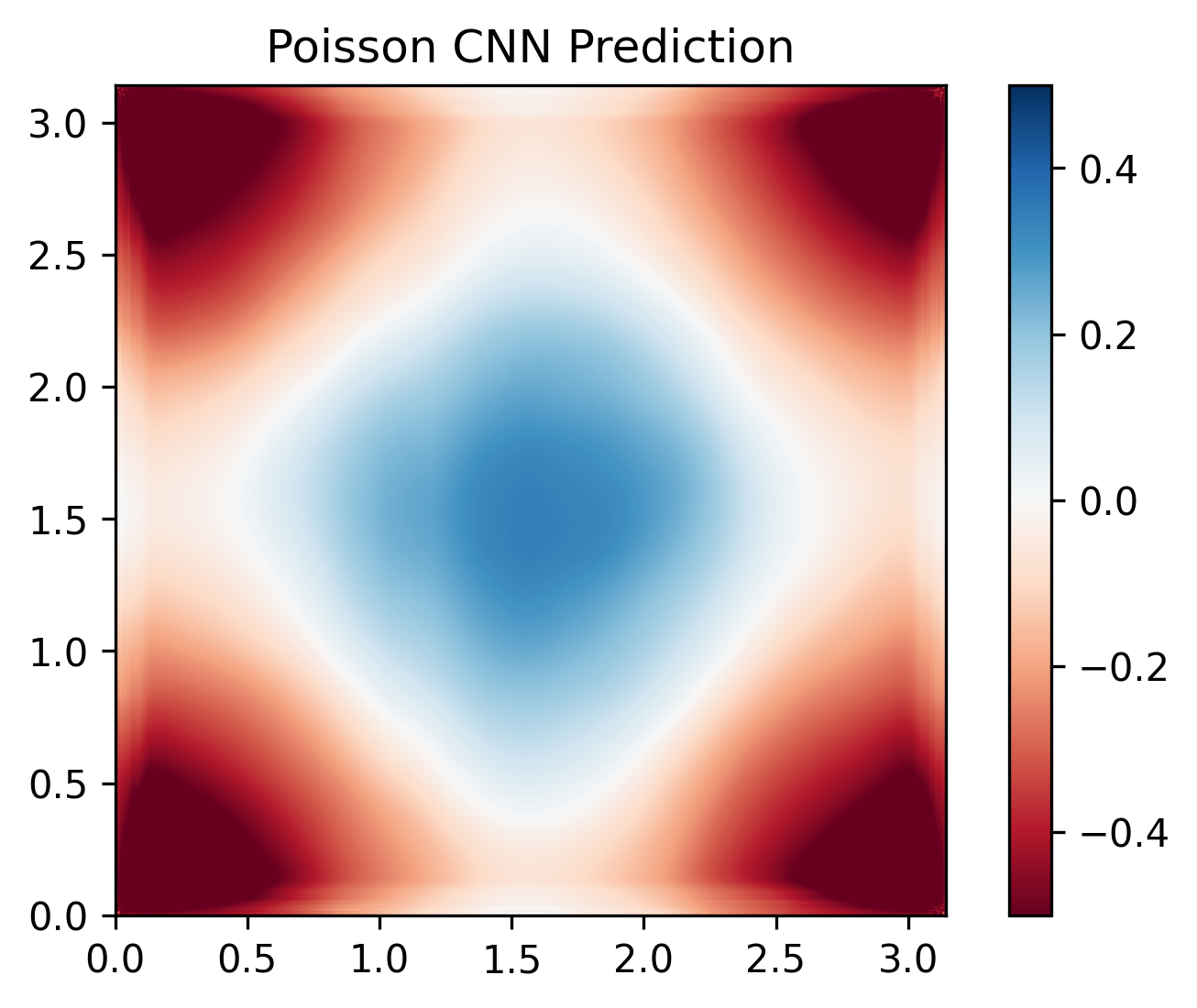}
    \end{subfigure}
    \caption{Poisson CNN predictions on $120 \times 120$ (left), $500 \times 500$ (middle) and $750 \times 750$ (right) grids for the TGV case. Although the model does not perform well for grids smaller than those seen during training, its predictive ability diminishes only gradually for larger grids}
    \label{fig:tgv_differentmeshpredictions}
\end{figure}

In the $120 \times 120$ case, the model gives an answer completely dissimilar to the analytical solution due to severe mispredictions by the HPNN sub-model as seen in the mid-section of the domain. For the $500 \times 500$ and $750 \times 750$ cases however, the general solution profile is retained despite gradually increasing under-prediction of the local maximum at $(\pi/2,\pi/2)$ and over-prediction of the local minima at the corners.

Although the model's performance when predicting on smaller grids than the training data is substantially worse, this is a use case with far narrower use cases than the converse. The fact that the model retains the ability to reproduce the general solution profile, albeit with lower accuracy, is very promising as this can enable the model to supply initial guesses for iterative algorithms for problems with larger grids than encountered during training. The capability of the model to do that for the multigrid algorithm will be explored in \secref{sec:post-smoothing}.

%\clearpage
\subsection{Post-smoothing and comparison vs. one multigrid cycle}
\label{sec:post-smoothing}
\begin{figure}[h!]
    \centering
    \begin{subfigure}[b]{0.45\textwidth}
        \includegraphics[width=0.99\textwidth]{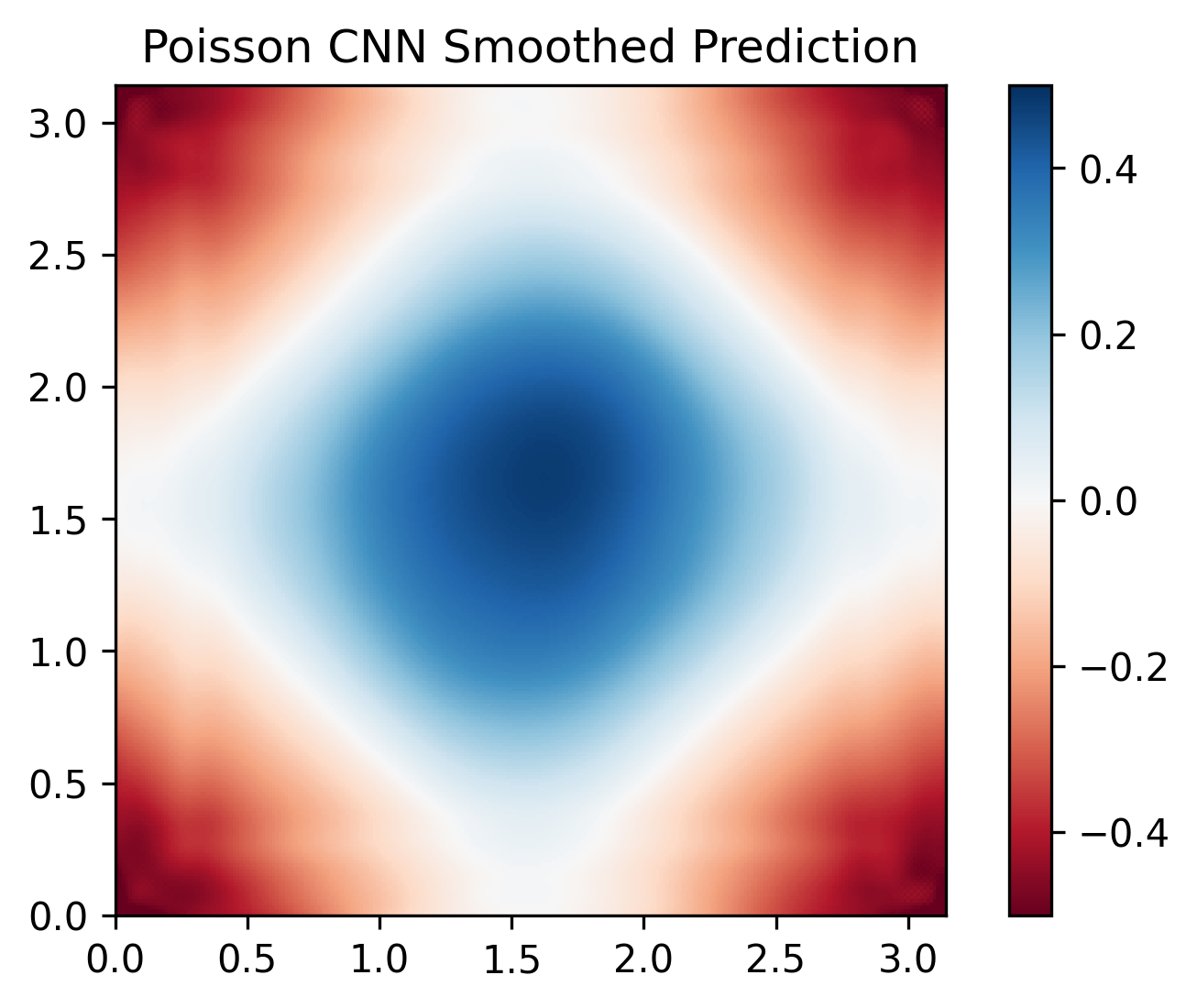}
    \end{subfigure}
    \begin{subfigure}[b]{0.45\textwidth}
        \includegraphics[width=0.99\textwidth]{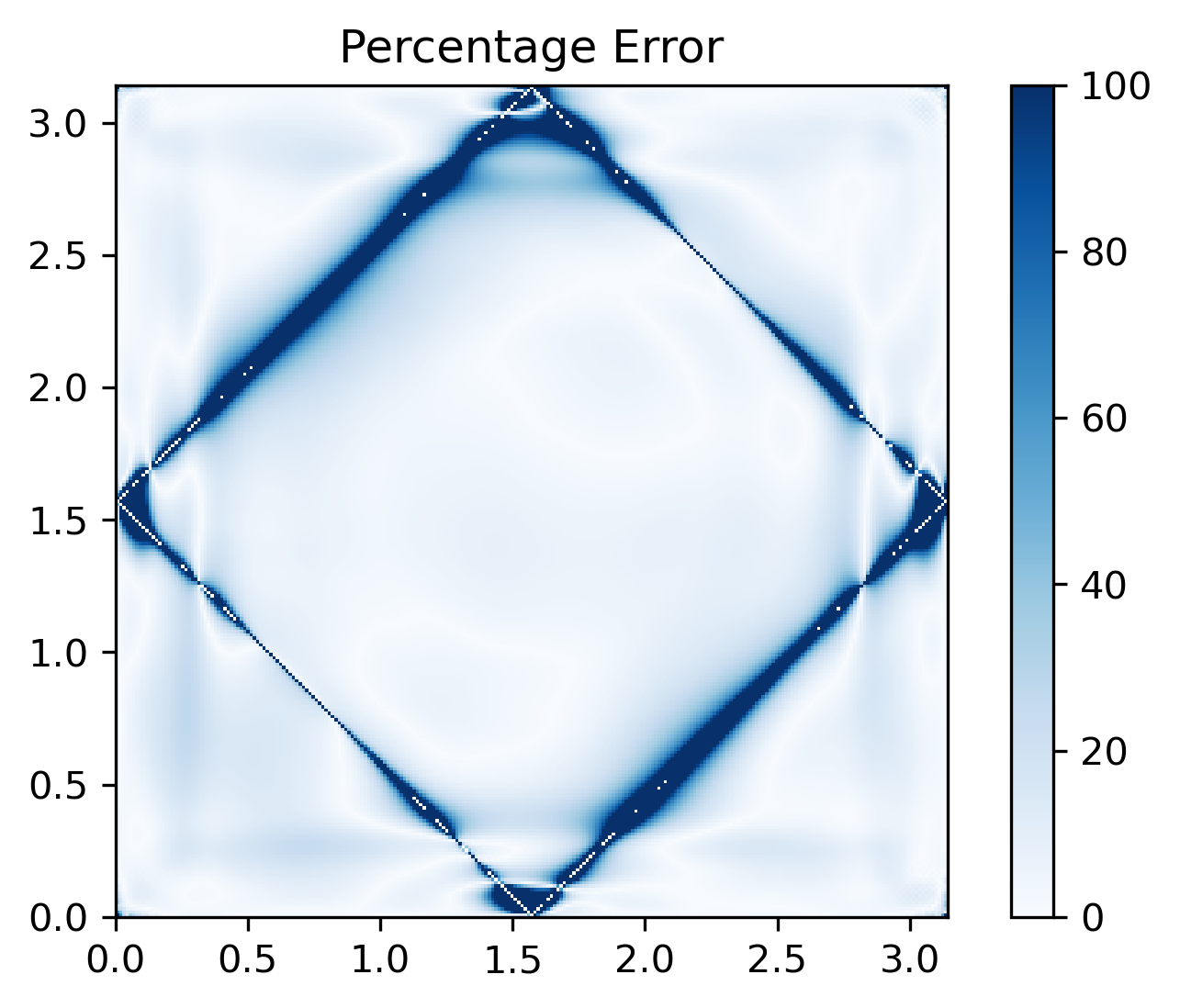}
    \end{subfigure}
    \begin{subfigure}[b]{0.45\textwidth}
        \includegraphics[width=0.99\textwidth]{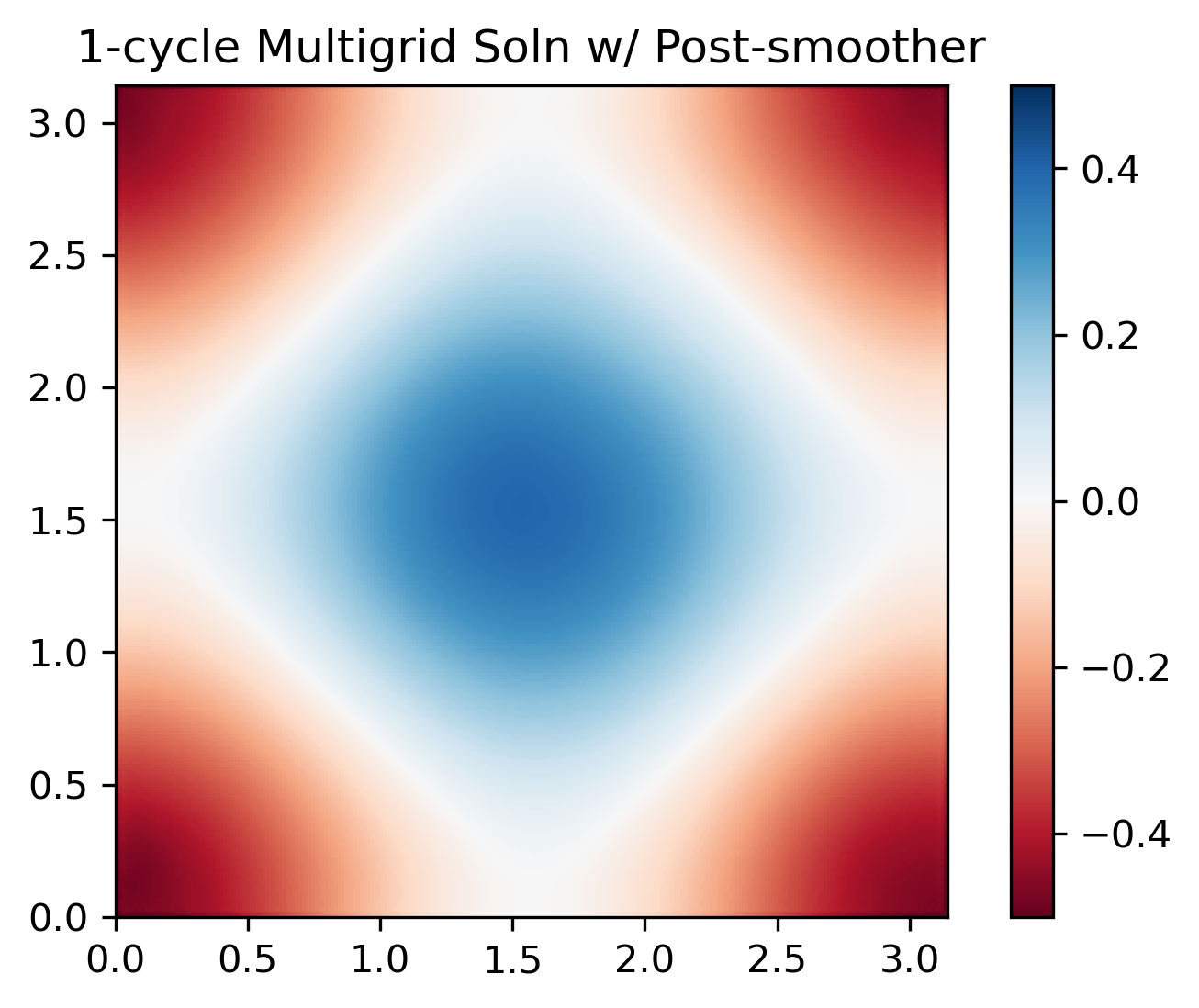}
    \end{subfigure}
    \begin{subfigure}[b]{0.45\textwidth}
        \includegraphics[width=0.99\textwidth]{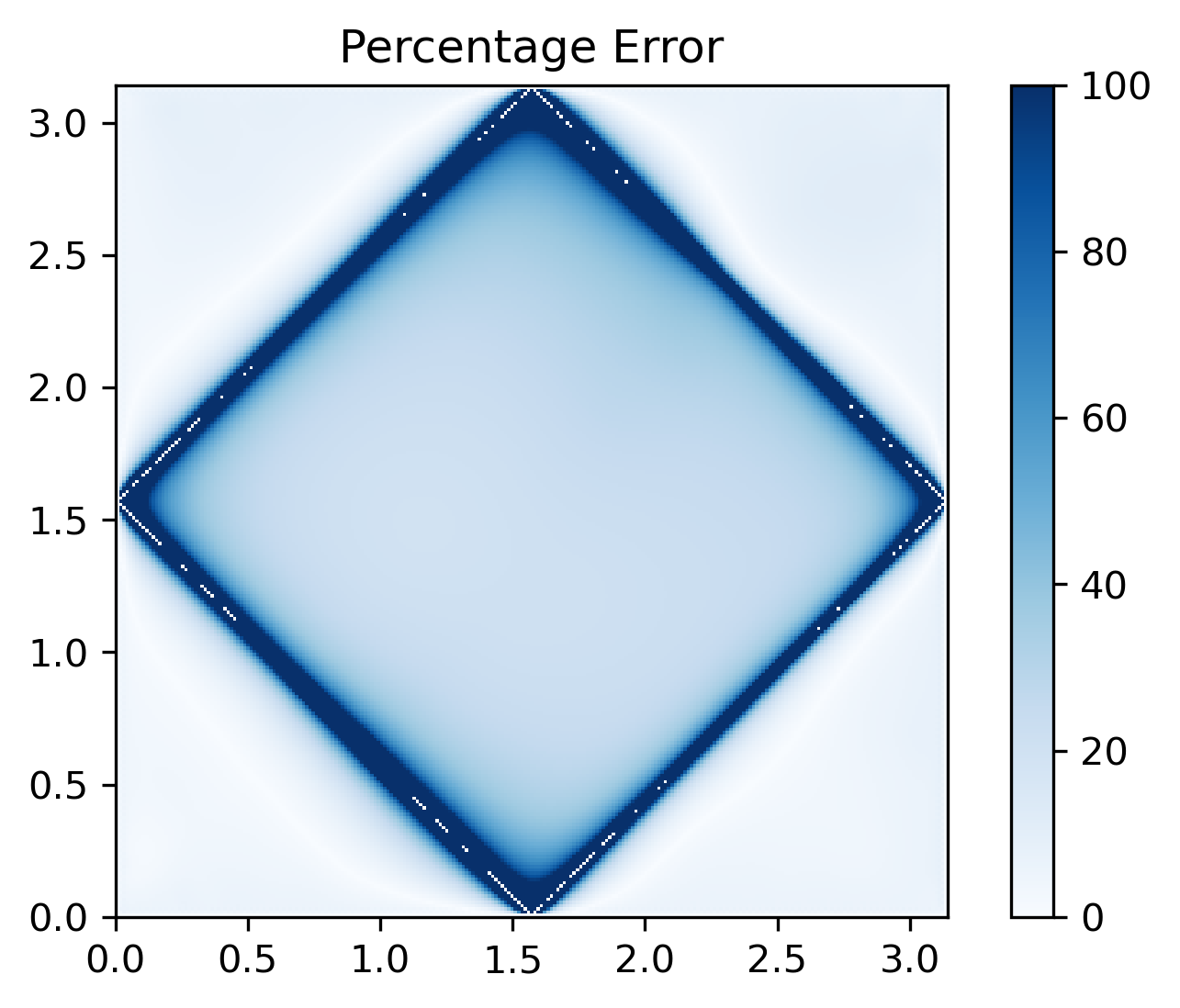}
    \end{subfigure}
    \begin{subfigure}[b]{0.45\textwidth}
        \includegraphics[width=0.99\textwidth]{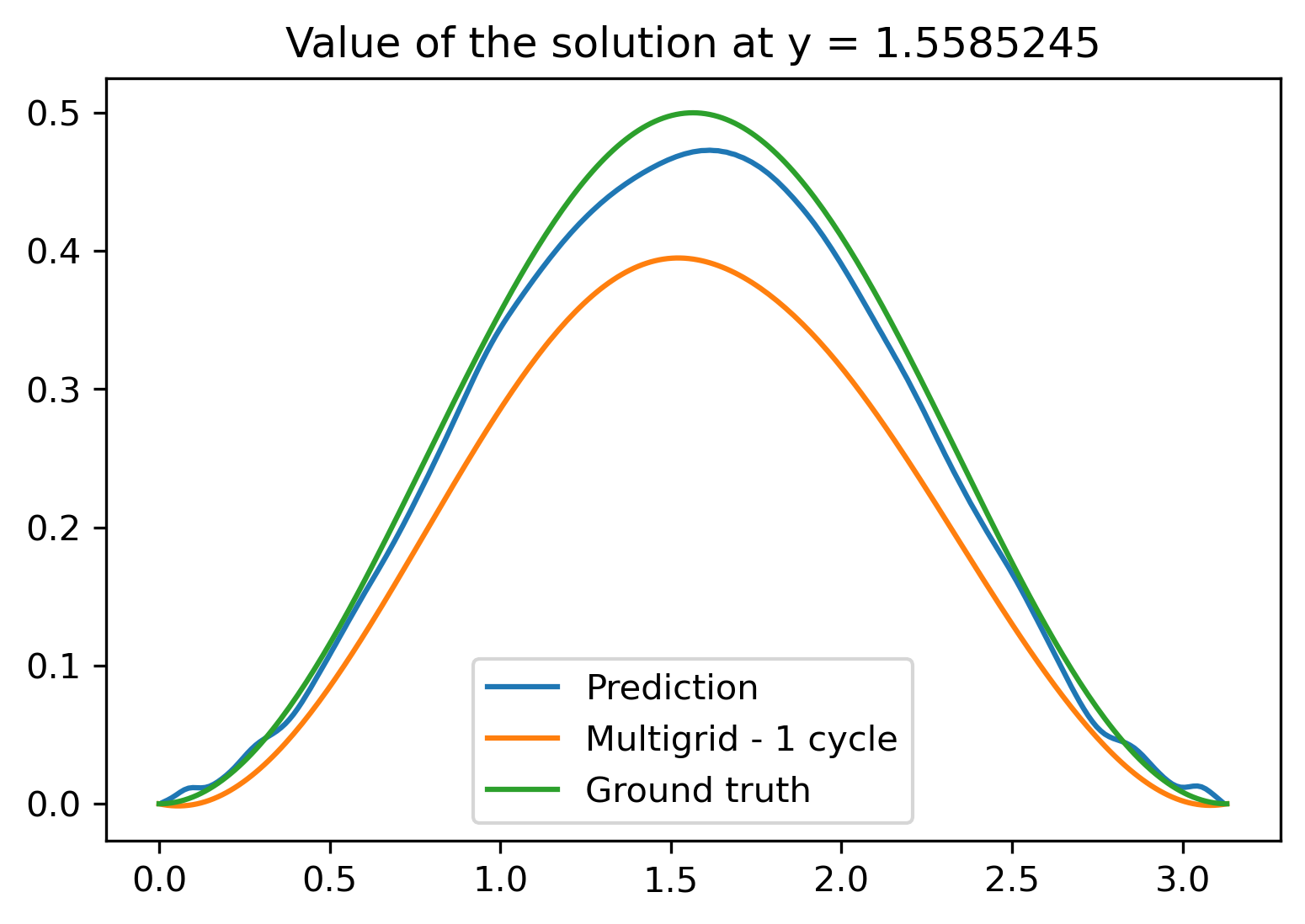}
    \end{subfigure}
    \caption{Comparison of the model's performance versus multigrid with a single cycle, with 5 Jacobi post-smoothing iterations applied to each. The Poisson CNN prediction clearly outperforms the single-cycle multigrid prediction}
    \label{fig:smoothedpred_vs_1cyclemgrid}
\end{figure}

Considering the primary envisioned use case for our model is accelerating iterative Poisson solvers, one problematic issue is the presence of high-frequency artefacting which can occasionally manifest itself as seen in \figref{fig:tgv_pcnn_pred}. Applying five Jacobi post-smoothing iterations to the predicted field eliminates most of the high frequency artefacting, greatly reducing the roughness of the produced solution surface as shown in \figref{fig:smoothedpred_vs_1cyclemgrid}.

The model shows remarkable capability in this task, being able to reduce the RMS error after a single multigrid iteration by a massive $94\%$ relative to an initial zero guess when utilised on a problem with a previously encountered $384 \times 384$ grid size. A reduction of $74\%$ is achieved on a $4500 \times 4500$ problem, a grid over an order of magnitude larger than the largest problem in the training set. It's further noteworthy that a single multigrid iteration with the model prediction as the initial guess is able to beat two multigrid iterations in accuracy even for the same $4500 \times 4500$ grid with $22\%$ lower RMS error. This suggests that the Poisson CNN model can provide very substantial increases of accuracy to the multigrid algorithm. Moreover, the model is capable of undertaking this task for problems much larger than the examples encountered during training.

This capability is especially important tn light of the results in \tabref{tab:runtime} which clearly display that our model increases the runtime gap versus multigrid as the problem size grows. Eventually, our model catches up to even a single cycle of multigrid for large problems while beating the accuracy of the said initial cycle, providing a substantial boost to the convergence rate of the multigrid method from the time perspective as well. 

The post-smoothing applied modestly improved the MAPE of the solution, lowering it to $11.61\%$. With the same post-smoothing applied, the multigrid method with 8 levels was able to achieve a MAPE of $21.08\%$. Hence, the post-smoothing resolves the issue of high-frequency oscillations created by the model near the edges as can be seen by comparing \figref{fig:tgv_pcnn_pred} and \figref{fig:smoothedpred_vs_1cyclemgrid}. 

\begin{figure}
    \centering
    \includegraphics[width=0.66\textwidth]{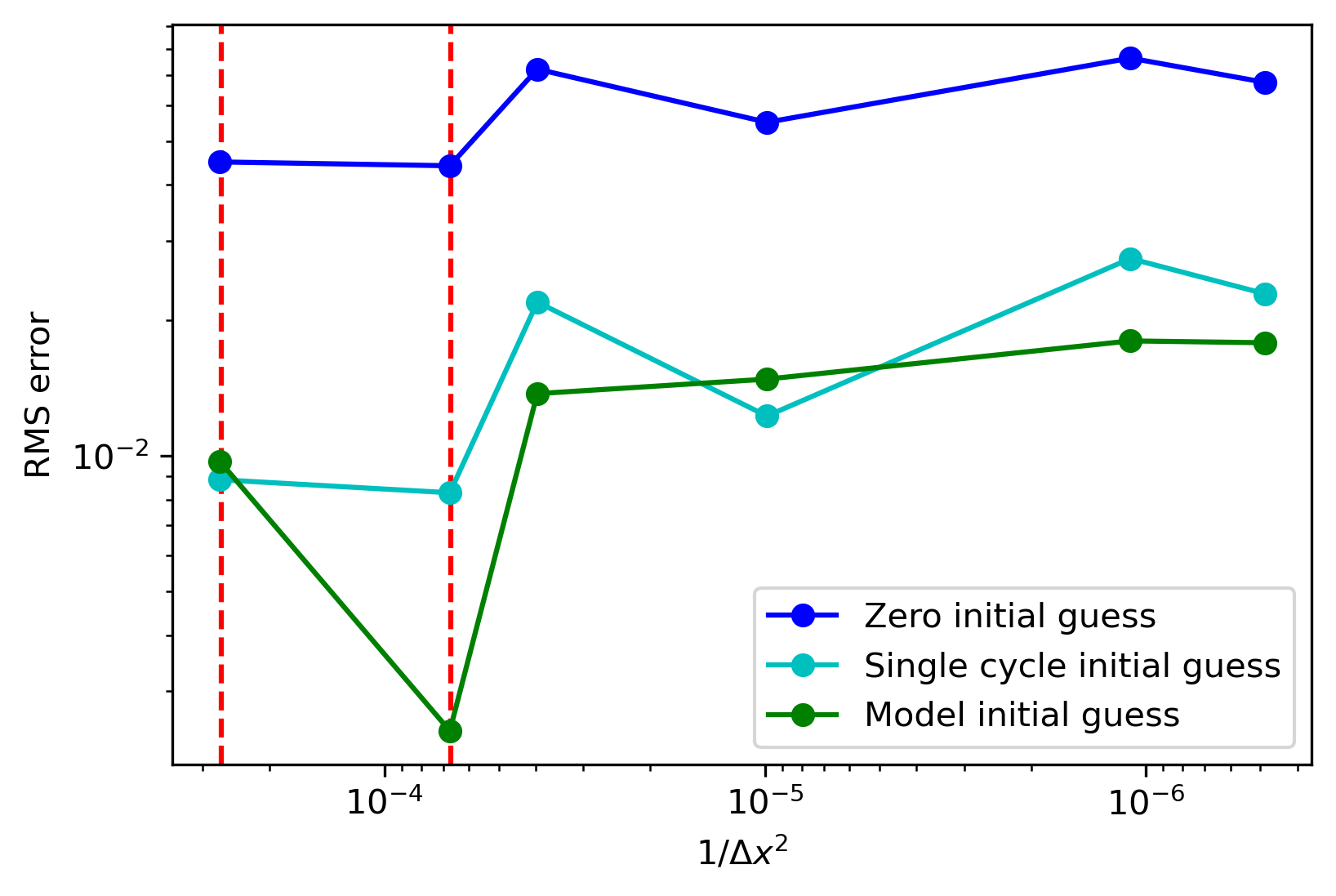}
    \caption{RMS error comparison for multigrid iterations with zero, single cycle multigrid and Poisson CNN prediction initial guesses on the TGV snapshot case. Red lines indicate the grid parameter range seen by the model during training}
    \label{fig:rms_multigrid_initial_guess}
\end{figure}

The preliminary results in \figref{fig:smoothedpred_vs_1cyclemgrid} demonstrate that it should be possible to increase the convergence rate of conventional iterative methods by using the present CNN architecture as the first step of an iterative strategy. \figref{fig:rms_multigrid_initial_guess} serves as a more detailed demonstration of that capability, juxtaposing the RMS error after a single multigrid iteration with a zero initial guess, two multigrid iterations with a zero initial guess and a single multigrid iteration with the model prediction as the initial guess for the TGV case over a number of grid resolutions, including those outside the training data range.

\clearpage

%\clearpage
\subsection{NN-assisted pressure projection method}
\label{sec:cfd}
Given the solid performance of the model on a single snapshot as shown in Sections \ref{sec:tgv} and \ref{sec:post-smoothing}, we present the results from a TGV simulation, where the pressure projection step is assisted by a variant of the proposed NN. To ensure boundary condition compatibility with this CFD problem, we require a variant of the model capable of handling Neumann as opposed to Dirichlet BCs. Considering the analytical solution for the TGV presented in \equref{equ:tgv_rhs}, sticking to the domain previously investigated in this section, we only require a model capable of handling homogeneous Neumann BCs.

Hence, the variant of our model used in this sub-section comprises of only the HPNN component, with the only change being the application of `symmetric' padding in the final convolution layer as opposed to zero padding to ensure that the boundary conditions are naturally applied by the model. The dataset used to train this model also incorporates a corresponding modification to accommodate the change in the BCs, switching to homogeneous Neumann BCs as opposed to Dirichlet BCs. In effect, this is achieved by replacing the sine series in \secref{sec:dataset_generation_hpnn} with a cosine series. The training on this dataset was done in a manner similar to the model with Dirichlet BCs in Sections \ref{sec:validationset} to \ref{sec:post-smoothing}, but with smaller grid sizes ranging between 80 to 120 on each side. In addition, the physics-informed loss from \cite{pinn} was found to reduce validation MSE with Neumann BCs, unlike the Dirichlet BC case, and thus was used as a loss function component in addition to \equref{eq:loss_func}.

The simulation itself was performed within the framework of the MIT licensed \texttt{Navier\_Stokes\_2D} codebase by \cite{Navier_Stokes_2D}, a simple two-dimensional finite-difference code incorporating a range of projection methods, including an implementation of the gauge method proposed by \cite{weinan2003gauge} which was utilised in this investigation. The codebase was modified to ensure inter-operability with the HPNN model and recent versions of Python3.

The non-dimensionalised Navier-Stokes equations require a choice of Reynolds number\footnote{The Reynolds number is a nondimensional quantity which is commonly interpreted as the ratio of inertial to viscous forces in a fluid, and determines the amount of diffusion in the non-dimensionalised equation system} $Re=1/\nu$, where $\nu$ is the kinematic viscosity of the fluid. Additionally, an appropriate Courant-Friedrichs-Lewy (CFL) number $C=\frac{\Delta t}{\Delta}(u+v)$ must be chosen to ensure that the time step $\Delta t$ is sufficiently small to prevent numerical instabilities given the grid spacing $\Delta$ and the $x-$ and $y-$direction velocities $u$ and $v$. The results presented in this section were obtained from a  simulation run with the default parameters of $Re = 1.0$ and $C=0.2$ in the \texttt{Navier\_Stokes\_2D} codebase. The domain was kept identical to the one in \secref{sec:tgv}, but was discretised with $100 \times 100$ grid points. Similar to the `hybrid' strategy applied by \cite{ajuria2020towards}, designed to reduce the accumulation of errors during the time marching process, traditional solver iterations are applied to the output of the model.

\begin{figure}[h!]
%\end{figure}
%\begin{figure}\ContinuedFloat
    \centering
    \begin{subfigure}[b]{0.30\textwidth}
        \centering
		\includegraphics[width=0.99\textwidth]{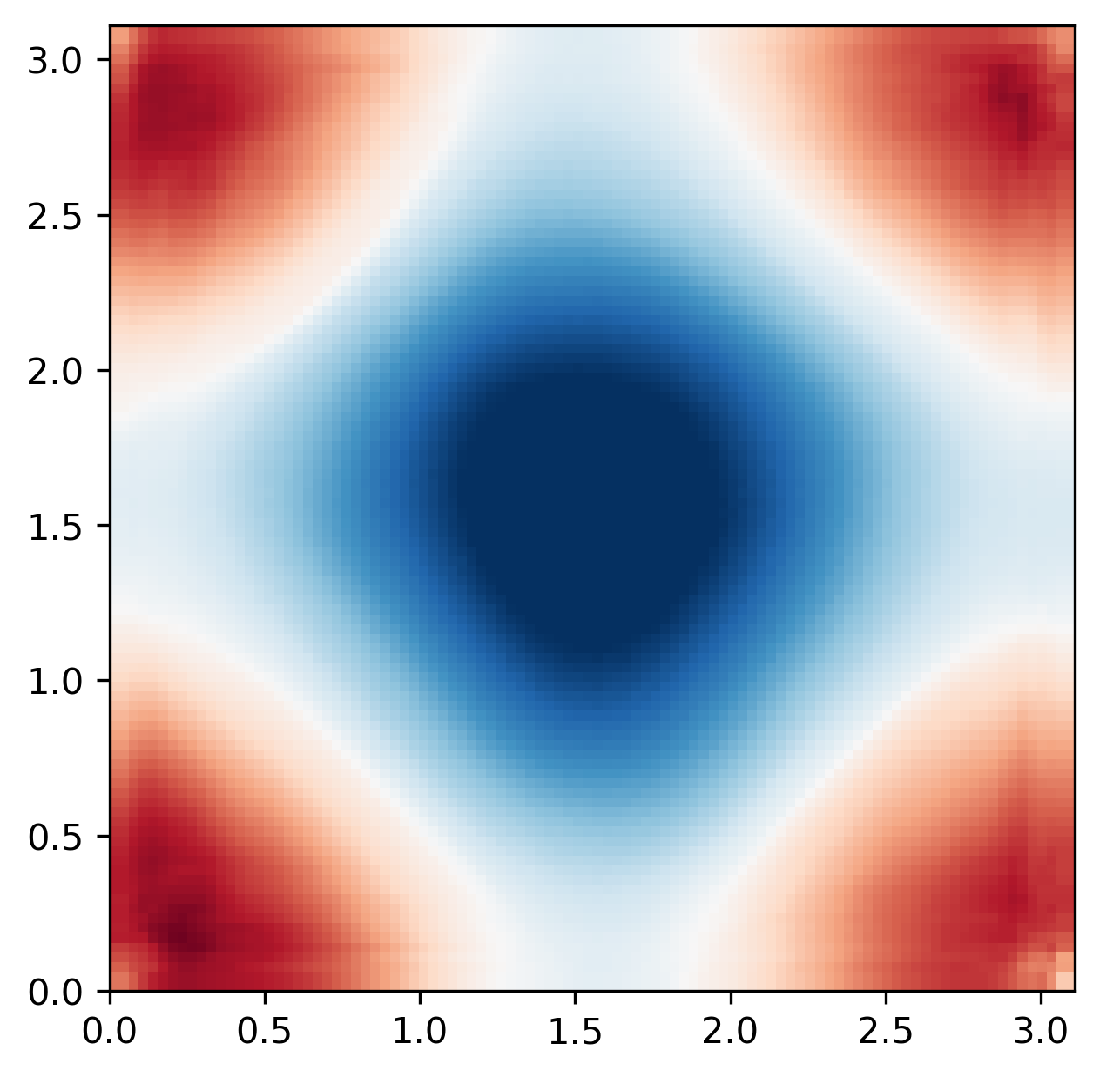}
    \end{subfigure}
    \begin{subfigure}[b]{0.30\textwidth}
        \centering
		\includegraphics[width=0.99\textwidth]{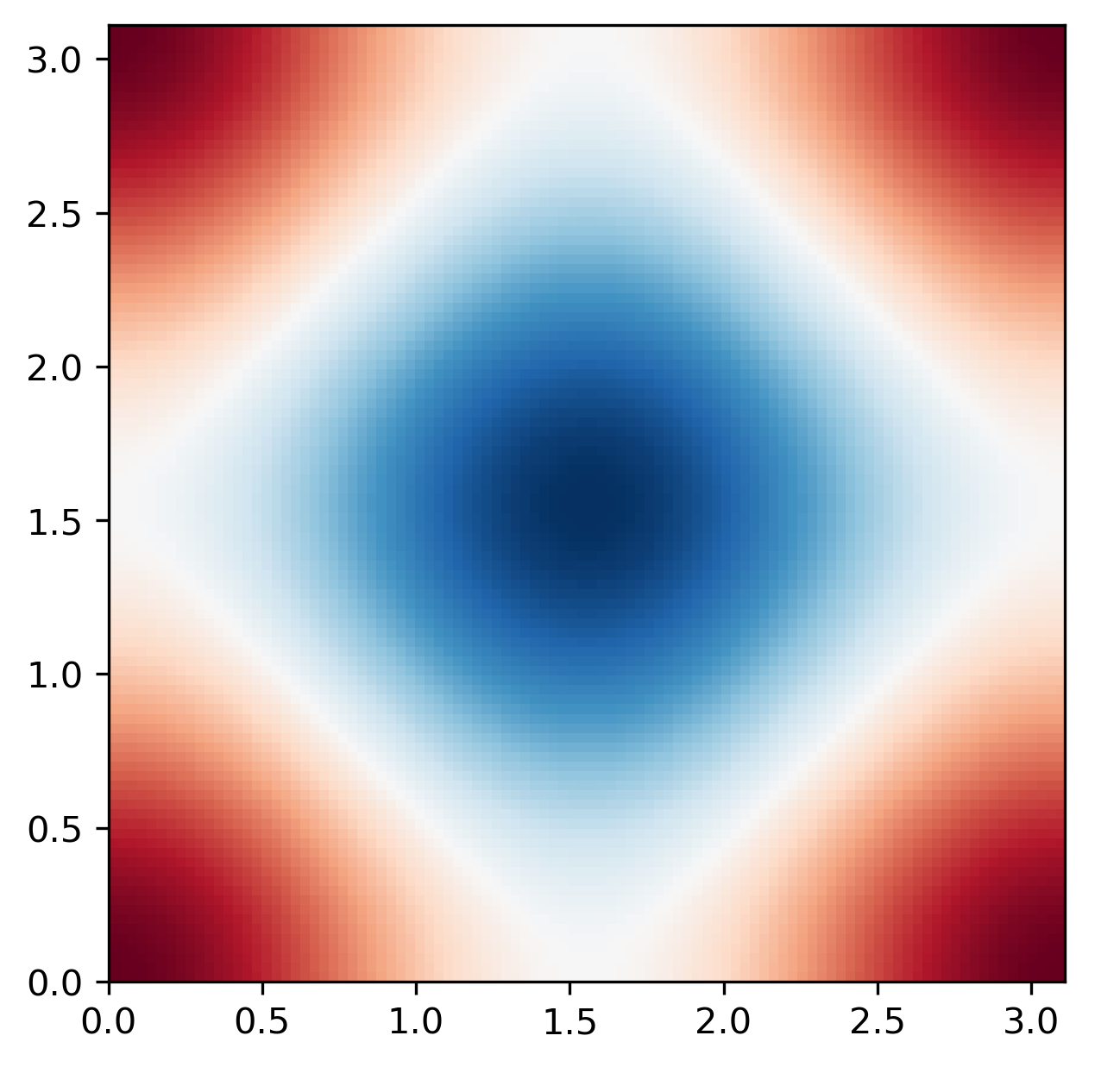}
    \end{subfigure}
    \begin{subfigure}[b]{0.37\textwidth}
        \centering
		\includegraphics[width=0.99\textwidth]{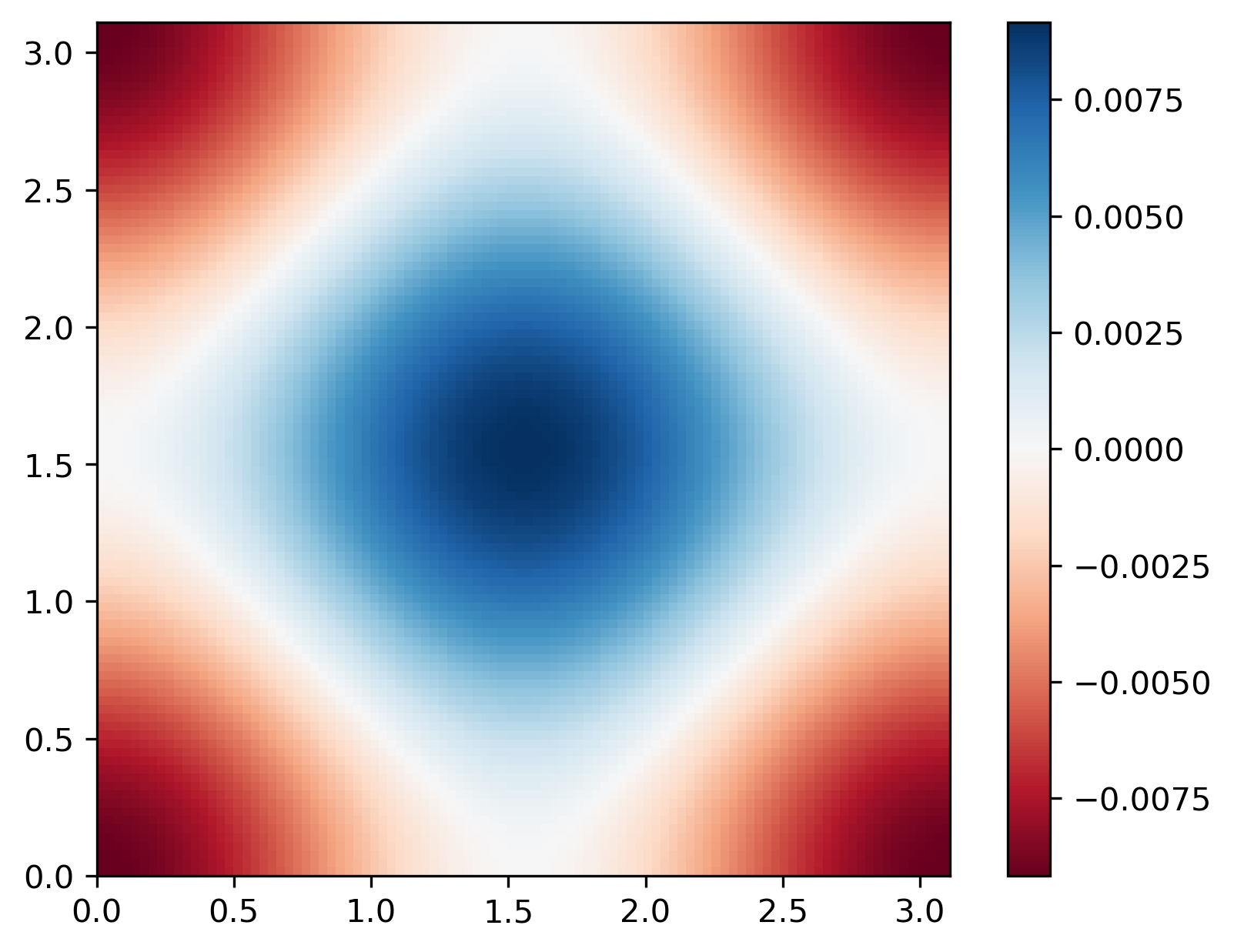}
    \end{subfigure}
    \caption{Raw Neumann BC HPNN prediction (left), Poisson CNN prediction with one traditional solver iteration (middle) and ground truth solution (right) of the TGV simulation at $t=1.0$ (time step 637). The raw HPNN prediction displays artefacting near the edges and overshoot at local extrema, however a single traditional solver iteration aids it to achieve a high degree of accuracy.}
    \label{fig:tgv_sim_img}
\end{figure}

\tabref{tab:tgv_sim_errnorm} displays the $L_2$ error norms of the velocities and the pressure compared to the analytical result at the end of the simulation ($t=1.0$). \figref{fig:tgv_sim_img} juxtaposes the the ground truth result with the predictions of the HPNN for the pressure at $t=1.0$, both raw and when assisted by a single iteration of the biconjugate gradient stabilized (BiCGSTAB) method.

\begin{table}[h!]
\centering
\begin{tabular}{@{}lcccc@{}}
\toprule
Case                                         & $||\mathrm{err}(u)||_2$   & $||\mathrm{err}(v)||_2$ & $||\mathrm{err}(p)||_2$                  & {$\%$ reduction in $||\mathrm{err}(p)||_2$ }\\ \midrule
{Zero initial pred. + 1 iter.} & {$3.66 \times 10^{-5}$ }&{ $3.53 \times 10^{-5}$ }& {$7.03 \times 10^{-3}$} & \multirow{2}{*}{{$99.3\%$}}        \\
{HPNN initial pred. + 1 iter.} & {$4.69 \times 10^{-6}$} & {$4.73 \times 10^{-6}$} & {$4.80 \times 10^{-5}$} &                                  \\ \cmidrule(l){2-5}
{Zero initial pred. + 2 iter.}     & {$3.93 \times 10^{-7}$} & {$3.69 \times 10^{-7}$ }& {$1.56 \times 10^{-5}$} & \multirow{2}{*}{{$52.2\%$}}        \\
{HPNN initial pred. + 2 iter.}     & $2.59 \times 10^{-7}$ & {$2.52 \times 10^{-7}$} & {$7.45 \times 10^{-6}$} &                                  \\ \bottomrule
\end{tabular}
\caption{{$L_2$ error norms for the velocities $(u,v)$ and the pressure $p$ at the end of the Navier-Stokes simulation ($t=1.0$).}}
\label{tab:tgv_sim_errnorm}
\end{table}

{Overall, the HPNN aided by a single BiCGSTAB prediction achieves a level of pressure error comparable to the level achieved by two iterations of BiCGSTAB with an initial zero guess, a massive $L_2$ error reduction compared to a single zero-initial-guess BiCGSTAB iteration. This is in line with our comparison with multigrid solvers in \secref{sec:post-smoothing} using the Dirichlet BC variant of the Poisson CNN model, showing strong evidence that our model performs well in practical applications for acceleration of Poisson problems.}

\subsection{{Ablation studies}}
\label{sec:ablation}
The proposed NN architecture incorporates many features which improve performance for the Poisson equation task relative to other CNN architectures commonly used in image-to-image translation tasks. This is clearly demonstrated by the order-of-magnitude difference in validation MAE achieved on various cases relative to the baseline models in \tabref{tab:results_summary_table}. To open a path for future advances in developing CNN architectures for this task, it is important to understand the contribution of each architectural feature to the performance of the model. One of the most common methods to investigate this in machine learning is via an ablation study, whereby certain features of the model under investigation are disabled in isolation and the performance is compared against the full model. \tabref{tab:ablation_study} presents the results of our ablation study conducted with five key improvements of our model compared to previous works by considering the rise in validation MSE as well as change in the inference speed of the model.

\begin{table}[h!]
\centering
\begin{tabular}{@{}lcc@{}}
\bottomrule
{Change to model architecture }     & {Validation MSE change} & {Inference runtime change} \\ \bottomrule
{No residual connections in HPNN   }     & { $+5.56\%$   }                  & $-1.26\%$                 \\ \midrule
{MAE loss only DBCNN training  }            &  {$+51.16\%$        }                      & \ N/A               \\ \midrule
3 fewer pooling blocks in HPNN       & $+56.81\%$                        & $-2.85\%$                   \\ \midrule
No positional embeddings in HPNN &  $+89.21\%$ &     $-0.25\%$ \\ \midrule
\makecell[l]{End-to-end training for\\ inhomogeneous problems} & $+544\%$ & N/A \\\bottomrule
\end{tabular}
\caption{Percentage change to the validation MSE and inference speed of the Poisson CNN when key model architecture features are removed.}
\label{tab:ablation_study}
\end{table}

{The ablation study clearly highlights the magnitude of the improvements Poisson CNN offers compared to previous architectures. Most significant advantage of the proposed model is the utilisation of the decomposition approach presented in \secref{sec:bc_strategy}, which enables it to achieve a more than six-fold reduction in validation MSE at zero cost in terms of inference time. Next in terms of importance comes the almost halving of validation MSE enabled by the inclusion of positional embeddings, a feature critical to enable good performance when handling variable input shapes to the model. Then, we see the approximately one-third reduction in validation MSE afforded by employing our novel loss function, the $L^p$ integral loss, and the addition of more pooling blocks compared to the Fluidnet architecture by \cite{fluidnet}. Finally, the inclusion of residual connections in convolutional layers enables a more modest further $5\%$ reduction in MSE values. Combined, these features enable our model to tackle the generalized form of the Poisson problem effectively, enabling a marked improvement over models previously employed on the Poisson problem or in more general image-to-image translation tasks.}

%\\~\\

\subsection{Wall-clock runtime}
\label{sec:runtime}
The practical applicability of Poisson CNN's ability to boost the accuracy of multigrid iterations is contingent on its wall-clock runtime. \tabref{tab:runtime} outlines the wall-clock runtime of both the sub-models and the full model using single precision run on an Nvidia V100 GPU, versus multigrid run on 64 threads on a 32-core/64-thread AMD Threadripper 2990WX CPU and the same GPU. Note that the model architecture and weights were not changed compared to the results presented above and the accuracy of the model was not evaluated.

\begin{table}[h!]
\caption{Wall clock runtimes of the Multigrid solver versus the DBCNN, HPNN and Poisson CNN models (in seconds)}
\centering
\begin{tabular}{@{}lccccc@{}}
\toprule
Grid size & Multigrid GPU & Multigrid CPU & DBCNN (4x) & HPNN    & Poiss. CNN \\ \midrule
100                      & 0.2193              & 0.0361        & 0.4419  & 0.1028  & 0.5811      \\
200                      & 0.3149              & 0.1101        & 0.4651  & 0.1208  & 0.6276      \\
384                      & 0.6974              & 0.5469        & 0.4897  & 0.1634  & 0.6936      \\
500                      & 0.8932              & 0.8580        & 0.5543  & 0.2777  & 0.8744      \\
1000                     & 4.3026              & 3.3701        & 0.7686  & 0.7917  & 1.6382      \\
3000                     & 21.2910             & 45.6587       & 2.9436  & 6.3047  & 9.6978     \\ 
4500                     & 45.4543             & 106.2516      & 6.2163  & 14.2176    & 21.4086     \\ \bottomrule
\end{tabular}
\label{tab:runtime}
\end{table}

The Poisson CNN model begins to outperform both CPU and GPU multigrid around grid sizes approaching $500 \times 500$, eventually building up to five times the speed of CPU multigrid at $4500 \times 4500$, and over twice that of GPU multigrid. While in the case of CPU multigrid there likely exists a severe memory bandwidth bottleneck\footnote{For comparison, the Nvidia V100 GPU reports over 800GB/s of memory bandwidth, compared to the 100GB/s supplied by the quad channel DDR4 3200MT/s memory available to the CPU.} for such large problems, as the performance of the multigrid algorithm was demonstrated to be memory bandwidth limited by various authors such as \cite{multigrid_memory_bandwidth}, our model performs favourably runtime-wise against GPU multigrid as well.

These results show that the NN model proposed in the current study has the potential to offer substantial speedups to solve the Poisson equation for practical applicatons especially in light of the results shown in Sections \ref{sec:post-smoothing} and \ref{sec:cfd}.

It is noteworthy that no hyperparameter optimization efforts were conducted for our model. It is likely that by tuning the number of hyperparameters (by \eg reducing the number of channels in the intermediate layers or the number convolution layers altogether) substantial speedups could be achieved while making little or no compromise in terms of the predictive accuracy of the model, albeit larger model sizes may be necessary for larger problems. The extreme similarity of many of the feature maps in the initial layers of the model is supportive of this possibility. Finally, further very substantial speedups are possible by taking full advantage of quickly developing deep learning acceleration hardware. For example, the `Tensor Cores' on the V100 GPU (which require half-precision operations and specific programming to properly utilise) theoretically offer up to 125TFlops of performance as opposed to 15.2 TFlops of standard single-precision performance as explained in \cite{v100_specs}.

\section{Conclusion and future work}
A CNN model to estimate the solution of the 2D Poisson equation with Dirichlet BCs was developed, based on splitting the problem into a homogeneous Poisson problem and four Laplace problems with one inhomogeneous BC each, with envisioned practical applications to accelerate iterative Poisson solvers by providing initial guesses that are both faster and more accurate than a single iteration of multigrid. Training was done on synthetic, random datasets generated to replicate the smooth functions a Poisson solver would be expected to handle in real world use cases. In order to achieve good convergence during training, the novel $L^p$ integral loss was developed and found to be superior for this task to the MSE.

The model developed can estimate solutions with pointwise deviations below 10\% even when given inputs that are materially different from both training and validation data, based on analytical test cases. {The predictions are accurate enough such that the model can be used in the pressure projection step of a Navier-Stokes simulation in conjunction with a traditional solver iteration, beating the accuracy of a classical simulation with more iterations per projection step.} Comparison of the runtime performance of the model indicates that the runtime at the grid sizes used to generate results for this work is similar to that of multigrid, though with superior accuracy compared to a single cycle. Theoretical speedups up to 5x were observed when using the model with larger inputs. Hence, our model provides a solid foundation to substantially accelerate the solution of the Poisson equation.

Based on the successes showcased in this work, in future work we intend to improve our model by incorporating the following features:

\textbf{Predicting on larger, previously unseen grids} By fixing the scaling issue demonstrated in the predictions made by the model when encountering larger grids than previously seen, the need for training on a specific range of grid sizes may be completely eliminated.

\textbf{BC combinations} Many applications require solving the Poisson equation with different combinations of BCs. Though all-Dirichlet and all-Neumann cases were investigated in this work, one possible approach for tackling more complicated combinations can be adopting a strategy similar to the one outlined in \secref{sec:bc_strategy}, but since the BCs must retain their respective types in the homogeneous problem as well, having multiple models to solve the homogeneous problem with each combination of BCs possible. The number of separate models can be minimized by grouping BC combinations that are identical under rotations/reflections.

\textbf{3D grids} The work presented focused on less memory- and time-intensive 2D problems, but most problems of interest in CFD are 3D. It is not expected that vast architectural changes will be necessary for a 3D version of this work since most convolutions {etc.\ }can be easily made 3D, however the greater number of parameters and much larger inputs can make training difficult and more memory intensive, necessitating work on model parallelism across GPUs.

\textbf{Domain adaptation/transfer learning for different domain sizes} As seen in \figref{fig:tgv_gridsize_vs_rms}, currently the Poisson CNN model's predictions degrade in quality for grids larger than those encountered in training. This problem is analogous to the domain mismatch problem encountered in many deep learning applications such as computer vision, as explored in a survey by \cite{deep_visual_domain_adaptation}. Adapting transfer learning techniques can be a way to overcome this weakness by training a Poisson CNN model on a very diverse range of data and then fine-tuning for more specific domain ranges as necessary for increased accuracy.

\textbf{Objects inside domain} Currently our model works only with BCs imposed on the edges of a rectangular domain. Objects inside the domain are commonly encountered in CFD and electrodynamics, requiring imposition of BCs on the surface of these objects. A possible way to tackle this problem is adding a `mask' channel to the inputs with 1.0 values for points lying inside the objects placed in the domain and 0.0 values for points outside, similar to the approach in \cite{fluidnet}.

\textbf{Hyperparameter optimization} The \textit{hyperparameters} of an NN model such as kernel sizes and number of convolution layer channels play an important role in the accuracy, inference speed and training time of deep NNs. Choosing an optimal size for the model will be crucial to obtain maximum performance especially in comparison to multigrid on GPUs. Gaussian Process based Bayesian hyperparameter search or Tree-structured Parzen Estimators as explored by \cite{hyperparam_opt} are two systematic approaches common in literature for this task.

\section*{Acknowledgements}

\paragraph{Funding statement}
This work was supported by a PhD studentship funded by the Department of Aeronautics, Imperial College London.

\paragraph{Code}
The code to replicate the results of this article may be found \textcolor{blue}{\href{https://github.com/aligirayhanozbay/poisson_CNN}{on Github}}

\clearpage

\printbibliography

\appendix
\section{Higher aspect ratio examples}
\label{sec:appendix_mesh}
{Below are further examples obtained from a HPNN model, trained with a range of aspect ratios ranging from 0.25 to 4.0. Despite slightly lower performance compared to the models the results of which are investigated in \secref{sec:tset_hpnn}, overall performance is still high, demonstrating that our model can adapt to a wide range of aspect ratios similar to those commonly encountered in many applications like computational fluid dynamics.}

\begin{figure}[h!]
    \centering
		\includegraphics[width=0.495\textwidth]{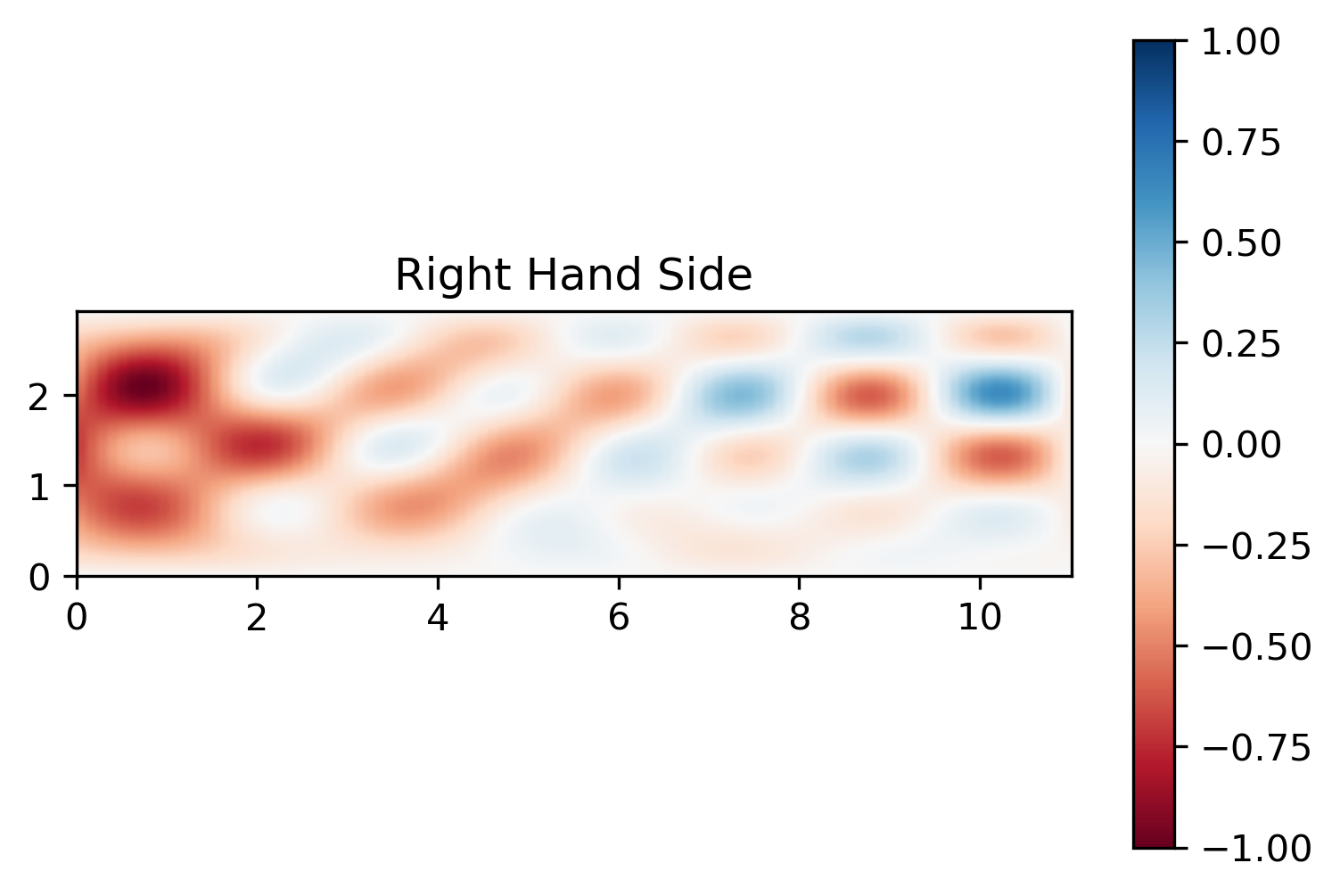}
		\includegraphics[width=0.495\textwidth]{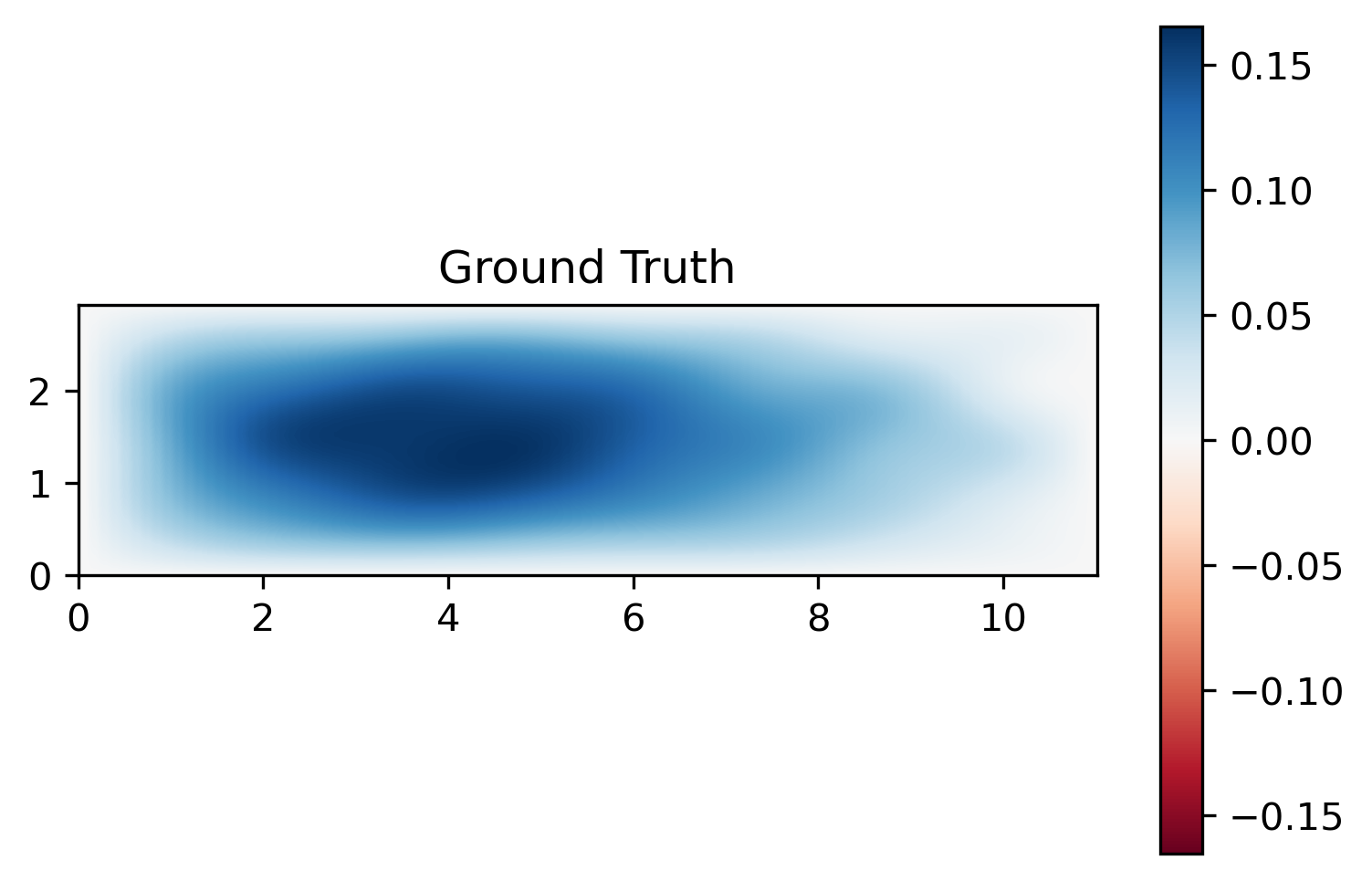}
		\includegraphics[width=0.495\textwidth]{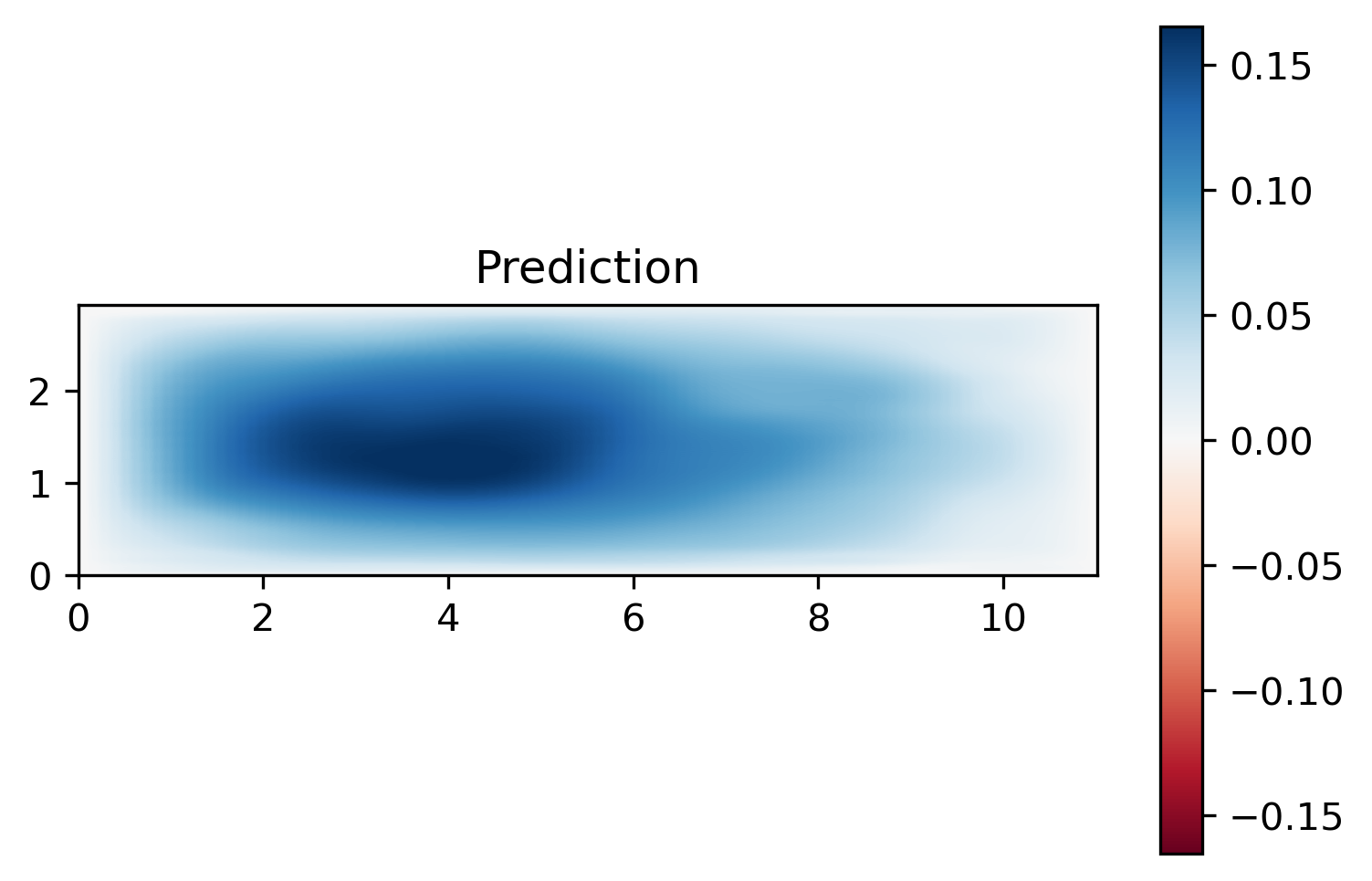}
		\includegraphics[width=0.495\textwidth]{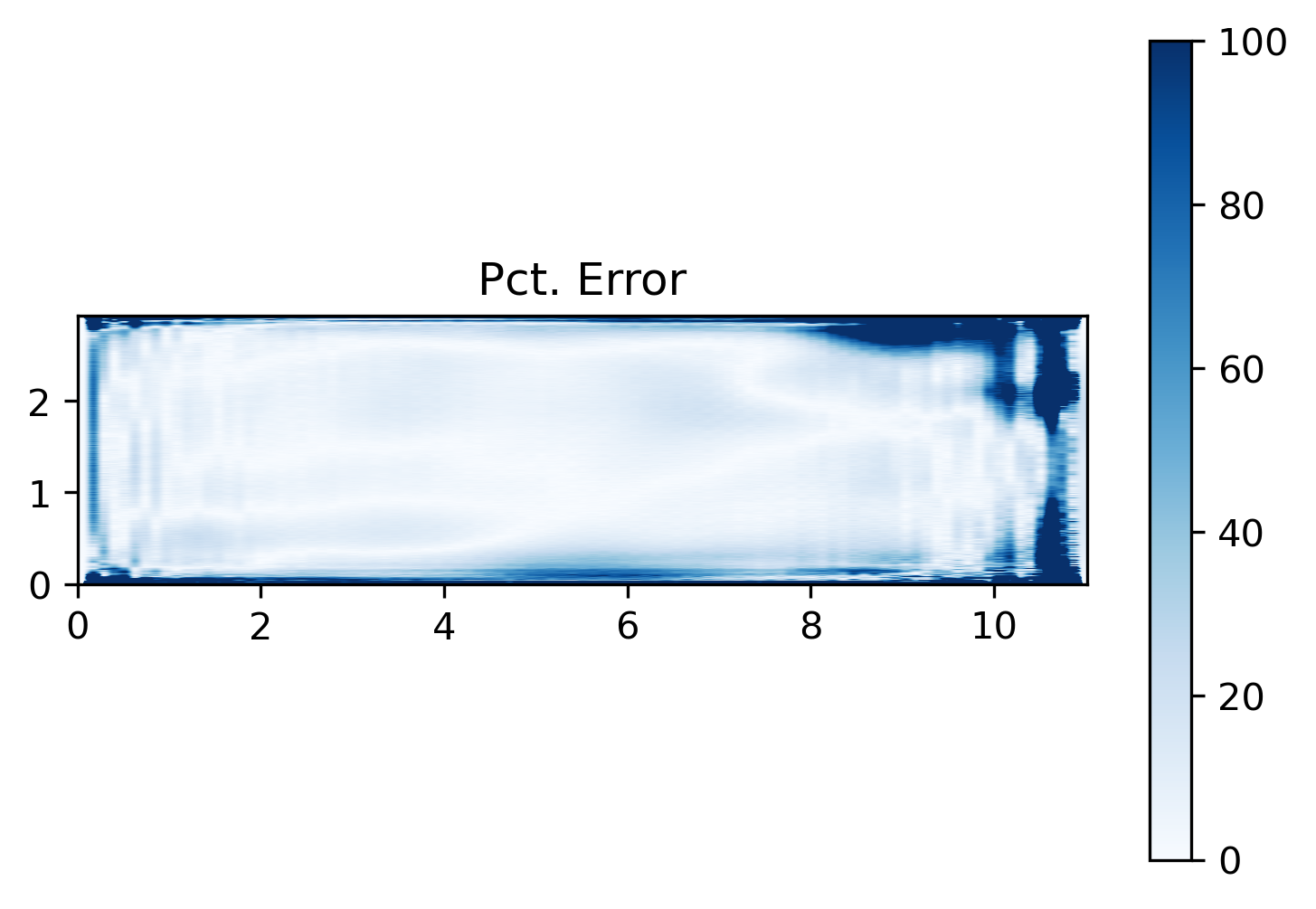}
    \caption{{Prediction of a Homogeneous Poisson NN model on an example with grid size $384 \times 96$ and $\Delta = 2.64 \times 10^{-2}$. As shown, when trained on an appropriate dataset, the model architecture is capable of handling aspect ratios well above the results in \secref{sec:results}}}
    \label{fig:app_highar_hpnn_2}
\end{figure}

\begin{figure}[h!]
    \centering
		\includegraphics[width=0.35\textwidth]{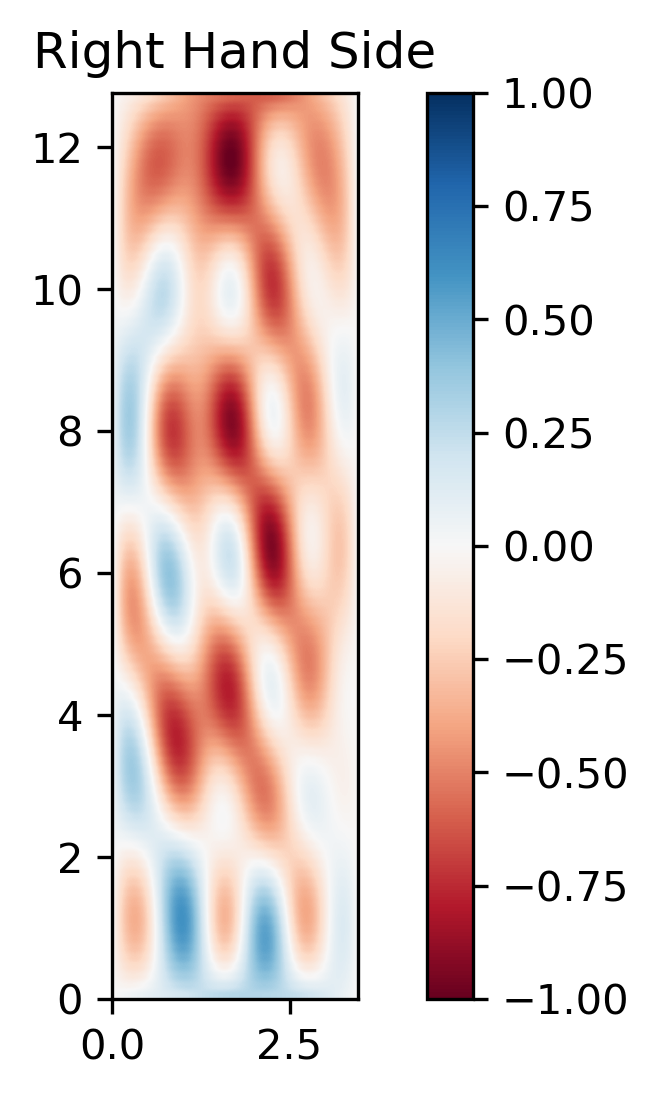}
		\includegraphics[width=0.35\textwidth]{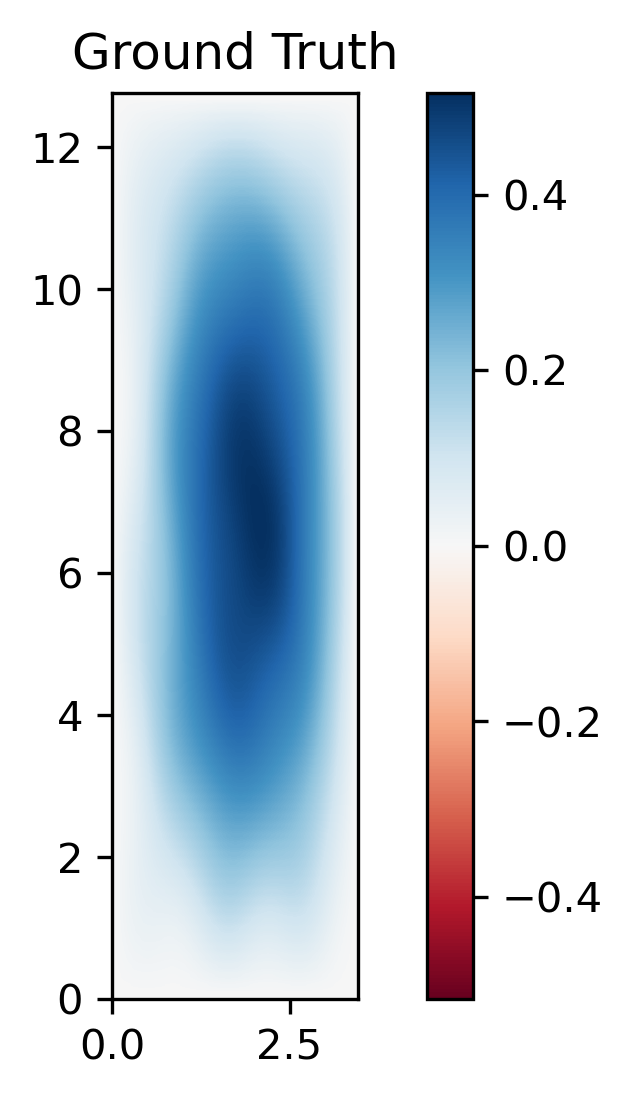}
		\includegraphics[width=0.35\textwidth]{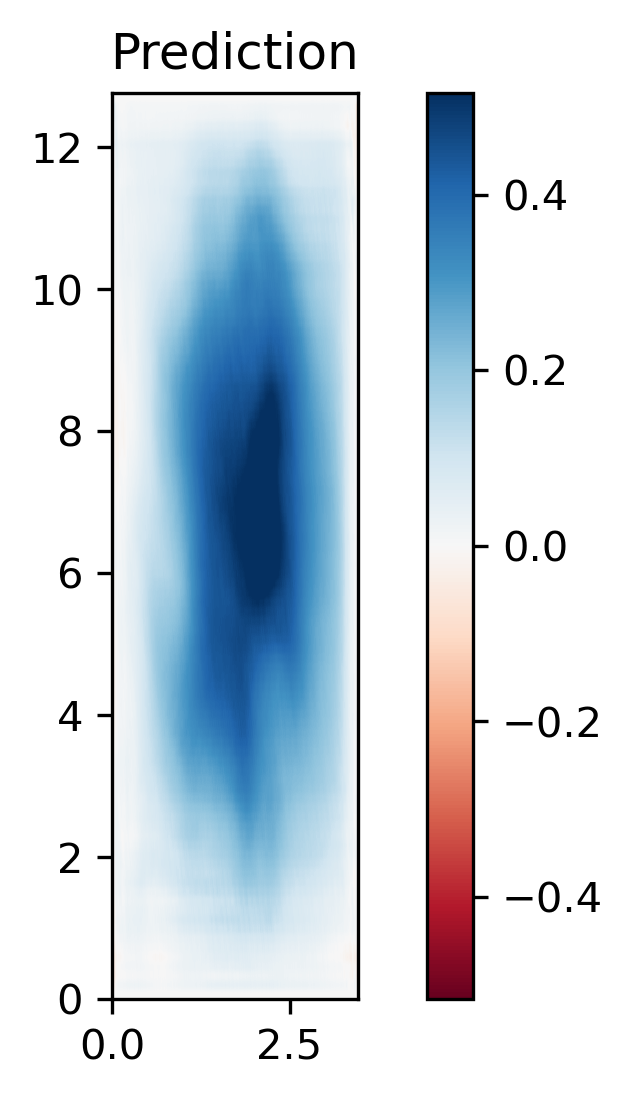}
		\includegraphics[width=0.35\textwidth]{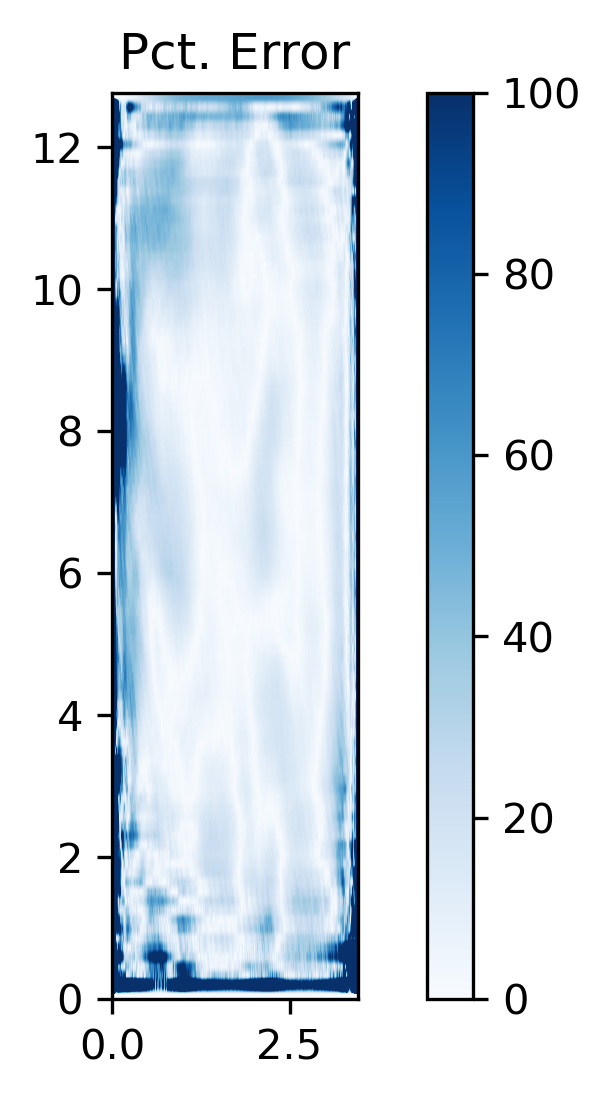}
    \caption{{Prediction of a Homogeneous Poisson NN model on an example with grid size $96 \times 384$ and $\Delta = 3.26 \times 10^{-2}$. The same model can handle domains that have low aspect ratios as well as those with large ones}}
    \label{fig:app_highar_hpnn_1}
\end{figure}

\end{document}